\documentclass[a4paper,11pt]{article}
\usepackage{jcappub} 
\usepackage{float}
\usepackage{subcaption}
\usepackage[none]{hyphenat}
\title{\boldmath Semi-Analytical Jeans Filtering Functions for Baryonic Density and Velocity Fields in Second-Order Cosmological Perturbation Theory}

\author[a,b]{Diego Fernando Fonseca,}
\author[a]{Leonardo Castañeda,}
\author[c]{and Luz Ángela García}
\affiliation[a]{Observatorio Astronómico Nacional, Universidad Nacional de Colombia, Carrera 45 No. 26-85, Bogotá 111321, Colombia.}
\affiliation[b]{Universidad Antonio Nariño, Calle 58A Bis No. 37-94, Bogotá 111321, Colombia.}
\affiliation[c]{Universidad ECCI, Cra. 19 No. 49-20, Bogotá 111311, Colombia.}

\emailAdd{dffonsecam@unal.edu.co}
\abstract{Cosmological perturbation theory provides the fundamental framework for describing the evolution of the matter-energy density field in an expanding Universe and serves as the basis for understanding the formation of large-scale structures within the $\Lambda$CDM paradigm. In this work, we present a semi-analytical approach to describe the evolution of fluctuations —small deviations from the mean density— in a mixed fluid composed of cold dark matter (CDM) and baryonic matter. Assuming that the Universe is governed by General Relativity (GR), we employ the Vlasov equation to derive the general equations of motion for this mixed cosmological fluid, incorporating baryonic effects through the stress tensor by considering only the contributions from baryonic pressure. We then introduce the Jeans Filtering Functions (JFF) as a biasing tool that allows us to describe baryonic fluctuations with CDM as a tracer, and obtain an semi-analytical description of the fluctuations —a novel and uncommon approach compared to the accepted computational advances currently available in this field. First- and second-order solutions are obtained using an semi-analytical method based on a single iteration of the equations of motion. These solutions can be evaluated at low computational cost to study the behavior of the filtering scale at second order relative to the linear case and its impact on the matter power spectrum, without computing it explicitly. Unlike previous works, which evaluate these effects at the level of the power spectrum, we rely solely on the fluctuation fields to quantify the impact of baryonic matter on the matter power spectrum. Specifically, we obtain semi-analytical expressions for baryonic fluctuations in the density and velocity fields, with our most significant result concerning to the velocity divergence field. These semi-analytical results represent a main contribution of this work, are readily evaluable, and provide valuable physical insight into the role of baryons in the large-scale structure (LSS) of the Universe. They also reveal how pressure effects shift the filtering scale and how the inclusion of this component could influence parameters such as the filtering mass and the temperature of pressure-supported components.}

\begin{document}
\maketitle
\flushbottom
\section{Introduction}
The development of cosmological perturbation theory has advanced significantly, employing both analytical approaches \cite{Mukhanov:1992, Mukhanov:2005, Bernardeau:2002, Somogyi:2010} and sophisticated computational techniques \cite{Bernardeau:2002, Angulo:2022}, with the purpose of enhancing our understanding of the formation and evolution of the cosmic web. However, many of these efforts have been oriented toward describing the evolution of the matter–energy density field, considering cold dark matter (CDM) for the most part, and their contribution has been significant in shaping our description of the Universe within the $\Lambda$CDM model \cite{Goroff:1986, Makino:1992, Jain:1994, Bernardeau:2002, Carlson:2009, Dodelson:2021, Angulo:2022}. For example, the study of the evolution of fluctuations in the density field has provided an indispensable tool in cosmology, known as the linear power spectrum, which is constructed from the two-point correlation function \cite{Peebles:2020, Dodelson:2021, Huterer:2023}. This tool constitutes a major achievement in the field, as it closely matches high-precision observations and has proven highly effective in describing structure formation \cite{Bernardeau:2002, Tegmark:2002, Huterer:2023}. In this sense, many other works \cite{Shoji:2009, Somogyi:2010, Schmidt:2016, Desjacques:2018} have shown that the description of fluctuations includes additional species in the cosmological fluid, such as baryons, which play an important role in structure formation. On small scales, their presence becomes significant for modeling the evolution of structures in our Universe, such as galaxies \cite{Desjacques:2018}, and consequently their implications for observations \cite{Huterer:2023} and their connection to the standard cosmological model may be important within cosmological perturbation theory. Therefore, this work aims to improve our understanding of the role of baryonic matter in the evolution of fluctuations in the matter–energy density field, with special emphasis on the velocity divergence field. In particular, we aim to demonstrate how the fundamental equations of cosmological perturbation theory can be applied to a cosmological fluid composed of baryonic matter and CDM, and how baryonic fluctuations across different scales can be treated using a semi-analytical method originally developed by Makino and Sasaki (1992) for CDM only \cite{Makino:1992}, but extended here to include both components. This extension constitutes a fundamental and distinguishing aspect of our work and differentiates it from the study by Shoji and Komatsu (2009). Additionally, we aim to complement to the work developed by Shoji and Komatsu (2009), where they calculated the nonlinear power spectrum without neglecting the pressure gradient up to third order —showing it as originating from baryons or neutrinos \cite{Naoz:2005, Shoji:2009}— in the fluid equations. These terms, associated with pressure, have significant implications for the nonlinear matter power spectrum and, consequently, for quantities such as the inferred temperature of the intergalactic medium and the filtering mass \cite{Shoji:2009}, which represents a characteristic mass scale determining which baryonic matter fluctuations fail to grow due to pressure effects in an expanding Universe. In this work, however, we describe these behaviors through the evolution of fluctuations themselves —without explicitly computing the power spectrum— and within the framework of a second-order theory. In addition, we aim to provide a theoretical motivation, indicating that the Jeans filtering function is not merely a tool for modeling the component associated with baryonic pressure \cite{Somogyi:2010}, but rather it allows us to extend our description of scale-dependent density bias parameters \cite{Schmidt:2016, Desjacques:2018}.

Therefore, in this work, we demonstrate how semi-analytical tools can be employed to describe second-order fluctuations in CDM and baryonic matter at low computational cost in both the density and velocity fields. The integrals involved can be readily evaluated using standard numerical methods, thereby allowing us to infer their potential effects on the matter and velocity power spectra without explicitly computing them. As a novel and distinctive aspect, we introduce a combined analytical–numerical approach to compute the baryonic velocity divergence filtering function, $h^{(0)}_{_2}(k)$ and $h^{(0)}_{_{1+2}}(k)$, revealing a significant shift in the Jeans scale. Another noteworthy feature of this work is its step-by-step analytical treatment, which distinguishes it from previous contributions and renders the analysis self-contained and readily reproducible.

This work is organized as follows: In Section 2, we present the standard cosmological perturbation theory. In Section 3, we introduce the fundamental equations of a fluid composed of CDM and baryonic matter, describing these equations both in the linear regime and at second order, following \cite{Shoji:2009}. In Section 4, we provide a semi-analytical description of baryonic fluctuations, where we connect the approaches developed by \cite{Shoji:2009} and \cite{Makino:1992} to derive the equations that describe the evolution of the Jeans filtering functions for the matter density and velocity fields, and evaluate them at low computational cost. Section 5 discusses the results and their implications within a cosmological framework. Finally, the Appendices provide additional mathematical details useful for following the developments throughout the paper and for reproducing the results. 
\section{Cosmological Pertubation Theory}
The most important equation that describes structure formation is the Boltzmann equation \cite{Bernardeau:2002, Peebles:2020}. This relation represents the evolution of matter and radiation in an expanding universe through changes in the distribution function $f\equiv f(t, \boldsymbol{x}, \boldsymbol{p})$ in phase space,
\begin{equation}\tag{2.1}\label{E1}
	\dfrac{df}{dt} = \dfrac{\partial{f}}{\partial{t}}+\dfrac{\partial{f}}{\partial{x^{i}}}\dfrac{dx^{i}}{dt}+\dfrac{\partial{f}}{\partial{p^{i}}}\dfrac{dp^{i}}{dt}=c[f], 
\end{equation}
where the term  $c[f]$, represents interactions between the species that constitute the components of matter and radiation in the Universe \cite{Dodelson:2021, Huterer:2023}. If we assume that the rate of change of particles entering and leaving a volume element is zero, this is equivalent to stating that structure formation is governed by dark matter, since it is the dominant source of matter density and decouples from radiation earlier than baryons do \cite{Dodelson:2021}. Therefore, equation \eqref{E1} is known as the Vlasov equation, and it can be written as \cite{Bernardeau:2002},
\begin{equation}\tag{2.2}\label{E2}
	\dfrac{df}{dt}=\dfrac{\partial{f}}{\partial{t}}+\dfrac{d{\boldsymbol{x}}}{dt}\cdot \nabla_{_{\boldsymbol{x}}}f+\dfrac{d{\boldsymbol{p}}}{dt}\cdot \nabla_{_{\boldsymbol{p}}}f  =0. 
\end{equation}
We assume that particles in the cosmological fluid interact only through gravity. If we describe space-time using coordinates $(t,{\boldsymbol{x}})$, where $t$ is identified as the cosmological time and $\boldsymbol{x}$ are the comoving coordinates such that the proper separation between particles varies as ${\boldsymbol{r}}=a(t){\boldsymbol{x}}$, these are known as Eulerian coordinates \cite{Bernardeau:2002}. As a reminder $a(t)$ is the scale factor. Therefore, the Vlasov equation \eqref{E2} can be written as in \cite{Peebles:2020,Fonseca:2024}:
\begin{equation}\tag{2.3}\label{E3}
	\dfrac{\partial{f}}{\partial{t}}+\dfrac{{\boldsymbol{p}}}{ma^{2}(t)}\cdot \nabla_{_{\boldsymbol{x}}}f-m\nabla_{_{\boldsymbol{x}}}\phi\cdot \nabla_{_{\boldsymbol{p}}}f=0,
\end{equation}  
where $m$ is the mass of a CDM particle, and $\phi(t,{\boldsymbol{x}})$ is a generalized potential \cite{Bernardeau:2002,Peebles:2020}. Our goal is to describe the distribution of matter (both dark and baryonic) and its dynamics in terms of deviations from the homogeneous and isotropic universe model. Consequently, we use the fundamental equations of fluid dynamics, given in Appendix \ref{app1}. Thus, in proper time $\tau$, we obtain the continuity equation:
\begin{equation}\tag{2.4}\label{E4}
	\dfrac{\partial}{\partial{\tau}}{\delta(\tau,{\boldsymbol{x}})}+\nabla_{_{\boldsymbol{x}}}\cdot \bigg[\bigg(1+\delta(\tau,{\boldsymbol{x}})\bigg){\boldsymbol{u}}(\tau,{\boldsymbol{x}})\bigg] = 0, 
\end{equation}
and Euler's equation
\begin{equation}\tag{2.5}\label{E5}
	\dfrac{\partial}{\partial \tau} \boldsymbol{u}(\tau, \boldsymbol{x}) 
	+ \mathcal{H}(\tau)\, \boldsymbol{u}(\tau, \boldsymbol{x}) 
	+ \bigg( \boldsymbol{u}(\tau, \boldsymbol{x}) \cdot \nabla_{_{\boldsymbol{x}}} \bigg) \boldsymbol{u}(\tau, \boldsymbol{x})
	+ \nabla_{_{\boldsymbol{x}}} \phi(\tau, \boldsymbol{x}) 
	= -\dfrac{\nabla_{_{\boldsymbol{x}}} P[\rho(\tau, \boldsymbol{x})]}{\rho(\tau, \boldsymbol{x})},
\end{equation}
with $\boldsymbol{u}(\tau, \boldsymbol{x})$, $P(\rho)$ and $\delta(\tau,{\boldsymbol{x}})$ representing the peculiar velocity, the pressure (assuming a barotropic equation of state), and the local density contrast, respectively. The latter is defined as
\begin{equation}\tag{2.6}\label{E6}
	\delta(\tau,{\boldsymbol{x}}) = \dfrac{\rho(\tau,{\boldsymbol{x}})}{\overline{\rho}(\tau)}-1,
\end{equation} 
where $\rho(\tau,{\boldsymbol{x}})$ is the local cosmic density and $\overline{\rho}(\tau)$ is the mean cosmic density. As a reminder $\mathcal{H}(\tau)=H(t)a(t)$ is the conformal expansion time \cite{Bernardeau:2002}. Therefore, the set of motion equations for a fluid composed of dark (C) and baryonic (B) matter are \cite{Shoji:2009}:
\begin{align}
	\dfrac{\partial}{\partial{\tau}}\delta_{\mbox{\tiny{C}}}{(\tau,{\boldsymbol{x}})}+\nabla_{_{\boldsymbol{x}}}\cdot \bigg[\bigg(1+\delta_{\mbox{\tiny{C}}}{(\tau,{\boldsymbol{x}})}\bigg){\boldsymbol{u}}_{\mbox{\tiny{C}}}(\tau,{\boldsymbol{x}})\bigg] & = 0, \tag{2.7}\label{E7}\\
	\dfrac{\partial}{\partial{\tau}}\delta_{\mbox{\tiny{B}}}{(\tau,{\boldsymbol{x}})}+\nabla_{_{\boldsymbol{x}}}\cdot \bigg[\bigg(1+\delta_{\mbox{\tiny{B}}}{(\tau,{\boldsymbol{x}})}\bigg){\boldsymbol{u}}_{\mbox{\tiny{B}}}(\tau,{\boldsymbol{x}})\bigg] & = 0. \tag{2.8}\label{E8}
\end{align}
Here, equations \eqref{E7} and \eqref{E8} represent the continuity and Euler equations for CDM, given by:
\begin{equation}\tag{2.9}\label{E9}
	\dfrac{\partial}{\partial{\tau}}{\boldsymbol{u}_{\mbox{\tiny{C}}}(\tau,{\boldsymbol{x}})}+\mathcal{H}(\tau){\boldsymbol{u}}_{\mbox{\tiny{C}}}(\tau,{\boldsymbol{x}})+\bigg({\boldsymbol{u}}_{\mbox{\tiny{C}}}(\tau,{\boldsymbol{x}})\cdot \nabla_{_{\boldsymbol{x}}}\bigg){\boldsymbol{u}}_{\mbox{\tiny{C}}}(\tau,{\boldsymbol{x}})
	=-\nabla_{_{\boldsymbol{x}}}\phi(\tau,{\boldsymbol{x}}),
\end{equation}
and baryonic matter
\begin{equation}\tag{2.10}\label{E10}
	\dfrac{\partial}{\partial{\tau}}{\boldsymbol{u}}_{\mbox{\tiny{B}}}(\tau,{\boldsymbol{x}})+\mathcal{H}(\tau){\boldsymbol{u}}_{\mbox{\tiny{B}}}(\tau,{\boldsymbol{x}})+\bigg({\boldsymbol{u}}_{\mbox{\tiny{B}}}(\tau,{\boldsymbol{x}})\cdot \nabla_{_{\boldsymbol{x}}}\bigg){\boldsymbol{u}}_{\mbox{\tiny{B}}}(\tau,{\boldsymbol{x}}) +\dfrac{\nabla_{_{\boldsymbol{x}}}[P_{\mbox{\tiny{B}}}(\rho_{\mbox{\tiny{B}}})]}{\rho_{\mbox{\tiny{B}}}(\tau,{\boldsymbol{x}})} =-\nabla_{_{\boldsymbol{x}}}\phi(\tau,{\boldsymbol{x}}).
\end{equation}
It is important to highlight that Eqs. \eqref{E7}–\eqref{E10} are coupled to the Poisson equation
\begin{equation}\tag{2.11}\label{E11}
    \nabla^{2}_{_{\boldsymbol{x}}}\phi=\dfrac{3}{2}\mathcal{H}(\tau)\Omega(\tau)\delta(\tau,{\boldsymbol{x}}),
\end{equation}
through the generalized non-relativistic gravitational potential $\phi$, where $\Omega(\tau)$ is a dimensionless parameter of General Relativity. Throughout the following developments, we assume an Einstein–de Sitter cosmology. Therefore,
\begin{equation}\tag{2.12}\label{E12}
    \nabla^{2}_{_{\boldsymbol{x}}}\phi= \dfrac{3}{2}\mathcal{H}(\tau)\delta(\tau,{\boldsymbol{x}}).
\end{equation}
Here, the fluctuation field is described by
\begin{equation}\tag{2.13}\label{E13}
    \delta(\tau,{\boldsymbol{x}})=\dfrac{\overline{\rho}_{\mbox{\tiny C}}}{\rho_{\mbox{\tiny m}}}\delta_{\mbox{\tiny C}}(\tau,{\boldsymbol{x}})+\dfrac{\overline{\rho}_{\mbox{\tiny B}}}{\rho_{\mbox{\tiny m}}}\delta_{\mbox{\tiny B}}(\tau,{\boldsymbol{x}}),
\end{equation}
where $\overline{\rho}_{\mbox{\tiny C}}$ and $\overline{\rho}_{\mbox{\tiny B}}$ denote the mean densities of CDM and baryonic matter, respectively, following \cite{Somogyi:2010}. This framework belongs to cosmological perturbation theory, where density perturbations in a mixed fluid (composed of CDM and baryonic matter) evolve with respect to a homogeneous and isotropic universe. Equations \eqref{E7}–\eqref{E12} represent the starting point of our work.

\section{Equations of Motion in Fourier Space}

In this section, we present equations \eqref{E7} to \eqref{E10} in Fourier space, in order to describe the evolution of the density contrast in terms of small and large scales, and to introduce the Jeans filtering function (JFF) \cite{Gnedin:1998}. Then, the Fourier transform is defined as:
\begin{align}
	\mathcal{F}\{\tilde{\delta}({\boldsymbol{k}},\tau)\} & = \delta({\boldsymbol{x}},\tau) = \int\limits_{-\infty}^{\infty}\dfrac{d^{3}{\boldsymbol{k}}}{(2\pi)^{3}}e^{i{\boldsymbol{k}}\cdot {\boldsymbol{x}}}\tilde{\delta}({\boldsymbol{k}},\tau) \tag{3.1}\label{E14},\\
	\mathcal{F}\{\delta({\boldsymbol{k}},\tau)\} & = \tilde{\delta}({\boldsymbol{k}},\tau) = \int\limits_{-\infty}^{\infty}d^{3}{\boldsymbol{x}}e^{i{\boldsymbol{k}}\cdot {\boldsymbol{x}}}\delta({\boldsymbol{k}},\tau) \tag{3.2}\label{E15}.
\end{align}
Therefore, the equations of motion in Fourier space are given by the following expressions, derived in Appendix \ref{app2}.
\begin{align}
    &\dfrac{\partial}{\partial{\tau}}\tilde{\delta}_{\mbox{\tiny{C}}}({\boldsymbol{k}},\tau) +\tilde{\theta}_{\mbox{\tiny{C}}}({\boldsymbol{k}},\tau)\nonumber \\
    & \qquad = -\dfrac{1}{(2\pi)^{3}} \int\limits_{-\infty}^{\infty}\int\limits_{-\infty}^{\infty}d^{3}{\boldsymbol{k}}_{_1}d^{3}{\boldsymbol{k}}_{_2}\dfrac{{\boldsymbol{k}}\cdot {\boldsymbol{k}}_{_2}}{k_{_2}^{2}}\delta^{\mbox{\tiny{D}}}({\boldsymbol{k}}_{_1}+{\boldsymbol{k}}_{_2}-{\boldsymbol{k}})\tilde{\delta}_{\mbox{\tiny{C}}}({\boldsymbol{k}}_{_1},\tau)\tilde{\theta}_{\mbox{\tiny{C}}}({\boldsymbol{k}}_{_2},\tau) ,\tag{3.3}\label{E16}\\
	&\dfrac{\partial}{\partial{\tau}}\tilde{\delta}_{\mbox{\tiny{B}}}({\boldsymbol{k}},\tau) +\tilde{\theta}_{\mbox{\tiny{B}}}({\boldsymbol{k}},\tau) \nonumber\\
	& \qquad = -\dfrac{1}{(2\pi)^{3}}\int\limits_{-\infty}^{\infty}\int\limits_{-\infty}^{\infty}d^{3}{\boldsymbol{k}}_{_1}d^{3}{\boldsymbol{k}}_{_2}\dfrac{{\boldsymbol{k}}\cdot {\boldsymbol{k}}_{_2}}{k_{_2}^{2}}\delta^{\mbox{\tiny{D}}}({\boldsymbol{k}}_{_1}+{\boldsymbol{k}}_{_2}-{\boldsymbol{k}})\tilde{\delta}_{\mbox{\tiny{B}}}({\boldsymbol{k}}_{_1},\tau)\tilde{\theta}_{\mbox{\tiny{B}}}({\boldsymbol{k}}_{_2},\tau)\tag{3.4}\label{E17},
\end{align}
\begin{align}
	& \dfrac{\partial}{\partial{\tau}}\tilde{\theta}_{\mbox{\tiny{C}}}({\boldsymbol{k}},\tau) +\mathcal{H}(\tau)\tilde{\theta}_{\mbox{\tiny{C}}}({\boldsymbol{k}},\tau)+\dfrac{6}{\tau^{2}}\tilde{\delta}({\boldsymbol{k}},\tau) \nonumber\\
	& \qquad  = -\dfrac{1}{(2\pi)^{3}}\int\limits_{-\infty}^{\infty}\int\limits_{-\infty}^{\infty}d^{3}{\boldsymbol{k}}_{_1}d^{3}{\boldsymbol{k}}_{_2}k^{2}\dfrac{{\boldsymbol{k}_{_1}}\cdot {\boldsymbol{k}}_{_2}}{2k_{_1}^{2}k_{_2}^{2}}\delta^{\mbox{\tiny{D}}}({\boldsymbol{k}}_{_1}+{\boldsymbol{k}}_{_2}-{\boldsymbol{k}})\tilde{\theta}_{\mbox{\tiny{C}}}({\boldsymbol{k}}_{_1},\tau)\tilde{\theta}_{\mbox{\tiny{C}}}({\boldsymbol{k}}_{_2},\tau)\tag{3.5}\label{E18},\\
	& \dfrac{\partial}{\partial{\tau}}\tilde{\theta}_{\mbox{\tiny{B}}}({\boldsymbol{k}},\tau) +\mathcal{H}(\tau)\tilde{\theta}_{\mbox{\tiny{B}}}({\boldsymbol{k}},\tau)+\dfrac{6}{\tau^{2}}\tilde{\delta}({\boldsymbol{k}},\tau) \nonumber\\
	& \qquad  = -\dfrac{1}{(2\pi)^{3}}\int\limits_{-\infty}^{\infty}\int\limits_{-\infty}^{\infty}d^{3}{\boldsymbol{k}}_{_1}d^{3}{\boldsymbol{k}}_{_2}k^{2}\dfrac{{\boldsymbol{k}_{_1}}\cdot {\boldsymbol{k}}_{_2}}{2k_{_1}^{2}k_{_2}^{2}}\delta^{\mbox{\tiny{D}}}({\boldsymbol{k}}_{_1}+{\boldsymbol{k}}_{_2}-{\boldsymbol{k}})\tilde{\theta}_{\mbox{\tiny{B}}}({\boldsymbol{k}}_{_1},\tau)\tilde{\theta}_{\mbox{\tiny{B}}}({\boldsymbol{k}}_{_2},\tau)+c^{2}_{_s}(\tau)k^{2}\nonumber\\ 
	& \qquad  \times \bigg[\tilde{\delta}_{\mbox{\tiny{B}}}({\boldsymbol{k}},\tau)-\dfrac{1}{2(2\pi)^{3}}\int\limits_{-\infty}^{\infty}\int\limits_{-\infty}^{\infty}d^{3}{\boldsymbol{k}}_{_1}d^{3}{\boldsymbol{k}}_{_2}\delta^{\mbox{\tiny{D}}}({\boldsymbol{k}}_{_1}+{\boldsymbol{k}}_{_2}-{\boldsymbol{k}})\tilde{\delta}_{\mbox{\tiny{B}}}({\boldsymbol{k}}_{_1},\tau)\tilde{\delta}_{\mbox{\tiny{B}}}({\boldsymbol{k}}_{_2},\tau)+\dfrac{1}{3(2\pi)^{6}}\bigg. \nonumber\\
	& \qquad \bigg.\times \int\limits_{-\infty}^{\infty}\int\limits_{-\infty}^{\infty}\int\limits_{-\infty}^{\infty}d^{3}{\boldsymbol{k}}_{_1}d^{3}{\boldsymbol{k}}_{_2}d^{3}{\boldsymbol{k}}_{_3}\delta^{\mbox{\tiny{D}}}({\boldsymbol{k}}_{_1}+{\boldsymbol{k}}_{_2}+{\boldsymbol{k}}_{_3}-{\boldsymbol{k}})\tilde{\delta}_{\mbox{\tiny{B}}}({\boldsymbol{k}}_{_1},\tau)\tilde{\delta}_{\mbox{\tiny{B}}}({\boldsymbol{k}}_{_2},\tau)\tilde{\delta}_{\mbox{\tiny{B}}}({\boldsymbol{k}}_{_3},\tau)\bigg]. \tag{3.6}\label{E19}
\end{align}
Here, $\boldsymbol{k}=\boldsymbol{k}_{_1}+\boldsymbol{k}_{_2}$, and $\delta^{\mbox{\tiny D}}$ denotes the three-dimensional Dirac delta function. Following \cite{Shoji:2009}, we can obtain solutions to any order for this set of equations. In this work, we highlight an analytical approach to second order to study the evolution of the density contrast, using the method developed by \cite{Makino:1992}.

\subsection{Analytical Approach to First-Order Solutions}
To first order, the equations of motion from \eqref{E7} to \eqref{E10} are written as
\begin{align}
    \dfrac{\partial}{\partial{\tau}}\delta_{\mbox{\tiny{C}}}{(\tau,{\boldsymbol{x}})}+\nabla_{_{\boldsymbol{x}}}\cdot {\boldsymbol{u}}_{\mbox{\tiny{C}}}(\tau,{\boldsymbol{x}}) & = 0, \tag{3.7}\label{E20}\\
    \dfrac{\partial}{\partial{\tau}}\delta_{\mbox{\tiny{B}}}{(\tau,{\boldsymbol{x}})}+\nabla_{_{\boldsymbol{x}}}\cdot {\boldsymbol{u}}_{\mbox{\tiny{B}}}(\tau,{\boldsymbol{x}}) & = 0, \tag{3.8}\label{E21}\\
    \dfrac{\partial}{\partial{\tau}}{\boldsymbol{u}_{\mbox{\tiny{C}}}(\tau,{\boldsymbol{x}})}+\mathcal{H}(\tau){\boldsymbol{u}}_{\mbox{\tiny{C}}}(\tau,{\boldsymbol{x}})+\nabla_{_{\boldsymbol{x}}}\phi(\tau,{\boldsymbol{x}}) & = 0\tag{3.9}\label{E22},\\
    \dfrac{\partial}{\partial{\tau}}{\boldsymbol{u}}_{\mbox{\tiny{B}}}(\tau,{\boldsymbol{x}})+\mathcal{H}(\tau){\boldsymbol{u}}_{\mbox{\tiny{B}}}(\tau,{\boldsymbol{x}}) +\dfrac{\nabla_{_{\boldsymbol{x}}}[P_{\mbox{\tiny{B}}}(\rho_{\mbox{\tiny{B}}})]}{\rho_{\mbox{\tiny{B}}}(\tau,{\boldsymbol{x}})} & = -\nabla_{_{\boldsymbol{x}}}\phi(\tau,{\boldsymbol{x}})\tag{3.10}\label{E23}.
\end{align}
By applying the divergence to the Euler equations, we find, in real space,
\begin{align}
    \dfrac{\partial}{\partial{\tau}}\theta_{\mbox{\tiny C}}(\tau,\boldsymbol{x})+\dfrac{2}{\tau}\theta_{\mbox{\tiny C}}(\tau,\boldsymbol{x}) +\nabla^{2}_{_{\boldsymbol{x}}}\phi & = 0, \tag{3.11}\label{E24}\\
    \dfrac{\partial}{\partial{\tau}}\theta_{\mbox{\tiny B}}(\tau,\boldsymbol{x})+\dfrac{2}{\tau}\theta_{\mbox{\tiny B}}(\tau,\boldsymbol{x}) +\nabla^{2}_{_{\boldsymbol{x}}}\phi & = -c_{_s}^{2}\nabla_{_{\boldsymbol{x}}}^{2}\delta_{\mbox{\tiny B}}(\tau,\boldsymbol{x}).\tag{3.12}\label{E25}
\end{align}
In Fourier space, the latter pair of equations can be written as:
\begin{align*}
    \dfrac{\partial}{\partial{\tau}}\tilde{\theta}_{\mbox{\tiny C}}(\boldsymbol{k},\tau)+\dfrac{2}{\tau}\tilde{\theta}_{\mbox{\tiny C}}(\boldsymbol{k},\tau)+\dfrac{6}{\tau^{2}}\tilde{\delta}(\boldsymbol{k},\tau) & = 0, \tag{3.13}\label{E26}\\
     \dfrac{\partial}{\partial{\tau}}\tilde{\theta}_{\mbox{\tiny B}}(\boldsymbol{k},\tau)+\dfrac{2}{\tau}\tilde{\theta}_{\mbox{\tiny B}}(\boldsymbol{k},\tau)+\dfrac{6}{\tau^{2}}\left[\tilde{\delta}(\boldsymbol{k},\tau)-\dfrac{k^{2}}{k^{2}_{_ J}}\tilde{\delta}_{\mbox{\tiny B}}(\boldsymbol{k},\tau)\right] & = 0. \tag{3.14}\label{E27}
\end{align*}
Here, we have used the Jeans wavenumber, defined as $k_{_J}(\tau)=\dfrac{\sqrt{6}}{c_{_s}(\tau)\tau}$ \cite{Shoji:2009}. It is obtained by equating the pressure term and the gravitational term; recall that we have assumed $\Omega(\tau)=1$, and it plays an important role in determining whether baryonic perturbations grow (growing modes) or oscillate (decaying modes). Then, to solve for the baryonic fluctuations, we introduce the Jeans filtering function.
\begin{equation}\tag{3.15}\label{E28}
    g(\boldsymbol{k},\tau) \equiv \dfrac{\tilde{\delta}_{\mbox{\tiny B}}(\boldsymbol{k},\tau)}{\tilde{\delta}_{\mbox{\tiny C}}(\boldsymbol{k},\tau)}.
\end{equation}
Assuming that CDM is the dominant source of gravity, $f_{\mbox{\tiny C}}=1$, and considering only the zeroth-order iteration. From equation \eqref{E26}, we find that CDM fluctuations scale as $\delta^{(0)}_{\mbox{\tiny C}}(\boldsymbol{k},\tau) \propto \tau^{2}$ for growing modes\footnote{The superscript $(0)$ denotes the zeroth-order iteration.}. Therefore, we obtain that the Jeans filtering function at zeroth order is \cite{Fonseca:2024, Shoji:2009}
\begin{equation*}\tag{3.16}\label{E29}
    g_{_1}^{(0)}(k,\tau) = \underbrace{c_{_1}\,\tau^{-\frac{5}{2}\left(1 + \sqrt{1 - \frac{24}{25}\left(1 + \frac{k^{2}}{k_{_J}^{2}}\right)}\right)}+ c_{_2}\,\tau^{-\frac{5}{2}\left(1 - \sqrt{1 - \frac{24}{25}\left(1 + \frac{k^{2}}{k_{_J}^{2}}\right)}\right)}}_{\mbox{Decaying mode}}+\underbrace{\dfrac{1}{1+\dfrac{k^{2}}{k^{2}_{_J}}}}_{_{\mbox{Growing mode}}},
\end{equation*}
where $c_{_1}$ and $c_{_2}$ are integrations constants. If we only consider the growing modes, baryonic fluctuations tend to zero when $k\gg k_{_J}$ and follow the same evolution as CDM when $k\ll k_{_J}$. Additional solutions can be found at any iteration order to first order \cite{Fonseca:2024}.

\subsection{Analytical Approach to Second-Order Solutions}
To obtain an analytical approach to second order, $n=2$, in the density field and velocity divergence, we introduce the Jeans filtering functions \cite{Shoji:2009}.
\begin{equation}\tag{3.17}\label{E30}
	g_{_n}(\boldsymbol{k},\tau) \equiv \dfrac{\tilde{\delta}_{_{n,\mbox{\tiny B}}}(\boldsymbol{k},\tau)}{\tilde{\delta}_{_{n,\mbox{\tiny C}}}(\boldsymbol{k},\tau)}, \hspace{0.8cm} h_{_n}(\boldsymbol{k},\tau) \equiv \dfrac{\tilde{\theta}_{_{n,\mbox{\tiny B}}}(\boldsymbol{k},\tau)}{\tilde{\theta}_{_{n,\mbox{\tiny C}}}(\boldsymbol{k},\tau)}.
\end{equation}
These functions trace the behavior of the baryonic density contrast and velocity divergence, through the functions $g_{_n}(\boldsymbol{k}, \tau)$ and $h_{_n}(\boldsymbol{k}, \tau)$, relative to the corresponding perturbations in CDM. They can be interpreted as representing a bias between baryonic matter and CDM. If we assume that $g_{_n}(\boldsymbol{k}, \tau) \rightarrow 1$ at large scales, baryonic matter is gravitationally coupled to CDM, and the evolution of both fields is the same. Conversely, if $ 0< g_{_n}(\boldsymbol{k}, \tau) < 1$, near the Jeans scale and at small scales, fluctuations in baryonic matter may be comparable to those of CDM. A similar interpretation can be made for the velocity field through the function $h_{_n}(\boldsymbol{k},\tau)$. If we assume that CDM is the dominant source of gravity, and using the general solutions in Einstein–de Sitter cosmology, the equations of motion can be solved through the following perturbative expansion \cite{Bernardeau:2002, Shoji:2009, Scoccimarro:1996, Jain:1994},
\begin{align}
	\tilde{\delta}_{\mbox{\tiny C}}(\boldsymbol{k},\tau) & = \displaystyle{\sum_{n=1}^{\infty}}a^{n}(\tau)\tilde{\delta}_{_{n,\mbox{\tiny C}}}(\boldsymbol{k}), \tag{3.18}\label{E31}\\
	\tilde{\delta}_{\mbox{\tiny B}}(\boldsymbol{k},\tau) & = \displaystyle{\sum_{n=1}^{\infty}}a^{n}(\tau)\tilde{\delta}_{_{n,\mbox{\tiny B}}}(\boldsymbol{k}) = \displaystyle{\sum_{n=1}^{\infty}}a^{n}(\tau)g_{_n}(\boldsymbol{k},\tau)\tilde{\delta}_{_{n,\mbox{\tiny C}}}(\boldsymbol{k}), \tag{3.19}\label{E32}\\
	\tilde{\theta}_{\mbox{\tiny C}}(\boldsymbol{k},\tau) & = \displaystyle{\sum_{n=1}^{\infty}}\dot{a}(\tau)a^{n-1}(\tau)\tilde{\theta}_{_{n,\mbox{\tiny C}}}(\boldsymbol{k}), \tag{3.20}\label{E33}\\
	\tilde{\theta}_{\mbox{\tiny B}}(\boldsymbol{k},\tau) & =
	\displaystyle{\sum_{n=1}^{\infty}}\dot{a}(\tau)a^{n-1}(\tau)\tilde{\theta}_{_{n,\mbox{\tiny B}}}(\boldsymbol{k})= \displaystyle{\sum_{n=1}^{\infty}}\dot{a}(\tau)a^{n-1}(\tau)h_{_n}(\boldsymbol{k},\tau)\tilde{\theta}_{_{n,\mbox{\tiny C}}}(\boldsymbol{k}). \tag{3.21}\label{E34}
\end{align}
Therefore, the equations for baryonic matter fluctuations, \eqref{E17} and \eqref{E19}, can be written as (see Appendix \ref{app3}):
\begin{align}
	& \sum_{n=1}^{\infty} \bigg\{ 
	\bigg[ n a^{n-1}(\tau)\dot{a}(\tau)g_{_n}(\boldsymbol{k},\tau) 
	+ a^{n}(\tau)\dot{g}(\boldsymbol{k},\tau) \bigg] 
	\delta_{_{n,\mbox{\tiny C}}}(\boldsymbol{k}) + \dot{a}(\tau) a^{n-1}(\tau) 
	\tilde{\theta}_{_{n,\mbox{\tiny C}}}(\boldsymbol{k}) h_{_n}(\boldsymbol{k},\tau) \bigg\} \nonumber \\
	& \qquad = -\dfrac{1}{(2\pi)^{3}} 
	\int\limits_{-\infty}^{\infty}\int\limits_{-\infty}^{\infty} d^{3}\boldsymbol{k}_{_1} d^{3}\boldsymbol{k}_{_2} \,
	\dfrac{\boldsymbol{k} \cdot \boldsymbol{k}_{_2}}{k_{_2}^{2}} 
	\delta^{\mbox{\tiny D}}(\boldsymbol{k}_{_1} + \boldsymbol{k}_{_2} - \boldsymbol{k}) 
	\nonumber \\
	& \qquad \times \sum_{\ell=1}^{\infty} \sum_{m=1}^{\infty} \bigg\{ a^{\ell + m - 1}(\tau) \dot{a}(\tau) 
	g_{_\ell}(\boldsymbol{k}_{_1},\tau) h_{_m}(\boldsymbol{k}_{_2},\tau)
	\tilde{\delta}_{_{\ell,\mbox{\tiny C}}}(\boldsymbol{k}_{_1}) 
	\tilde{\theta}_{_{m,\mbox{\tiny C}}}(\boldsymbol{k}_{_2}) \bigg\}. \tag{3.22}\label{E35}
\end{align}
In the case of the Euler equation, we obtain a similar expansion, which can be expressed as \cite{Shoji:2009}:
\begin{align}
	& \sum_{n=1}^{\infty} \bigg\{ 
	\bigg[\ddot{a}(\tau)a^{n-1}(\tau) + \dot{a}^{2}(\tau)a^{n-2}(\tau)(n-1)\bigg]
	\tilde{\theta}_{_{n,\mbox{\tiny C}}}(\boldsymbol{k})h_{_n}(\boldsymbol{k},\tau)+ \dot{a}(\tau)a^{n-1}(\tau)\tilde{\theta}_{_{n,\mbox{\tiny C}}}(\boldsymbol{k}) 
	\dot{h}_{_n}(\boldsymbol{k},\tau) \nonumber 
	\\ & \qquad + \dfrac{2}{\tau} \dot{a}(\tau)a^{n-1}(\tau)\tilde{\theta}_{_{n,\mbox{\tiny C}}}(\boldsymbol{k})h_{_n}(\boldsymbol{k},\tau) + \dfrac{6}{\tau^{2}} a^{n}(\tau) 
	\bigg[f_{\mbox{\tiny C}} + f_{\mbox{\tiny B}}g_{_n}(\boldsymbol{k},\tau)\bigg]
	\delta_{_{n,\mbox{\tiny C}}}(\boldsymbol{k}) \bigg\} \nonumber \\
	&  \qquad = -\dfrac{1}{(2\pi)^{3}}\int\limits_{-\infty}^{\infty}\int\limits_{-\infty}^{\infty} d^{3}\boldsymbol{k}_{_1} d^{3}\boldsymbol{k}_{_2} \,
	\delta^{\mbox{\tiny D}}(\boldsymbol{k}_{_1} + \boldsymbol{k}_{_2} - \boldsymbol{k}) 
	k^{2} \dfrac{\boldsymbol{k}_{_1} \cdot \boldsymbol{k}_{_2}}{2 k_{_1}^{2} k_{_2}^{2}}\sum_{\ell=1}^{\infty} \sum_{m=1}^{\infty} \bigg\{\dot{a}^{2}(\tau)a^{\ell + m - 2}(\tau) 
	\nonumber \\
	& \qquad \times 
	h_{_\ell}(\boldsymbol{k}_{_1},\tau) h_{_m}(\boldsymbol{k}_{_2},\tau)
	\tilde{\theta}_{_{\ell,\mbox{\tiny C}}}(\boldsymbol{k}_{_1}) 
	\tilde{\theta}_{_{m,\mbox{\tiny C}}}(\boldsymbol{k}_{_2}) \bigg\} + k^{2}c_{_s}^{2}(\tau) 
	\sum_{n=1}^{\infty} a^{n}(\tau) 
	\delta_{_{n,\mbox{\tiny C}}}(\boldsymbol{k})g_{_n}(\boldsymbol{k},\tau)- \dfrac{k^{2}c_{_s}^{2}(\tau)}{2(2\pi)^{3}}\nonumber \\
	& \qquad  \times \int\limits_{-\infty}^{\infty}\int\limits_{-\infty}^{\infty} d^{3}\boldsymbol{k}_{_1} d^{3}\boldsymbol{k}_{_2} \,
	\delta^{\mbox{\tiny D}}(\boldsymbol{k}_{_1} + \boldsymbol{k}_{_2} - \boldsymbol{k}) \sum_{\ell=1}^{\infty} \sum_{m=1}^{\infty} \bigg\{a^{\ell + m}(\tau) 
	\delta_{_{\ell,\mbox{\tiny C}}}(\boldsymbol{k}_{_1}) 
	\delta_{_{m,\mbox{\tiny C}}}(\boldsymbol{k}_{_2})\bigg.
	\nonumber \\
	& \qquad \times \bigg.g_{_\ell}(\boldsymbol{k}_{_1},\tau) 
	g_{_m}(\boldsymbol{k}_{_2},\tau) \bigg\}+ \dfrac{k^{2}c_{_s}^{2}(\tau)}{3(2\pi)^{6}} 
	\int\limits_{-\infty}^{\infty}\int\limits_{-\infty}^{\infty}\int\limits_{-\infty}^{\infty} d^{3}\boldsymbol{k}_{_1} d^{3}\boldsymbol{k}_{_2} d^{3}\boldsymbol{k}_{_3} \,
	\delta^{\mbox{\tiny D}}(\boldsymbol{k}_{_1} + \boldsymbol{k}_{_2} + \boldsymbol{k}_{_3} - \boldsymbol{k}) 
    \nonumber \\
	& \qquad \times \sum_{\ell=1}^{\infty} \sum_{m=1}^{\infty} \sum_{p=1}^{\infty} \bigg\{ a^{\ell + m + p}(\tau)\delta_{_{\ell,\mbox{\tiny C}}}(\boldsymbol{k}_{_1})
	\delta_{_{m,\mbox{\tiny C}}}(\boldsymbol{k}_{_2})\delta_{p,\mbox{\tiny C}}(\boldsymbol{k}_{_3})
	g_{_\ell}(\boldsymbol{k}_{_1},\tau) g_{_m}(\boldsymbol{k}_{_2},\tau)
	g_{_p}(\boldsymbol{k}_{_3},\tau) \bigg\}. \tag{3.23}\label{E36}
\end{align}
Following \cite{Naoz:2005, Shoji:2009}, we consider $c_{_s}^{2}=\dfrac{dP}{d\rho}$ and $c_{_s}=\dfrac{\sqrt{6}}{k_{_J}(\tau)\tau}$. However, to simplify the problem, we assume that  $k_{_J}(\tau)$ is independent of $\tau$ in order to extract physical insight into structure formation and the evolution of the baryonic density contrast; that is, we assume that the matter temperature evolves as if the matter were coupled to radiation \cite{ Shoji:2009}. Equations \eqref{E35} and \eqref{E36} describe the general behavior of baryonic matter fluctuations. In this work, we focus on second-order solutions for growing modes; therefore, we begin by recovering the first-order solution for the function $g_{_n}(\boldsymbol{k},\tau)$ \cite{Shoji:2009, Fonseca:2024}. Then, with $n=2$, we have (see Appendix \ref{app4}):
\begin{align}
	& \dot{g}_{_2}(\boldsymbol{k},\tau)\tilde{\delta}_{_{2,\mbox{\tiny C}}}(\boldsymbol{k})
	+ \dfrac{4}{\tau}g_{_2}(\boldsymbol{k},\tau)\tilde{\delta}_{_{2,\mbox{\tiny C}}}(\boldsymbol{k})
	+ \dfrac{2}{\tau}\tilde{\theta}_{_{2,\mbox{\tiny C}}}(\boldsymbol{k})h_{_2}(\boldsymbol{k},\tau) = \dfrac{2}{\tau} \dfrac{1}{(2\pi)^{3}} \nonumber \\
	& \quad \times
	\int\limits_{-\infty}^{\infty}\int\limits_{-\infty}^{\infty} d^{3}\boldsymbol{k}_{_1} d^{3}\boldsymbol{k}_{_2} \,
	\dfrac{\boldsymbol{k}\cdot\boldsymbol{k}_{_2}}{k_{_2}^{2}} 
	\delta^{\mbox{\tiny D}}(\boldsymbol{k}_{_1} + \boldsymbol{k}_{_2} - \boldsymbol{k})
	\, g_{_1}(\boldsymbol{k}_{_1}) g_{_1}(\boldsymbol{k}_{_2}) \tilde{\delta}_{_{1,\mbox{\tiny C}}}(\boldsymbol{k}_{_1})
	\tilde{\delta}_{_{1,\mbox{\tiny C}}}(\boldsymbol{k}_{_2})\equiv A_{_2}(\boldsymbol{k}), \tag{3.24}\label{E37} \\
	& \qquad\quad \dfrac{10}{\tau^{2}} \tilde{\theta}_{_{2,\mbox{\tiny C}}}(\boldsymbol{k}) h_{_2}(\boldsymbol{k},\tau)
	+ \dfrac{2}{\tau^{2}} \tilde{\theta}_{_{2,\mbox{\tiny C}}}(\boldsymbol{k}) \dot{h}_{_2}(\boldsymbol{k},\tau)
	+ \dfrac{6}{\tau^{2}} \tilde{\delta}_{_{2,\mbox{\tiny C}}}(\boldsymbol{k})
	- \dfrac{k^{2}}{k_{_J}^{2}} \tilde{\delta}_{_{2,\mbox{\tiny C}}}(\boldsymbol{k}) g_{_2}(\boldsymbol{k},\tau) \nonumber \\
	& = \dfrac{4}{\tau^{2}} \dfrac{1}{(2\pi)^{3}}
	\int\limits_{-\infty}^{\infty}\int\limits_{-\infty}^{\infty} d^{3}\boldsymbol{k}_{_1} d^{3}\boldsymbol{k}_{_2} \,
	\delta^{\mbox{\tiny D}}(\boldsymbol{k}_{_1} + \boldsymbol{k}_{_2} - \boldsymbol{k}) \nonumber\\
    & \qquad\quad \times\left[-\dfrac{3}{4} \dfrac{k^{2}}{k^{2}_{_J}}
	- k^{2} \dfrac{\boldsymbol{k}_{_1} \cdot \boldsymbol{k}_{_2}}{2k_{_1}^{2} k_{_2}^{2}} \right] g_{_1}(\boldsymbol{k}_{_1}) g_{_1}(\boldsymbol{k}_{_2})
	\tilde{\delta}_{_{1,\mbox{\tiny C}}}(\boldsymbol{k}_{_1}) 
	\tilde{\delta}_{_{1,\mbox{\tiny C}}}(\boldsymbol{k}_{_2}) \equiv B_{_2}(\boldsymbol{k})\tag{3.25} \label{E38}.
\end{align}
By combining the latter equations, we obtain,
\begin{equation}
	\ddot{g}_{_2}(\boldsymbol{k},\tau) +\dfrac{10}{\tau}\dot{g}_{_2}(\boldsymbol{k},\tau)+\dfrac{1}{\tau^{2}}\left[20+6\dfrac{k^{2}}{k^{2}_{_J}}\right]g_{_2}(\boldsymbol{k},\tau)
	-\dfrac{1}{\tau^{2}}\left[6+10\dfrac{A_{_2}(\boldsymbol{k})}{\tilde{\delta}_{_{2,\mbox{\tiny C}}}(\boldsymbol{k})}-\dfrac{4B_{_2}(\boldsymbol{k})}{\tilde{\delta}_{_{2,\mbox{\tiny C}}}(\boldsymbol{k})}\right] = 0. \tag{3.26}\label{E39}
\end{equation}
By solving the homogeneous version of the last equation, we obtain, where $c_{_1}$ and $c_{_2}$ are integrations constants $g_{_2}(\boldsymbol{k}, \tau)$ (see Appendix \ref{app5}),
\begin{align}
	g_{_2}^{(0)}(\boldsymbol{k},\tau) & = c_{_1}\, \tau^{-\frac{9}{2} \left(1 + \sqrt{1 - \frac{4}{81} \left(20 + 6 \frac{k^{2}}{k^{2}_{_J}} \right)}\right)} + c_{_2}\, \tau^{-\frac{9}{2} \left(1 - \sqrt{1 - \frac{4}{81} \left(20 + 6 \frac{k^{2}}{k^{2}_{_J}} \right)}\right)}, \tag{3.27}\label{E40}\\
    g_{_2}^{(0)}(\boldsymbol{k},\tau) & \propto \tau^{-\frac{9}{2} \left(1 \pm \sqrt{1 - \frac{4}{81} \left(20 + 6 \frac{k^{2}}{k^{2}_{_J}} \right)}\right)}\propto\mathcal{O}(\tau^{-9/2}).\tag{3.28}\label{E41}
\end{align}
The homogeneous solution for $g_{_2}(\boldsymbol{k},\tau)$, represents an oscillatory component that decays for any choice of $k/k_{_J}\geqslant 0$ \cite{Shoji:2009}. By incorporating the non-homogeneous term, we can show that the general solution for the JFF at second order for the density contrast is:
\begin{multline*}
	g_{_2}^{(0)}(\boldsymbol{k},\tau) = \dfrac{6+10\dfrac{A_{_2}(\boldsymbol{k})}{\tilde{\delta}_{_{2,\mbox{\tiny C}}}(\boldsymbol{k})}-\dfrac{4B_{_2}(\boldsymbol{k})}{\tilde{\delta}_{_{2,\mbox{\tiny C}}}(\boldsymbol{k})}}{20+6\frac{k^{2}}{k^{2}_{_J}}}
	+c_{_1}\, \tau^{-\frac{9}{2} \left(1 + \sqrt{1 - \frac{4}{81} \left(20 + 6 \frac{k^{2}}{k^{2}_{_J}} \right)}\right)} \\+ c_{_2}\, \tau^{-\frac{9}{2} \left(1 - \sqrt{1 - \frac{4}{81} \left(20 + 6 \frac{k^{2}}{k^{2}_{_J}} \right)}\right)},\tag{3.29}\label{E42}
\end{multline*}
with $\tilde{\delta}_{_{2,\mbox{\tiny C}}}(\boldsymbol{k})=\dfrac{1}{(2\pi)^{3}}\displaystyle{\int}d{\boldsymbol{q}}F_{_2}^{(s)}(\boldsymbol{q},\boldsymbol{k-\boldsymbol{q}})\tilde{\delta}_{_{1,\mbox{\tiny C}}}(\boldsymbol{q})\tilde{\delta}_{_{1,\mbox{\tiny C}}}(\boldsymbol{k}-\boldsymbol{q})$ \cite{Bernardeau:2002, Shoji:2009,Goroff:1986, Jain:1994}, where $F_{_2}$ is a symmetric, homogeneous function of the wave vectors $\{\boldsymbol{q}_{_1},\boldsymbol{q}_{_2}\}$ with degree zero\footnote{It is important to note that $\boldsymbol{k}_{_1}\equiv \boldsymbol{q}_{_1}$ and $\boldsymbol{k}_{_2}\equiv \boldsymbol{q}_{_2}$ in the second-order theory ($n=2$).}, constructed from the fundamental mode–coupling functions given in \eqref{E16} and \eqref{E18}, following \cite{Goroff:1986, Reimberg:2016}. 
\begin{equation}
    \alpha (\boldsymbol{k}_{_1},\boldsymbol{k}_{_2}) = \dfrac{\boldsymbol{k}\cdot \boldsymbol{k}_{_2}}{k_{_2}^{2}}, \hspace{0.5cm} \beta(\boldsymbol{k}_{_1},\boldsymbol{k}_{_2}) = k^{2}\dfrac{\boldsymbol{k}_{_1}\cdot \boldsymbol{k}_{_2}}{2k_{_1}^{2}k_{_2}^{2}}. \tag{3.30}\label{E43}
\end{equation}
Now, using the expressions in \eqref{E37} and \eqref{E38} applied to the particular solution, we obtain (see Appendix \ref{app5}):
\begin{equation}
	g_{_2}^{(0)}(\boldsymbol{k},\tau) = \underbrace{\mathcal{O}(\tau^{-9/2})}_{\mbox{Decaying mode}}+\underbrace{\dfrac{\dfrac{10}{3}-\dfrac{7}{3}\left[1-\dfrac{\tilde{\delta}'_{_{2,\mbox{\tiny C}}}(\boldsymbol{k})}{\tilde{\delta}_{_{2,\mbox{\tiny C}}}(\boldsymbol{k})}\right]}{\dfrac{10}{3}+\dfrac{k^{2}}{k^{2}_{_J}}}}_{\mbox{Growing mode}},\tag{3.31}\label{E44}
\end{equation}
where
\begin{equation}
    \tilde{\delta}'_{_{2,\mbox{\tiny C}}}(\boldsymbol{k}) = \dfrac{1}{(2\pi)^{3}}\int\limits_{-\infty}^{\infty}\int\limits_{-\infty}^{\infty} d^{3}\boldsymbol{k}_{_1} d^{3}\boldsymbol{k}_{_2}
	\delta^{\mbox{\tiny D}}(\boldsymbol{k}_{_1} + \boldsymbol{k}_{_2} - \boldsymbol{k})\mathcal{F}^{(s)}_{_2}(\boldsymbol{k}_{_1},\boldsymbol{k}_{_2})
    \tilde{\delta}_{_{1,\mbox{\tiny C}}}(\boldsymbol{k}_{_1})\tilde{\delta}_{_{1,\mbox{\tiny C}}}(\boldsymbol{k}_{_2}),\tag{3.32}\label{E45}
\end{equation}
and $\mathcal{F}^{(s)}_{_2}(\boldsymbol{k}_{_1},\boldsymbol{k}_{_2})\equiv \left[F_{_2}^{(s)}(\boldsymbol{q},\boldsymbol{k-\boldsymbol{q}})+\dfrac{3}{14}\dfrac{k^{2}}{k_{_J}^{2}}\right]g_{_1}(\boldsymbol{k}_{_1}) g_{_1}(\boldsymbol{k}_{_2})$. When $k_{\mbox{\tiny J}}\rightarrow \infty$, we have for the function $g_{_2}(\boldsymbol{k},\tau)\rightarrow 1$. For the Jeans filtering function of the velocity divergence field, we can demostrate that 
\begin{equation*}
    h_{_2}(\boldsymbol{k}) = \dfrac{A_{_2}(\boldsymbol{k})-2\tilde{\delta}_{_{2,\mbox{\tiny B}}}(\boldsymbol{k})}{\tilde{\theta}_{_{2,\mbox{\tiny C}}}(\boldsymbol{k})} = \dfrac{A_{_2}(\boldsymbol{k})-2\tilde{\delta}_{_{2,\mbox{\tiny C}}}(\boldsymbol{k})g_{_2}(\boldsymbol{k})}{\tilde{\theta}_{_{2,\mbox{\tiny C}}}(\boldsymbol{k})}.\tag{3.33}\label{E46}
\end{equation*}
It is possible to substitute \eqref{E32} and \eqref{E34} into \eqref{E17}, according to \cite{Jain:1994}. Then, by using \eqref{E30} together with the expression for $A(\boldsymbol{k})$ in \eqref{E35}, we obtain 
\begin{multline}
    h_{_2}^{(0)}(\boldsymbol{k}) = \dfrac{1}{\tilde{\theta}^{(0)}_{_{2,\mbox{\tiny C}}}(\boldsymbol{k})}\bigg[\dfrac{1}{(2\pi)^{3}}\int\limits_{-\infty}^{\infty}\int\limits_{-\infty}^{\infty} d^{3}\boldsymbol{k}_{_1} d^{3}\boldsymbol{k}_{_2}
	\delta^{\mbox{\tiny D}}(\boldsymbol{k}_{_1} + \boldsymbol{k}_{_2} - \boldsymbol{k}) \underbrace{\bigg[2F_{_2}(\boldsymbol{k}_{_1},\boldsymbol{k}_{_2})-G_{_2}(\boldsymbol{k}_{_1},\boldsymbol{k}_{_2})\bigg]}_{\boldsymbol{k}\cdot\boldsymbol{k}_{_2}/k_{_2}^{2}}\bigg.\\
    \bigg. \times g^{(0)}_{_1}(\boldsymbol{k}_{_1})g^{(0)}_{_1}(\boldsymbol{k}_{_2})\tilde{\delta}^{(0)}_{\mbox{\tiny 1,C}}(\boldsymbol{k}_{_1})\tilde{\delta}^{(0)}_{\mbox{\tiny 1,C}}(\boldsymbol{k}_{_2})\bigg]
    -2\dfrac{\tilde{\delta}^{(0)}_{_{2,\mbox{\tiny C}}}(\boldsymbol{k})}{\tilde{\theta}^{(0)}_{_{2,\mbox{\tiny C}}}(\boldsymbol{k})}g^{(0)}_{_2}(\boldsymbol{k}),\tag{3.34}\label{E47}
\end{multline}
where $G_{_2}$ has the same nature as the function $F_{_2}$ described previously. Therefore, using the symmetric forms of the kernels $F_{_2}$ and $G_{_2}$ \cite{Bernardeau:2002}, we can write $h_{_2}(\boldsymbol{k})$ as follows \cite{Shoji:2009}:
\begin{multline}
    h_{_2}^{(0)}(\boldsymbol{k}) = \dfrac{1}{\tilde{\theta}^{(0)}_{_{2,\mbox{\tiny C}}}(\boldsymbol{k})}\left[\dfrac{1}{(2\pi)^{3}}\int\limits_{-\infty}^{\infty}\int\limits_{-\infty}^{\infty} d^{3}\boldsymbol{k}_{_1} d^{3}\boldsymbol{k}_{_2}
	\delta^{\mbox{\tiny D}}(\boldsymbol{k}_{_1} + \boldsymbol{k}_{_2} - \boldsymbol{k}) \right.\\\left.\bigg[2F^{(s)}_{_2}(\boldsymbol{k}_{_1},\boldsymbol{k}_{_2})-G^{(s)}_{_2}(\boldsymbol{k}_{_1},\boldsymbol{k}_{_2})\bigg]g^{(0)}_{_1}(\boldsymbol{k}_{_1})g^{(0)}_{_1}(\boldsymbol{k}_{_2})\tilde{\delta}^{(0)}_{\mbox{\tiny 1,C}}(\boldsymbol{k}_{_1})\tilde{\delta}^{(0)}_{\mbox{\tiny 1,C}}(\boldsymbol{k}_{_2})\right]
    -2\dfrac{\tilde{\delta}^{(0)}_{_{2,\mbox{\tiny C}}}(\boldsymbol{k})}{\tilde{\theta}^{(0)}_{_{2,\mbox{\tiny C}}}(\boldsymbol{k})}g^{(0)}_{_2}(\boldsymbol{k}), \tag{3.35}\label{E48}
\end{multline}
with $2F^{(s)}_{_2}(\boldsymbol{k}_{_1},\boldsymbol{k}_{_2})-G^{(s)}_{_2}(\boldsymbol{k}_{_1},\boldsymbol{k}_{_2})\equiv 1+\dfrac{\boldsymbol{k}_{_1}\cdot\boldsymbol{k}_{_2}(\boldsymbol{k}_{_1}+\boldsymbol{k}_{_2})}{2\boldsymbol{k}_{_1}\boldsymbol{k}_{_2}}$. This pair of equations, \eqref{E44} and \eqref{E48}, allows us to describe the fluctuations in both the density and velocity fields at any scale. These equations serve as our starting point for applying the method proposed by Makino \cite{Makino:1992} to obtain semi-analytical solutions that can be computed at low computational cost, as we will show in the next section.

\section{Semi-Analytical Description of Baryonic Fluctuations}
In this section, we describe the behavior of the field $\tilde{\delta}_{_{2,\mbox{\tiny B}}}(\boldsymbol{k})$ using $\tilde{\delta}_{_{2,\mbox{\tiny C}}}(\boldsymbol{k})$ as a tracer at the zeroth-order iteration, through the JFF $g_{_2}(\boldsymbol{k},\tau)$. We then compare its evolution with the corresponding first-order solution, considering only the growing modes.
\begin{equation}
    \dfrac{\tilde{\delta}^{(0)}_{_{2,\mbox{\tiny B}}}(\boldsymbol{k})}{\tilde{\delta}^{(0)}_{_{2,\mbox{\tiny C}}}(\boldsymbol{k})} = g_{_2}^{(0)}(\boldsymbol{k}) = \dfrac{\dfrac{10}{3}-\dfrac{7}{3}\left(1-\dfrac{\tilde{\delta}'^{(0)}_{_{2,\mbox{\tiny C}}}(\boldsymbol{k})}{\tilde{\delta}^{(0)}_{_{2,\mbox{\tiny C}}}(\boldsymbol{k})}\right)}{\dfrac{10}{3}+\dfrac{k^{2}}{k^{2}_{_J}}}. \tag{4.1}\label{E49}
\end{equation}
Using the fact that fluctuations in the density field in the linear regime can be written as
$\tilde{\delta}^{(0)}_{_{1,\mbox{\tiny C}}}(\boldsymbol{k}) \propto \tau^{2}\delta_{_0}(\boldsymbol{k})\propto \tau^{2}\sqrt{P_{_0}(k)}\propto \sqrt{P_{\mbox{\tiny lin}}(k)}$,
where we use the linear power spectrum defined as $P_{\mbox{\tiny lin}}(k)=A k T^{2}(k)$, with $A=2.19\times 10^{4}\,\mbox{Mpc}^{4}$ and the transfer function for linear perturbations
$T(k)=\dfrac{\ln(1+9.36k)}{9.36k}\left[1+15.6k+(64.4k)^{2}+(21.8k)^{3}+(26.8k)^{4}\right]^{-1/4}$ \cite{Bardeen:1986, Jain:1994},
we can therefore write the following pair of equations for $\tilde{\delta}^{(0)}_{_{2,\mbox{\tiny C}}}(\boldsymbol{k})$ and $\tilde{\delta}'^{(0)}_{_{2,\mbox{\tiny C}}}(\boldsymbol{k})$:
\begin{align*}
    \tilde{\delta}^{(0)}_{_{2,\mbox{\tiny C}}}(\boldsymbol{k}) & \propto \dfrac{1}{(2\pi)^{3}}\int d\boldsymbol{q}
    F_{_2}^{(s)}(\boldsymbol{q}, \boldsymbol{k}-\boldsymbol{q})\sqrt{P_{\mbox{\tiny lin}}(q)}\sqrt{P_{\mbox{\tiny lin}}(\vert \boldsymbol{k}-\boldsymbol{q}\vert)}, \tag{4.2}\label{E50}\\
    \tilde{\delta}'^{(0)}_{_{2,\mbox{\tiny C}}}(\boldsymbol{k}) & \propto \dfrac{1}{(2\pi)^{3}} \int d\boldsymbol{q}
    \mathcal{F}_{_2}^{(s)}(\boldsymbol{q}, \boldsymbol{k}-\boldsymbol{q})\sqrt{P_{\mbox{\tiny lin}}(q)}\sqrt{P_{\mbox{\tiny lin}}(\vert \boldsymbol{k}-\boldsymbol{q}\vert)}. \tag{4.3}\label{E51}
\end{align*}
We employ the method proposed by \cite{Makino:1992} to treat the kernels $F_{_2}^{(s)}(\boldsymbol{q}, \boldsymbol{k}-\boldsymbol{q})$ and $\mathcal{F}_{_2}^{(s)}(\boldsymbol{q}, \boldsymbol{k}-\boldsymbol{q})$, as well as the linear power spectrum. Accordingly, expressions \eqref{E50} and \eqref{E51} can be rewritten as follows: first, introducing spherical coordinates for the wavevector $\boldsymbol{q}$, such that $\boldsymbol{q} = (q_{_1}, q_{_2}, q_{_3}) = (q \sin{\theta} \cos{\varphi},\, q \sin{\theta} \sin{\varphi},\, q \cos{\theta})$, the integral over the volume element $d\boldsymbol{q}$ can be expressed as \cite{Jain:1994}:
\begin{align*}
    \int d\boldsymbol{q} = \dfrac{1}{(2\pi)^{3}}\int\int\int dq_{_1}\,dq_{_2}\,dq_{_3}q^{2}\sin{\theta}\,dq\,d\theta \,d\varphi = \dfrac{1}{(2\pi)^{2}}\int\int q^{2}dq\, d(\cos{\theta}).
\end{align*}
As a second step, we rewrite the kernels in these coordinates. We choose the wavevector $\boldsymbol{k}$ to be aligned with the $z-$axes. For the mode $\boldsymbol{k}-\boldsymbol{q}$, we then have
\begin{align*}
    \boldsymbol{k}-\boldsymbol{q} & = (-q\sin{\theta}\cos{\varphi},-q\sin{\theta}\sin{\varphi}, k-q\cos{\theta}),\\
    \vert \boldsymbol{k}-\boldsymbol{q} \vert & = \sqrt{q^{2}+k^{2}-2kq\cos{\theta}} = k\sqrt{1+\left(\dfrac{q}{k}\right)^{2}-2\dfrac{q}{k}\cos{\theta}} = k\sqrt{1+r^{2}-2rx},\tag{4.4}\label{E52}
\end{align*}
where we have defined the new variables $r\equiv q/k$ and $\cos{\theta}\equiv x$ \cite{Makino:1992}. Thus, we can compute the kernel $F^{(s)}_{_2}(\boldsymbol{q}, \boldsymbol{k}-\boldsymbol{q})$,
\begin{align*}
    F^{(s)}_{_2}(\boldsymbol{q}, \boldsymbol{k}-\boldsymbol{q}) & = \dfrac{5}{7}+\dfrac{1}{2}\dfrac{\boldsymbol{q}\cdot (\boldsymbol{k}-\boldsymbol{q})}{q^{2}\vert \boldsymbol{k}-\boldsymbol{q}\vert ^{2}}\left[q^{2}+\vert \boldsymbol{k}-\boldsymbol{q}\vert^{2} \right]+\dfrac{2}{7}\dfrac{\left[\boldsymbol{q}\cdot (\boldsymbol{k}-\boldsymbol{q})\right]^{2}}{q^{2}\vert \boldsymbol{k}-\boldsymbol{q}\vert^{2}},\\
    & = \dfrac{5}{7} + \dfrac{1}{2}\dfrac{qkx-q^{2}}{q^{2}k^{2}(1+r^{2}-2rx)}\left[q^{2}+k^{2}(1+r^{2}-2rx)\right]+\dfrac{2}{7}\dfrac{[kqx-q^{2}]}{q^{2}k^{2}(1+r^{2}-2rx)},\\
    & = \dfrac{10qk^{2}(1+r^{2}-2rx)+(kx-q)\left[3q^{2}+7k^{2}+7k^{2}r^{2}-14k^{2}rx+4qkx\right]}{14qk^{2}(1+r^{2}-2rx)},\\
    F^{(s)}_{_2}(\boldsymbol{q}, \boldsymbol{k}-\boldsymbol{q}) & = \dfrac{10r+10r^{3}-20r^{2}x+(x-r)(10r^{2}+7-10rx)}{14r(1+r^{2}-2rx)},
\end{align*}
where $\boldsymbol{k}\cdot \boldsymbol{q}=qk\cos{\theta}=qkx$ and $q=rk$. Therefore,
\begin{equation} \tag{4.5}\label{E53}
    F^{(s)}_{_2}(\boldsymbol{q}, \boldsymbol{k}-\boldsymbol{q}) = \dfrac{3r+7x-10rx^{2}}{14r(1+r^{2}-2rx)}.
\end{equation}
Thus, the fluctuations can be described as follows, where the volume element has also been expressed in terms of the new variables,
\begin{align*}
    \tilde{\delta}^{(0)}_{_{2,\mbox{\tiny C}}}(\boldsymbol{k}) & \propto \dfrac{1}{(2\pi)^{2}}\int \int r^{2}k^{2}\,d(rk)\,dx\,\dfrac{3r+7x-10rx^{2}}{14r(1+r^{2}-2rx)}\sqrt{P_{\mbox{\tiny lin}}(kr)}\sqrt{P_{\mbox{\tiny lin}}(k\sqrt{1+r^{2}-2rx})},\\
    \tilde{\delta}^{(0)}_{_{2,\mbox{\tiny C}}}(\boldsymbol{k}) & \propto \dfrac{k^{4}A}{14(2\pi)^{2}}\int\limits_{0}^{\infty}\int\limits_{-1}^{1}dx\,dr\,  r^{3/2}\dfrac{3r+7x-10rx^{2}}{(1+r^{2}-2rx)^{3/4}}T(kr)T(k\sqrt{1+r^{2}-2rx}).\tag{4.6}\label{E54}
\end{align*}
For $\tilde{\delta}'^{(0)}_{_{2,\mbox{\tiny C}}}(\boldsymbol{k})$, we follow a similar treatment. We begin with the expression in \eqref{E51}:
\begin{align*}
    \tilde{\delta}'^{(0)}_{_{2,\mbox{\tiny C}}}(\boldsymbol{k}) & \propto \dfrac{1}{(2\pi)^{3}}\int d\boldsymbol{q}\left[F^{(s)}_{_2}(\boldsymbol{q}, \boldsymbol{k}-\boldsymbol{q}) +\dfrac{3}{14}\dfrac{k^{2}}{k^{2}_{_J}}\right]g^{(0)}_{_1}(q)g^{(0)}_{_1}(\vert \boldsymbol{k}-\boldsymbol{q}\vert)\sqrt{P_{\mbox{\tiny lin}}(q)}\sqrt{P_{\mbox{\tiny lin}}(\vert \boldsymbol{k}-\boldsymbol{q}\vert)}.
    \intertext{Using the growing solution to first order $g_{_1}^{(0)}(k)=\left[1+\dfrac{k^{2}}{k_{_J}^2}\right]^{-1}$,}
    \tilde{\delta}'^{(0)}_{_{2,\mbox{\tiny C}}}(\boldsymbol{k}) & \propto \dfrac{1}{(2\pi)^{3}}\int d\boldsymbol{q}\left[F^{(s)}_{_2}(\boldsymbol{q}, \boldsymbol{k}-\boldsymbol{q}) +\dfrac{3}{14}\dfrac{k^{2}}{k^{2}_{_J}}\right]\dfrac{1}{1+\dfrac{q^{2}}{k_{_J}^{2}}}\dfrac{1}{1+\dfrac{\vert \boldsymbol{k}-\boldsymbol{q}\vert^{2}}{k_{_J}^{2}}}\sqrt{P_{\mbox{\tiny lin}}(q)}\sqrt{P_{\mbox{\tiny lin}}(\vert \boldsymbol{k}-\boldsymbol{q}\vert)},\\
    & \propto \dfrac{1}{(2\pi)^{3}}\int d\boldsymbol{q}\left[F^{(s)}_{_2}(\boldsymbol{q}, \boldsymbol{k}-\boldsymbol{q}) +\dfrac{3}{14}\dfrac{k^{2}}{k^{2}_{_J}}\right]\dfrac{\sqrt{P_{\mbox{\tiny lin}}(kr)}\sqrt{P_{\mbox{\tiny lin}}(k\sqrt{1+r^{2}-2rx})}}{\left(1+r^{2}\dfrac{k^{2}}{k_{_J}^{2}}\right)\left(1+(1+r^{2}-2rx)\dfrac{k^{2}}{k_{_J}^{2}}\right)},\\
    & \propto \dfrac{1}{(2\pi)^{2}}\int\limits_{0}^{\infty}\int\limits_{-1}^{1}r^{2}k^{3}dx\,dr\,\dfrac{\left(\dfrac{3r+7x-10rx^{2}}{14r(1+r^{2}-2rx)}+\dfrac{3}{14}\dfrac{k^{2}}{k^{2}_{_J}}\right)\sqrt{P_{\mbox{\tiny lin}}(kr)}\sqrt{P_{\mbox{\tiny lin}}(k\sqrt{1+r^{2}-2rx})}}{\left(1+r^{2}\dfrac{k^{2}}{k_{_J}^{2}}\right)\left(1+(1+r^{2}-2rx)\dfrac{k^{2}}{k_{_J}^{2}}\right)}.
\end{align*}
Subsequently, defining the product $\overline{T}(k,r,x)\equiv T(kr)\ T\big(k\sqrt{1+r^{2}-2rx}\big)$, we can write
\begin{equation}\tag{4.7}\label{E55}
    \tilde{\delta}'^{(0)}_{_{2,\mbox{\tiny C}}}(k) \propto \dfrac{k^{4}A}{14(2\pi)^{2}}\int\limits_{0}^{\infty}\int\limits_{-1}^{1}dx\,dr\,\dfrac{\left(r^{3/2}\dfrac{3r+7x-10rx^{2}}{1+r^{2}-2rx}+3\dfrac{k^{2}}{k^{2}_{_J}}r^{5/2}\right)\overline{T}(k,r,x)}{\left(1+r^{2}\dfrac{k^{2}}{k_{_J}^{2}}\right)\left(1+(1+r^{2}-2rx)\dfrac{k^{2}}{k_{_J}^{2}}\right)}. 
\end{equation}
Therefore, the JFF at second order, equation \eqref{E49}, can be written as 
\begin{multline}
    g_{_2}^{(0)}(k)\propto \dfrac{1}{\dfrac{10}{3}+\dfrac{k^{2}}{k^{2}_{_J}}}+\dfrac{7}{3}\times\\ \dfrac{\displaystyle{\int\limits_{0}^{\infty}}\displaystyle{\int\limits_{-1}^{1}}dx\,dr\,\dfrac{\left(r^{3/2}\dfrac{3r+7x-10rx^{2}}{(1+r^{2}-2rx)^{3/4}}+3\dfrac{k^{2}}{k^{2}_{_J}}(1+r^{2}-2rx)^{1/4}r^{5/2}\right)\overline{T}(k,r,x)}{\left(1+r^{2}\dfrac{k^{2}}{k^{2}_{_J}}\right)\left(1+(1+r^{2}-2rx)\dfrac{k^{2}}{k_{_J}^{2}}\right)}}{\left(\dfrac{10}{3}+\dfrac{k^{2}}{k^{2}_{_J}}\right)\displaystyle{\int\limits_{0}^{\infty}}\displaystyle{\int\limits_{-1}^{1}}dx\,dr\,  r^{3/2}\dfrac{3r+7x-10rx^{2}}{(1+r^{2}-2rx)^{3/4}}\overline{T}(k,r,x)}.  \tag{4.8}\label{E56}
\end{multline}

\begin{figure}
    \centering
    \includegraphics[width=0.97\linewidth]{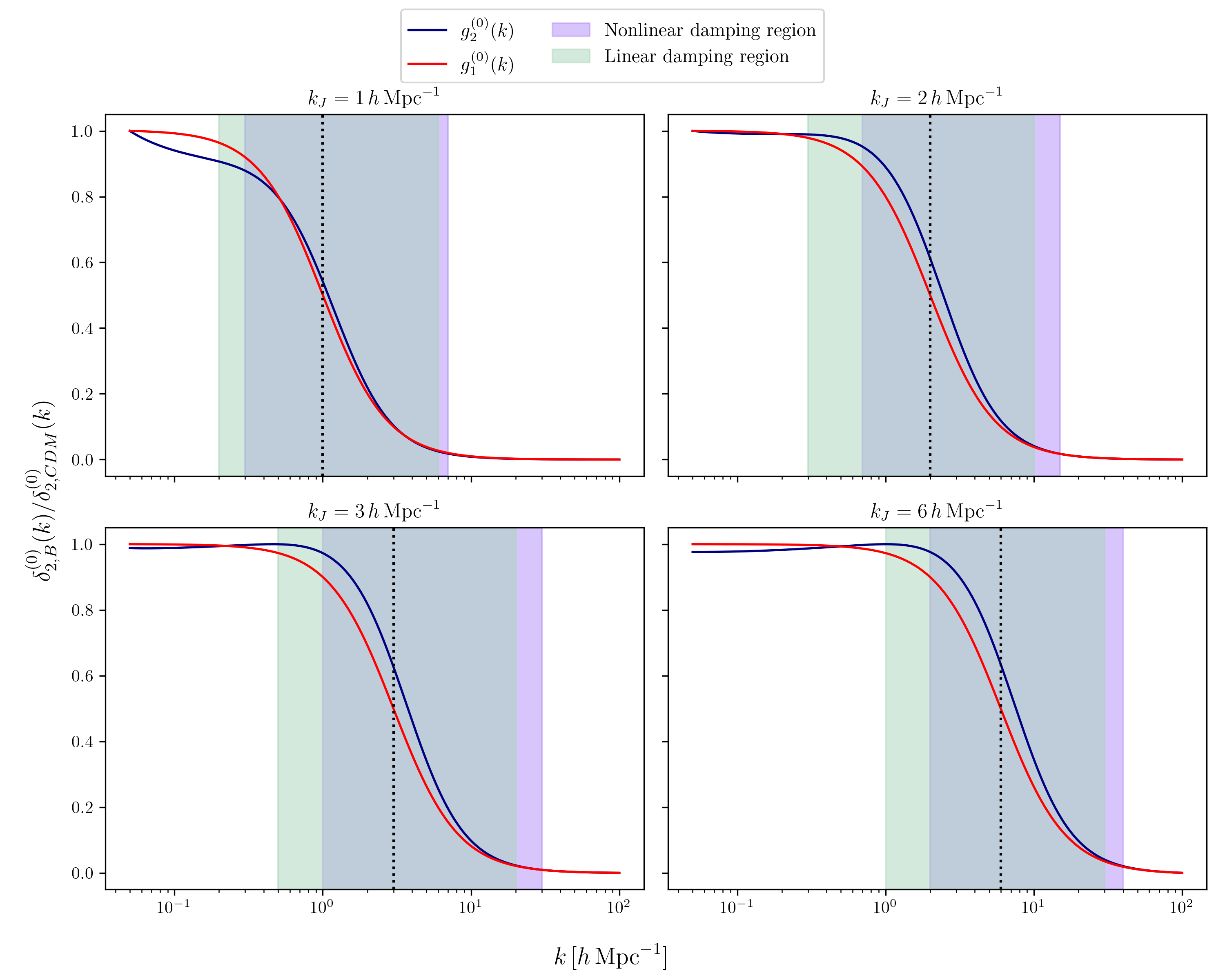}
    \caption{First- and second-order solutions for the baryonic filtering function in the density field at zeroth iteration. The red line corresponds to the first-order solution, $g^{(0)}_{_1}(k)$, while the blue line shows the second-order solution, $g^{(0)}_{_2}(k)$. The black dashed lines indicate the characteristic scale $k_{_J}$ for $k_{_J} = 1, 2, 3,$ and $6\, h\mathrm{Mpc}^{-1}$ in each panel. The shaded regions denote the linear (green) and nonlinear (purple) damping regimes.}
    \label{fig:1}
\end{figure}

The latter expression describes the behavior of baryonic matter fluctuations, using CDM fluctuations as a tracer only for the contribution to $\tilde{\delta}_{_2}(\boldsymbol{k},\tau)$. Thus, we are able to study their evolution at any scale, as shown in Figure~\ref{fig:1}, where we present the second-order solution for the evolution of the JFF, $g^{(0)}_{_2}(k)$, in the density field at zeroth iteration. 

\begin{figure}
    \centering
    \includegraphics[width=0.97\linewidth]{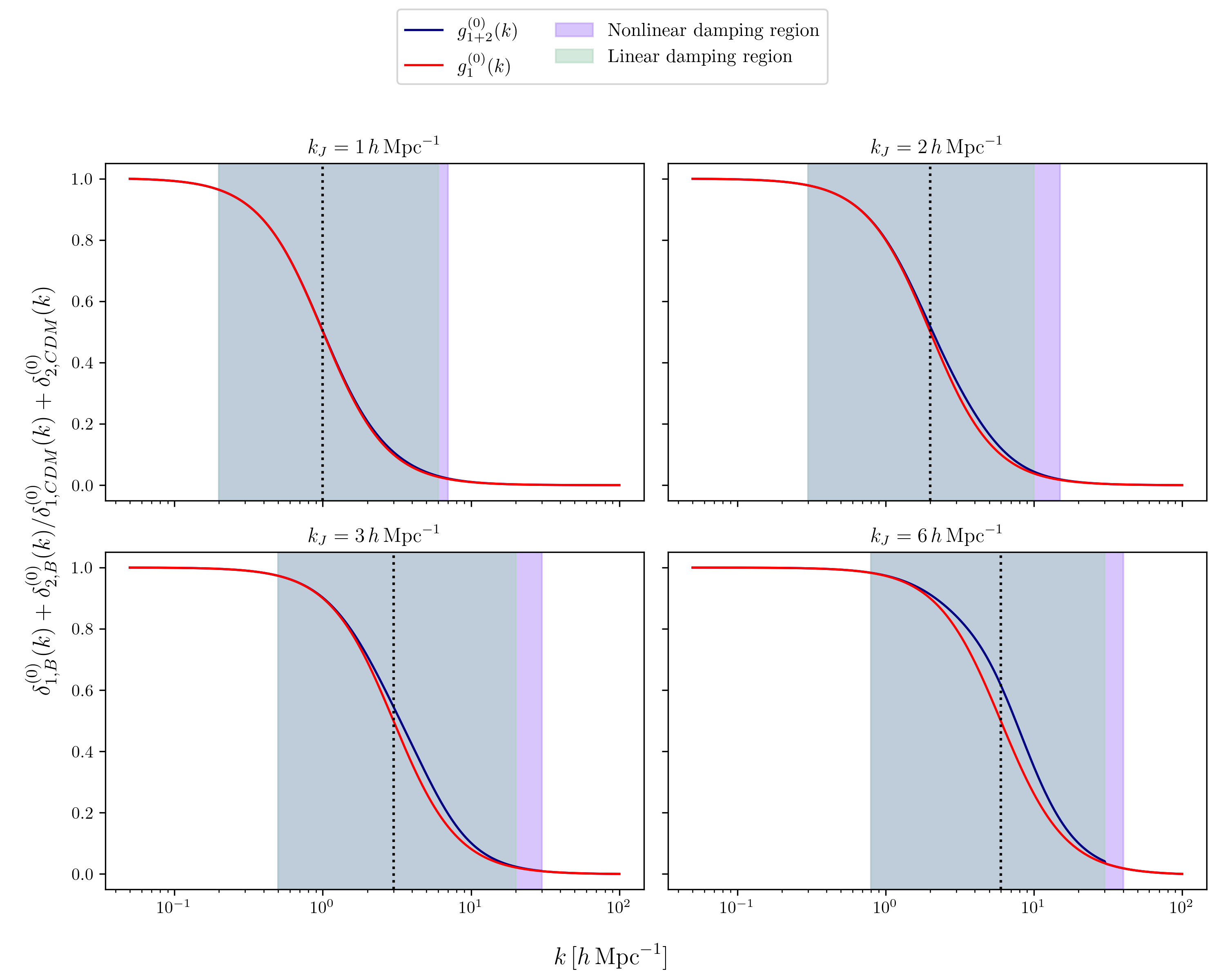}
    \caption{First- and second-order solutions for the baryonic filtering function in the density field at zeroth iteration. The red line corresponds to the first-order solution, $g^{(0)}_{_1}(k)$, while the blue line shows the second-order solution, $g^{(0)}_{_{1+2}}(k)$. The black dashed lines indicate the characteristic scale $k_{_J}$ for $k_{_J} = 1, 2, 3,$ and $6\, h\mathrm{Mpc}^{-1}$ in each panel. The shaded regions denote the linear (green) and nonlinear (purple) damping regimes.}
    \label{fig:2}
\end{figure}

As a preliminary assessment of the behavior of the contribution $g^{(0)}_{_2}(k)$, we focus on the maximum difference, $\vert g^{(0)}_{_2}(k)-g^{(0)}_{_1}(k)\vert$, between the solution for the growing modes in the linear regime, given by \eqref{E29} --red line--, and the second-order growing-mode solution, given by \eqref{E44} --blue line--, as shown in Figure \ref{fig:1}. This difference reaches approximately $5.8\,\%$ for $k_{_J}=1\ h\mathrm{Mpc}^{-1}$, which may imply a modest impact on the matter power spectrum and on the observables inferred from it, as also reported in \cite{Shoji:2009} for the total contribution to the density field. For $k_{_J} = 2, 3$, and $6\,h\mathrm{Mpc}^{-1}$, the difference remains at the $\sim 13\,\%$ level, indicating that as the filtering scale shifts toward smaller scales, its impact on the density field gradually increases, reaching approximately $13\text{--}15\,\%$. Although $\vert g^{(0)}_{_2}(k)-g^{(0)}_{_1}(k)\vert$ is not itself an observable and does not represent the total contribution, $\tilde{\delta}(\boldsymbol{k},\tau)=\tilde{\delta}_{_1}(\boldsymbol{k},\tau)+\tilde{\delta}_{_2}(\boldsymbol{k},\tau)$, it provides a direct and robust diagnostic of the scales at which nonlinear corrections to the filtering operator become relevant. Significant deviations signal the regime in which second-order mode couplings strongly modify the baryonic response and are therefore expected to leave a measurable imprint on the matter power spectrum. Therefore, to estimate its impact on the matter power spectrum through the JFF, we must compute the nonlinear density field,
\begin{align}
    \tilde{\delta}(\boldsymbol{k},\tau)=\tilde{\delta}_{_1}(\boldsymbol{k},\tau)+\tilde{\delta}_{_2}(\boldsymbol{k},\tau) \sim \dfrac{\tilde{\delta}_{_{1,\mbox{\tiny B}}}(\boldsymbol{k})+\tilde{\delta}_{_{2,\mbox{\tiny B}}}(\boldsymbol{k})}{\tilde{\delta}_{_{1,\mbox{\tiny C}}}(\boldsymbol{k})+\tilde{\delta}_{_{2,\mbox{\tiny C}}}(\boldsymbol{k})} \propto g_{_{1+2}}(\boldsymbol{k}). \tag{4.9}\label{E57}
\end{align}
Using \eqref{E17}, we can write
\begin{align}
    & g^{(0)}_{_{1+2}}(\boldsymbol{k}) \propto \dfrac{g^{(0)}_{_1}(\boldsymbol{k})\tilde{\delta}^{(0)}_{_{1,\mbox{\tiny C}}}(\boldsymbol{k})+g^{(0)}_{_2}(\boldsymbol{k})\tilde{\delta}^{(0)}_{_{2,\mbox{\tiny C}}}(\boldsymbol{k})}{\tilde{\delta}^{(0)}_{_{1,\mbox{\tiny C}}}(\boldsymbol{k})+\tilde{\delta}^{(0)}_{_{2,\mbox{\tiny C}}}(\boldsymbol{k})}, \nonumber \\
    & g^{(0)}_{_{1+2}}(k) \propto \dfrac{g^{(0)}_{_1}(k)\sqrt{P_{\mbox{\tiny lin}}(k)}+g^{(0)}_{_2}(k)C(k)\displaystyle{\int_{0}^{\infty}}\displaystyle{\int_{-1}^{1}}dxdr\,r^{3/2}\dfrac{3r+7x-10rx^{2}}{(1+r^{2}-2rx)^{3/4}}\overline{T}(k,r,x)}{\sqrt{P_{\mbox{\tiny lin}}(k)}+C(k)\displaystyle{\int_{0}^{\infty}}\displaystyle{\int_{-1}^{1}}dxdr\,r^{3/2}\dfrac{3r+7x-10rx^{2}}{(1+r^{2}-2rx)^{3/4}}\overline{T}(k,r,x)}
    \tag{4.10}\label{E58}
\end{align}
with $C(k)=k^{4}A/14(2\pi)^{2}$. This latter expression was obtained with the aid of the linear power spectrum, together with equations ~\eqref{E51} and \eqref{E53}, yielding the behavior shown in Figure~\ref{fig:2}. The corresponding difference, $\vert g^{(0)}_{_{1+2}}(k)-g^{(0)}_{_1}(k)\vert$, is refined to $0.08$, $3.47$, $6.32$ and $11.8\,\%$ for $k_{_J}=1, 2, 3$ and $6 \, h\mathrm{Mpc}^{-1}$, respectively. Finally, for the density field, as shown in Figures~\ref{fig:1} and \ref{fig:2}, indicate that on large scales ($k \ll k_{_J}$), baryonic fluctuations are strongly coupled to CDM, and the evolution of the density field is dominated by the dark matter component, with $g^{(0)}_{_{1+2}}(k) \approx 1$ $(\mbox{or} \, g^{(0)}_{_{2}}(k))\approx 1$, allowing density perturbations to grow without significant resistance. On scales comparable to $k_{_J}$, the baryonic matter contribution is non-negligible, and its effect on the total density contrast becomes significant. At these scales, the thermal pressure of the baryonic fluid becomes comparable to the gravitational attraction. The green and purple shaded regions represent the damping regimes, where baryonic pressure begins to counteract gravity, leading to the suppression of small-scale fluctuations in the linear and nonlinear solutions, respectively. Therefore, in the small-scale regime ($k \gg k_{_J}$), baryonic fluctuations asymptotically approach zero relative to the CDM fluctuations $g^{(0)}_{_{1+2}}(k) \approx 0$ $(\mbox{or} \, g^{(0)}_{_{2}}(k))\approx 0$.

At this stage, we emphasize that our analysis focuses on the divergence of the velocity field. Recall that, to first order, $h^{(0)}_{_1}(\boldsymbol{k}) = g^{(0)}_{_1}(\boldsymbol{k})$. By applying Makino’s method once again, we express the second-order velocity divergence, as established in equation \eqref{E47}, namely $\tilde{\theta}_{\mbox{\tiny 2,C}}(\boldsymbol{k}) = -\dfrac{1}{(2\pi)^{3}}\displaystyle{\int}d{\boldsymbol{k}_{_1}}d{\boldsymbol{k}_{_2}}\delta^{\mbox{\tiny D}}(\boldsymbol{k}_{_1}+\boldsymbol{k}_{_2}-\boldsymbol{k})G^{(s)}_{_2}(\boldsymbol{k}_{_1},\boldsymbol{k}_{_2})\delta_{\mbox{\tiny{1,C}}}(\boldsymbol{k}_{_1})\delta_{\mbox{\tiny{2,C}}}(\boldsymbol{k}_{_2})$ as discussed in \cite{Bernardeau:2002, Shoji:2009,Goroff:1986, Jain:1994}. Following the same procedure used for the kernel $F^{(s)}_{_2}(\boldsymbol{q},\boldsymbol{k}-\boldsymbol{q})$, shown in \eqref{E53}, we similarly obtain the corresponding expression for the kernel $G^{(s)}_{_2}(\boldsymbol{q},\boldsymbol{k}-\boldsymbol{q})$.
\begin{align*}
    G^{(s)}_{_2}(\boldsymbol{q}, \boldsymbol{k}-\boldsymbol{q}) & = \dfrac{3}{7}+\dfrac{1}{2}\dfrac{\boldsymbol{q}\cdot (\boldsymbol{k}-\boldsymbol{q})}{q^{2}\vert \boldsymbol{k}-\boldsymbol{q}\vert ^{2}}\left[q^{2}+\vert \boldsymbol{k}-\boldsymbol{q}\vert^{2} \right]+\dfrac{4}{7}\dfrac{\left[\boldsymbol{q}\cdot (\boldsymbol{k}-\boldsymbol{q})\right]^{2}}{q^{2}\vert \boldsymbol{k}-\boldsymbol{q}\vert^{2}},\\
     & = \dfrac{3}{7} + \dfrac{1}{2}\dfrac{qkx-q^{2}}{q^{2}k^{2}(1+r^{2}-2rx)}\left[q^{2}+k^{2}(1+r^{2}-2rx)\right]+\dfrac{4}{7}\dfrac{[kqx-q^{2}]}{q^{2}k^{2}(1+r^{2}-2rx)},\\
     G^{(s)}_{_2}(\boldsymbol{q}, \boldsymbol{k}-\boldsymbol{q}) & = \dfrac{42qk^{2}(1+r^{2}-2rx)+7(kx-q)\left[-q^{2}+7k^{2}+7k^{2}r^{2}-14k^{2}rx+8qkx\right]}{98qk^{2}(1+r^{2}-2rx)},
\end{align*}
therefore,
\begin{align*}
    G^{(s)}_{_2}(\boldsymbol{q}, \boldsymbol{k}-\boldsymbol{q}) & = \dfrac{6r+6r^{3}-12r^{2}x+(x-r)(6r^{2}+7-6rx)}{14r(1+r^{2}-2rx)}=\dfrac{7x-r-6rx^{2}}{14r(1+r^{2}-2rx)}.\tag{4.11}\label{E59}
\end{align*}
The features of the fluctuations are encoded in equations \eqref{E53} and \eqref{E59}. For this reason, in the same spirit as \eqref{E54}, the fluctuations in the velocity field can be described by
\begin{align*}
    \tilde{\theta}^{(0)}_{_{2,\mbox{\tiny C}}}(\boldsymbol{k}) & \propto -\dfrac{k^{4}A}{14(2\pi)^{2}}\int\limits_{0}^{\infty}\int\limits_{-1}^{1}dx\,dr\, r^{3/2}\,\dfrac{7x-r-6rx^{2}}{(1+r^{2}-2rx)^{3/4}}\overline{T}(k,r,x).\tag{4.12}\label{E60}
\end{align*}
Now, we evaluate the integral enclosed within $\left[\cdots\right]$ in \eqref{E48}. Using the solution for the growing modes $g^{(0)}_{_1}(k)$ and the linear-regime field $\tilde{\delta}{\mbox{\tiny 1,C}}(\boldsymbol{k})$, we obtain:
\begin{equation*}
    \dfrac{1}{(2\pi)^{3}}\int\limits_{-\infty}^{\infty}\int\limits_{-\infty}^{\infty} d^{3}\boldsymbol{k}_{_1} d^{3}\boldsymbol{k}_{_2}
	\delta^{\mbox{\tiny D}}(\boldsymbol{k}_{_1} + \boldsymbol{k}_{_2} - \boldsymbol{k})\bigg[1+\dfrac{\boldsymbol{k}_{_1}\cdot\boldsymbol{k}_{_2}(\boldsymbol{k}_{_1}+\boldsymbol{k}_{_2})}{2\boldsymbol{k}_{_1}\boldsymbol{k}_{_2}}\bigg]g_{_1}(\boldsymbol{k}_{_1})g_{_1}(\boldsymbol{k}_{_2})\tilde{\delta}_{\mbox{\tiny 1,C}}(\boldsymbol{k}_{_1})\tilde{\delta}_{\mbox{\tiny 1,C}}(\boldsymbol{k}_{_2})
\end{equation*}
\begin{align*}
& \propto \dfrac{1}{(2\pi)^{3}}\displaystyle{\int}d\boldsymbol{q}\left[1+\dfrac{\boldsymbol{q}\cdot (\boldsymbol{k}-\boldsymbol{q})}{2q^{2}\vert \boldsymbol{k}-\boldsymbol{q}\vert^{2}}\left[q^{2}+\vert \boldsymbol{k}-\boldsymbol{q}\vert^{2}\right]\right]\dfrac{\sqrt{P_{\mbox{\tiny lin}}(q)}}{1+\dfrac{q^{2}}{k_{_J}^{2}}}\dfrac{\sqrt{P_{\mbox{\tiny lin}}(\vert \boldsymbol{k}-\boldsymbol{q}\vert)}}{1+\dfrac{\vert \boldsymbol{k}-\boldsymbol{q}\vert^{2}}{k_{_J}^{2}}},\\
& \propto \dfrac{1}{(2\pi)^{3}}\displaystyle{\int}\displaystyle{\int}(kr)^{2}dq\,d(\cos{\theta})\,\dfrac{(r+x-2rx^{2})\sqrt{P_{\mbox{\tiny lin}}(kr)}\sqrt{P_{\mbox{\tiny lin}}(k\sqrt{1+r^{2}-2rx})}}{2r\left(1+r^{2}-2rx\right)\left(1+r^{2}\dfrac{k^{2}}{k_{_J}^{2}}\right)\left(1+(1+r^{2}-2rx)\dfrac{k^{2}}{k_{_J}^{2}}\right)},\\
& \propto \dfrac{k^{3}}{2(2\pi)^{2}}\int\limits_{0}^{\infty}\int\limits_{-1}^{1}dx\,dr\, r \dfrac{(r+x-2rx^{2})\sqrt{P_{\mbox{\tiny lin}}(kr)}\sqrt{P_{\mbox{\tiny lin}}(k\sqrt{1+r^{2}-2rx})}}{\left(1+r^{2}-2rx\right)\left(1+r^{2}\dfrac{k^{2}}{k_{_J}^{2}}\right)\left(1+(1+r^{2}-2rx)\dfrac{k^{2}}{k_{_J}^{2}}\right)},\\
& \propto \dfrac{k^{4}A}{2(2\pi)^{2}}\int\limits_{0}^{\infty}\int\limits_{-1}^{1}dx\,dr\, r^{3/2} \dfrac{(r+x-2rx^{2})\overline{T}(k,r,x)}{(1+r^{2}-2rx)^{3/4}\left(1+r^{2}\dfrac{k^{2}}{k_{_J}^{2}}\right)\left(1+(1+r^{2}-2rx)\dfrac{k^{2}}{k_{_J}^{2}}\right)}.
\end{align*}
Hence, by applying this latter expression in \eqref{E48},
\begin{multline*}
    h^{(0)}_{_2}(k) \propto \dfrac{\dfrac{k^{4}A}{2(2\pi)^{2}}\displaystyle{\int\limits_{0}^{\infty}}\displaystyle{\int\limits_{-1}^{1}}dx\,dr\,r^{3/2}\,\dfrac{(r+x-2rx^{2})\overline{T}(k,r,x)}{(1+r^{2}-2rx)^{3/4}\left(1+r^{2}\dfrac{k^{2}}{k_{_J}^{2}}\right)\left(1+(1+r^{2}-2rx)\dfrac{k^{2}}{k_{_J}^{2}}\right)}}{-\dfrac{k^{4}A}{14(2\pi)^{2}}\displaystyle{\int\limits_{0}^{\infty}}\displaystyle{\int\limits_{-1}^{1}}dx\,dr\, r^{3/2}\,\dfrac{7x-r-6rx^{2}}{(1+r^{2}-2rx)^{3/4}}\overline{T}(k,r,x)}\\
    -2\dfrac{\dfrac{k^{4}A}{14(2\pi)^{2}}\displaystyle{\int\limits_{0}^{\infty}}\displaystyle{\int\limits_{-1}^{1}}dx\,dr\,r^{3/2}\,\dfrac{3r+7x-10rx^{2}}{(1+r^{2}-2rx)^{3/4}}\overline{T}(k,r,x)}{-\dfrac{k^{4}A}{14(2\pi)^{2}}\displaystyle{\int\limits_{0}^{\infty}}\displaystyle{\int\limits_{-1}^{1}}dx\,dr\, r^{3/2}\,\dfrac{7x-r-6rx^{2}}{(1+r^{2}-2rx)^{3/4}}\overline{T}(k,r,x)}g^{(0)}_{_2}(k).\tag{4.13}\label{E61}
\end{multline*}
Simplifying and substituting \eqref{E56} into \eqref{E61}, we obtain,
\begin{multline}
     h^{(0)}_{_2}(k) \propto -7\dfrac{\displaystyle{\int\limits_{0}^{\infty}}\displaystyle{\int\limits_{-1}^{1}}dx\,dr\,r^{3/2}\,\dfrac{(r+x-2rx^{2})\overline{T}(k,r,x)}{(1+r^{2}-2rx)^{3/4}\left(1+r^{2}\dfrac{k^{2}}{k_{_J}^{2}}\right)\left(1+(1+r^{2}-2rx)\dfrac{k^{2}}{k_{_J}^{2}}\right)}}{\displaystyle{\int\limits_{0}^{\infty}}\displaystyle{\int\limits_{-1}^{1}}dx\,dr\, r^{3/2}\,\dfrac{7x-r-6rx^{2}}{(1+r^{2}-2rx)^{3/4}}\overline{T}(k,r,x)}\\
    +\dfrac{2}{\dfrac{10}{3}+\dfrac{k^{2}}{k^{2}_{_J}}}\dfrac{\displaystyle{\int\limits_{0}^{\infty}}\displaystyle{\int\limits_{-1}^{1}}dx\,dr\,r^{3/2}\,\dfrac{3r+7x-10rx^{2}}{(1+r^{2}-2rx)^{3/4}}\overline{T}(k,r,x)}{\displaystyle{\int\limits_{0}^{\infty}}\displaystyle{\int\limits_{-1}^{1}}dx\,dr\, r^{3/2}\,\dfrac{7x-r-6rx^{2}}{(1+r^{2}-2rx)^{3/4}}\overline{T}(k,r,x)} \times \\
    \left[1+\dfrac{7}{3}\dfrac{\displaystyle{\int\limits_{0}^{\infty}}\displaystyle{\int\limits_{-1}^{1}}dx\,dr\,\dfrac{\left(r^{3/2}\dfrac{3r+7x-10rx^{2}}{(1+r^{2}-2rx)^{3/4}}+3\dfrac{k^{2}}{k^{2}_{_J}}(1+r^{2}-2rx)^{1/4}r^{5/2}\right)\overline{T}(k,r,x)}{\left(1+r^{2}\dfrac{k^{2}}{k^{2}_{_J}}\right)\left(1+(1+r^{2}-2rx)\dfrac{k^{2}}{k_{_J}^{2}}\right)}}{\displaystyle{\int\limits_{0}^{\infty}}\displaystyle{\int\limits_{-1}^{1}}dx\,dr\,  r^{3/2}\dfrac{3r+7x-10rx^{2}}{(1+r^{2}-2rx)^{3/4}}\overline{T}(k,r,x)}\right].\tag{4.14}\label{E62}
\end{multline}

\begin{figure}
    \centering
    \includegraphics[width=0.97\linewidth]{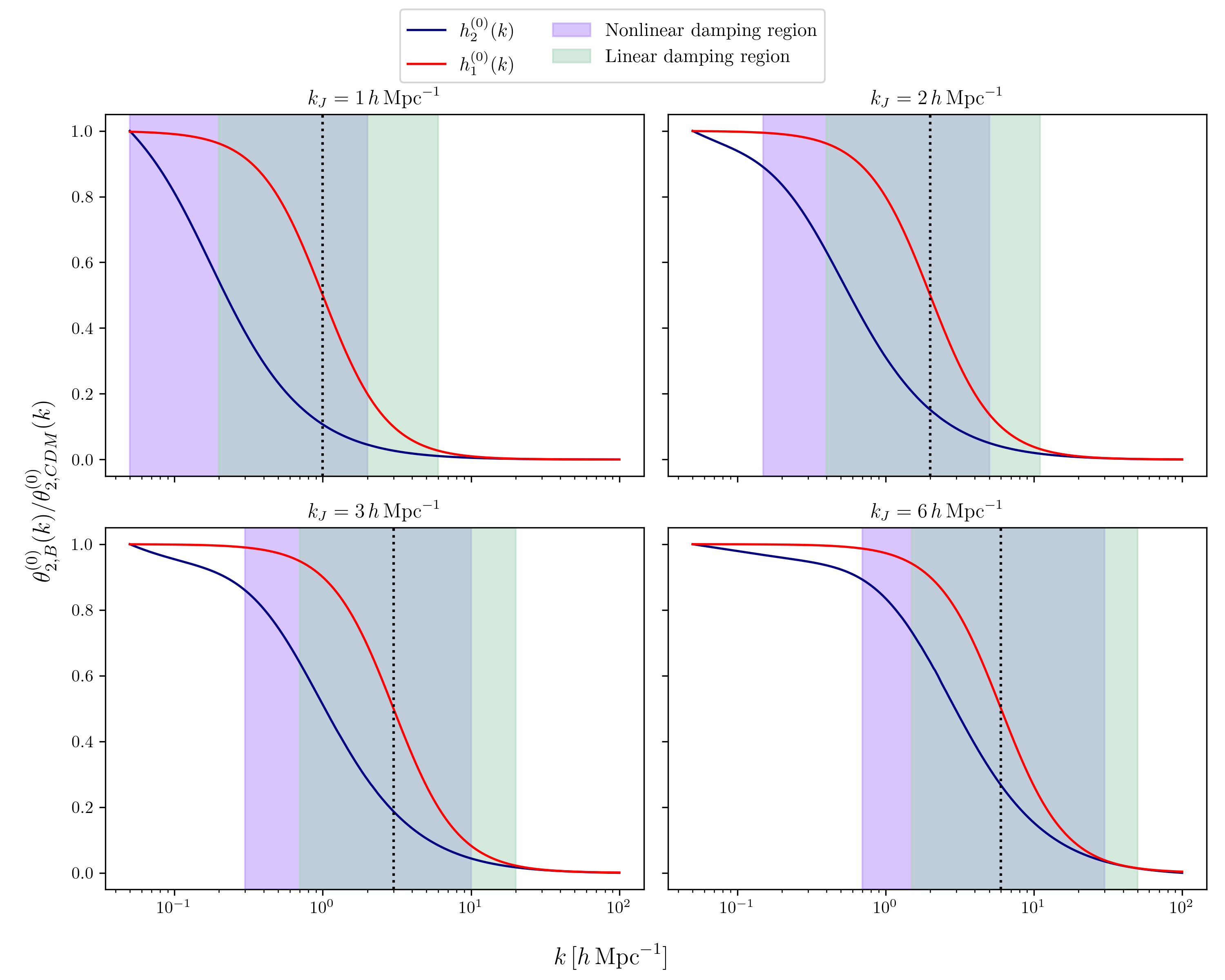}
    \caption{First- and second-order solutions for the baryonic filtering function of the velocity field at zeroth iteration. The red line corresponds to the first-order solution, $h^{(0)}_{_1}(k)$, while the blue line shows the second-order solution, $h^{(0)}_{_2}(k)$. The black dashed lines indicate the characteristic scale $k_{_J}$ for $k_{_J} = 1, 2, 3,$ and $6\, h\mathrm{Mpc}^{-1}$ in each panel. The shaded regions denote the linear (green) and nonlinear (purple) damping regimes.}
    \label{fig:3}
\end{figure}

\begin{figure}
    \centering
    \includegraphics[width=0.97\linewidth]{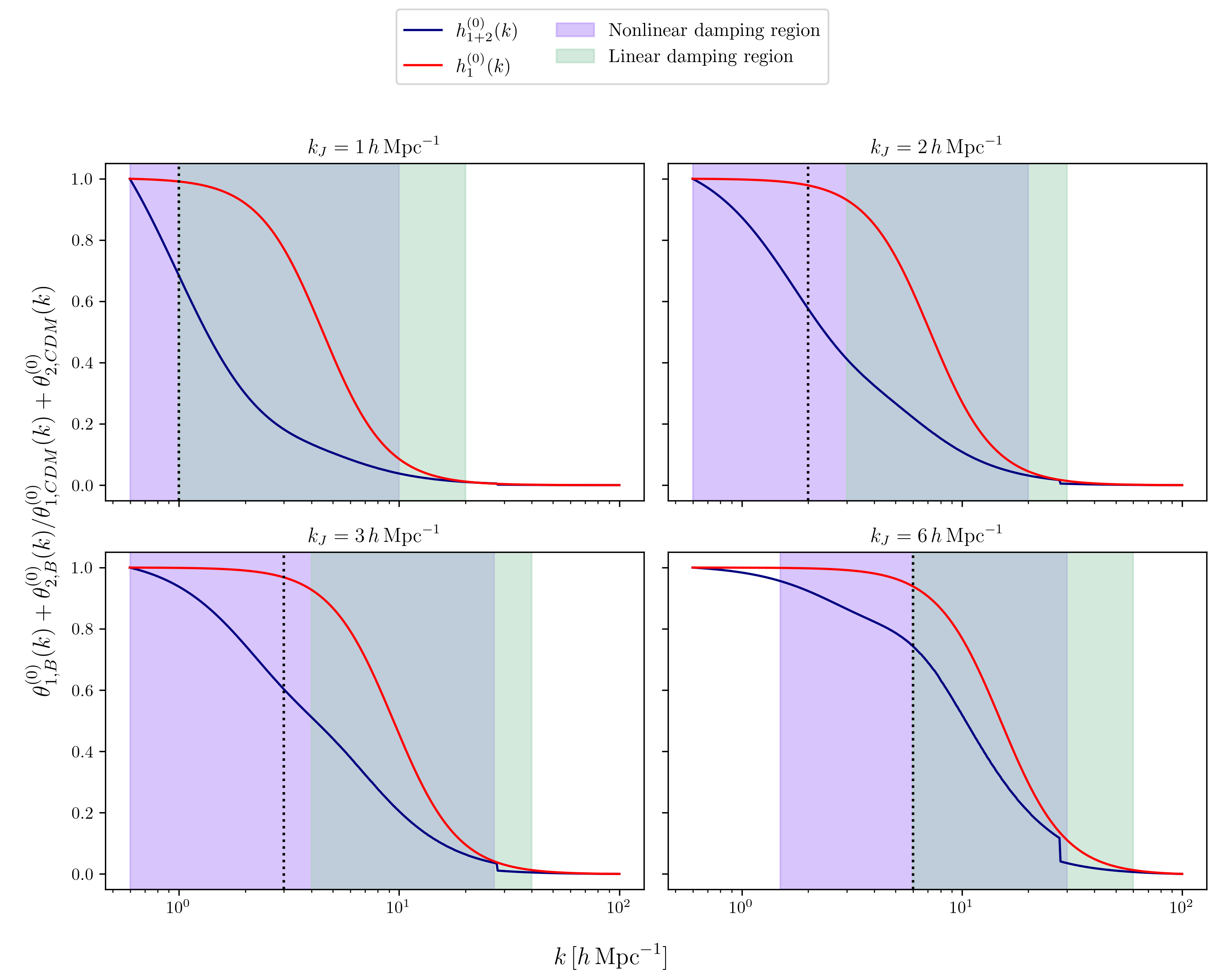}
    \caption{First- and second-order solutions for the baryonic filtering function of the velocity field at zeroth iteration. The red line corresponds to the first-order solution, $h^{(0)}_{_1}(k)$, while the blue line shows the second-order solution, $h^{(0)}_{_{1+2}}(k)$. The black dashed lines indicate the characteristic scale $k_{_J}$ for $k_{_J} = 1, 2, 3,$ and $6\, h\mathrm{Mpc}^{-1}$ in each panel. The shaded regions denote the linear (green) and nonlinear (purple) damping regimes.}
    \label{fig:4}
\end{figure}

Solving the latter expression, \eqref{E62}, numerically—one of the central results of this study—we obtain the behavior shown in Figure \ref{fig:3}, which describes only the evolution of $\tilde{\theta}_{_2}({\boldsymbol{k}},\tau)$ through the velocity filtering function $h^{(0)}_{_2}(k)$. However, to estimate the impact on the velocity power spectrum, we must consider the total contribution.

\begin{align}
    \tilde{\theta}(\boldsymbol{k},\tau)=\tilde{\theta}_{_1}(\boldsymbol{k},\tau)+\tilde{\theta}_{_2}(\boldsymbol{k},\tau) \sim \dfrac{\tilde{\theta}_{_{1,\mbox{\tiny B}}}(\boldsymbol{k})+\tilde{\theta}_{_{2,\mbox{\tiny B}}}(\boldsymbol{k})}{\tilde{\theta}_{_{1,\mbox{\tiny C}}}(\boldsymbol{k})+\tilde{\theta}_{_{2,\mbox{\tiny C}}}(\boldsymbol{k})} \propto h_{_{1+2}}(\boldsymbol{k}). \tag{4.15}\label{E63}
\end{align}
Using \eqref{E17}, we can write
\begin{align}
    & h^{(0)}_{_{1+2}}(\boldsymbol{k}) \propto \dfrac{h^{(0)}_{_1}(\boldsymbol{k})\tilde{\theta}^{(0)}_{_{1,\mbox{\tiny C}}}(\boldsymbol{k})+h^{(0)}_{_2}(\boldsymbol{k})\tilde{\theta}^{(0)}_{_{2,\mbox{\tiny C}}}(\boldsymbol{k})}{\tilde{\theta}^{(0)}_{_{1,\mbox{\tiny C}}}(\boldsymbol{k})+\tilde{\theta}^{(0)}_{_{2,\mbox{\tiny C}}}(\boldsymbol{k})}, \nonumber \\
    & h^{(0)}_{_{1+2}}(k) \propto \dfrac{g^{(0)}_{_1}(k)\sqrt{P_{\mbox{\tiny lin}}(k)}+h^{(0)}_{_2}(k)C(k)\displaystyle{\int_{0}^{\infty}}\displaystyle{\int_{-1}^{1}}dxdr\,r^{3/2}\dfrac{7x-r-6rx^{2}}{(1+r^{2}-2rx)^{3/4}}\overline{T}(k,r,x)}{\sqrt{P_{\mbox{\tiny lin}}(k)}+C(k)\displaystyle{\int_{0}^{\infty}}\displaystyle{\int_{-1}^{1}}dxdr\,r^{3/2}\dfrac{7x-r-6rx^{2}}{(1+r^{2}-2rx)^{3/4}}\overline{T}(k,r,x)} \tag{4.16}\label{E64}
\end{align}
Here, we have employed the relations $h^{(0)}_{_1}(k)=g^{(0)}_{_1}(k)$, $\tilde{\theta}^{(0)}_{_{1,\mbox{\tiny C}}}(\boldsymbol{k})=-\tilde{\delta}^{(0)}_{_{1,\mbox{\tiny C}}}(\boldsymbol{k})$ in the linear regime, together with equations \eqref{E60} and \eqref{E62}. Solving \eqref{E64} numerically, we obtain the result shown in Figure~\ref{fig:4}. A simple inspection of Figures~\ref{fig:3} and \ref{fig:4} shows that, on large scales, the divergence of the baryonic velocity field closely tracks that of the CDM component, indicating that baryons effectively trace the underlying dark matter distribution. Toward smaller scales, however, the velocity divergence becomes increasingly suppressed and asymptotically approaches zero, reflecting the growing dominance of baryonic pressure over gravitational clustering.

As shown in Figures~\ref{fig:3} and \ref{fig:4}, the behavior around the Jeans scale closely parallels that found for density fluctuations (see Figures~\ref{fig:1} and \ref{fig:2}): the divergence of the baryonic velocity field departs from the CDM evolution, and its contribution to the total velocity divergence becomes significant. Notably, the onset of the damping regime occurs at larger scales than in the density field, indicating that fluctuations in the velocity divergence field are more sensitive to pressure effects than their density-field counterparts. This shift underscores the earlier impact of baryonic pressure on velocity fields relative to density perturbations, marking the transition toward a pressure-dominated regime in which structures begin to form in the Universe. In the small-scale regime ($k \gg k_{_J}$), this transition is fully established: baryons no longer trace the CDM component, as baryonic pressure suppresses gravitational clustering, efficiently erasing baryonic modes and driving the velocity divergence toward zero.

Finally, a simple comparison between the linear and second-order growing-mode solutions, $\vert h^{(0)}_{_{1+2}}(k)-h^{(0)}_{_1}(k)\vert$, inferred from Figure~\ref{fig:4}, shows that nonlinear contributions to the coupled baryon--CDM dynamics introduce a maximum difference of approximately $\sim 63.2, 53.2, 42.7$ and $25.0\, \%$ for $k_{_J} =1, 2, 3,$ and $6\, h\mathrm{Mpc}^{-1}$, respectively. These values exceed the corresponding differences inferred from the behavior of $h^{(0)}_{_2}(k)$ by approximately $1\,\% - 6\,\%$. These results, derived from a semi-analytical treatment, suggest that nonlinear effects could play a non-negligible role in shaping the velocity power spectrum across these scales.

\section{Discussion and Outcomes}
\label{Discussion and Outcomes}
This work builds on the methods developed by \cite{Makino:1992} and \cite{Shoji:2009}, extending their contributions to derive an semi-analytical approximation that helps to elucidate the physics of baryonic matter. This approach reveals the role of baryonic pressure in the evolution of fluctuations in the matter–energy field and, in particular, in the velocity field, which is a central focus of this work. This result is obtained through the expressions given in equations \eqref{E56} and \eqref{E62}. In relation to the above, although equations \eqref{E42} and \eqref{E47} are reproduced in a form equivalent to that of \cite{Shoji:2009} for describing the second-order filtering functions, the derivation of equations \eqref{E56} and \eqref{E62} required the implementation of the method proposed by \cite{Makino:1992}, applied to equations \eqref{E42} and \eqref{E47} as described in Section 4, considering only the growing modes and explicitly incorporating the effects of baryonic matter in contrast to \cite{Shoji:2009}. This procedure constitutes a distinctive and novel aspect of our work, as it leads to equations \eqref{E56} and \eqref{E62}, which enable the evolution of the density and velocity fields to be evaluated at low computational cost in the calculation of the associated integrals, as well as to assess their potential impact on observables in both the density and velocity power spectra.

From a preliminary perspective, the results obtained in this study are consistent with theoretical expectations: at large scales, fluctuations can be described by linear theory, while on scales close to the Jeans scale, the competition between baryonic pressure and cosmic expansion becomes relevant to describe the evolution of fluctuations, ultimately leading to the suppression of fluctuations on small scales, as shown in Figures \ref{fig:2} and \ref{fig:4}. We highlight this result as a significant achievement, as it was obtained using a model in which the Jeans wavenumber remains independent of time, and consequently, the sound speed decreases proportionally to $a^{-1/2}$ \cite{Shoji:2009}, yielding results consistent with theoretical predictions. In future work, this approach could be extended to models in which the characteristic scales depend on temperature and, consequently, on time, reflecting the intrinsic connection between the filtering scale and the thermal history of baryonic matter. To better quantify the effects of baryonic temperature on the evolution of fluctuations at small scales, it will be necessary to account for the decoupling between baryonic matter and radiation. This observation suggests that the evolution equations for the JFF, such as equation \eqref{E39}, are highly nontrivial to solve. Indeed, once the filtering scale acquires an explicit time dependence, the semi-analytical treatment developed in this work is no longer straightforward to implement, and the analysis may need to be carried out within a fully numerical framework.

Using the method developed by \cite{Makino:1992}, we initially find that, for baryonic matter, the Jeans filtering function, $g^{(0)}_{_2}(k)$, varies by approximately $5.8\,\%$ at the filtering scale for $k_{_J} = 1\, h\mathrm{Mpc}^{-1}$ due to second-order nonlinear effects induced by baryonic pressure, as shown in the top-left panel of Figure~\ref{fig:1}. Such effects cannot be captured within the framework of linear theory, as illustrated in Figure~\ref{fig:1} for the range of $k_{_J}$ values considered in this work, following the methodology of \cite{Shoji:2009}. However, if we wish to compare our results with those inferred from the matter power spectrum, we must consider the total contribution given in \eqref{E58} and represented in Figure~\ref{fig:2}. In this case, we find that the maximum difference is reduced relative to that obtained when only the contribution $g^{(0)}_{_{2}}(k)$ is considered. For instance, for a filtering scale of $k_{_J} = 1\, h\mathrm{Mpc}^{-1}$, the difference is approximately $0.08\,\%$, while the corresponding values of the maximum difference for the other $k_{_J}$ cases are presented in Section $4$. Consequently, if the amplitude of $\delta(\boldsymbol{k})$ changes by this amount, for $k_{_J} = 1\,h\,\mathrm{Mpc}^{-1}$, the matter power spectrum —defined as the quadratic mean of these amplitudes \cite{Bernardeau:2002}—would scale as $\sim (1.01)^2$, corresponding to an approximately $2.0\,\%$ enhancement on the relevant scales. Furthermore, variations in $\delta(\boldsymbol{k})$ not only alter the local amplitude of $P(\boldsymbol{k})$, but also distort its shape around $k \sim k_{_J}$, as pressure modifies the perturbative kernels $F(\boldsymbol{q}, \boldsymbol{k}-\boldsymbol{q})$ and $G(\boldsymbol{q}, \boldsymbol{k}-\boldsymbol{q})$ —see equations \eqref{E53} and \eqref{E59}— and induces mode coupling across different scales. In this work, these effects are evaluated at low computational cost using equations \eqref{E56} and \eqref{E62}. This implementation builds upon the methods developed by \cite{Makino:1992} and \cite{Shoji:2009}, but differs from both approaches in that the nonlinear baryonic effects are evaluated directly from the fluctuations themselves, providing an alternative framework for estimating such effects.

Another important aspect of this analysis, illustrated in Figures~\ref{fig:1} and \ref{fig:2}, is that the nonlinear solution undergoes damping at smaller spatial scales (i.e., larger wavenumbers) than the linear solution. This behavior indicates that nonlinear mode coupling enhances the impact of baryonic pressure around the Jeans scale, leading to a delayed suppression of second-order fluctuations relative to the linear case. Consequently, the effective filtering scale associated with the nonlinear solution differs from the characteristic linear Jeans scale, $k_{_J}$ \cite{Gnedin:1998, Shoji:2009}. 

Based on our results, we estimate that the ratio $k_{_F}/k_{_J}$ associated with the total contribution $g^{(0)}_{_{1+2}}(k)$ is approximately $1.21, 1.26, 1.69$ and $1.48$ for $k_{_J}=1,2,3,$ and $6\, h\mathrm{Mpc}^{-1}$, respectively \footnote{The effective filtering scale $k_{_F}$ is obtained from an exponential fit to the full scale-dependent suppression function and therefore characterizes the overall damping profile rather than the precise onset of suppression \cite{Gnedin:1998}.}. Our results are qualitatively consistent with the behavior reported in \cite{Shoji:2009} at low redshift, in the sense that the suppression induced by baryonic pressure occurs over a scale-dependent transition region around the Jeans scale rather than exactly at $k_{_J}$. In particular, we find that the effective filtering scale satisfies $k_{_F}/k_{_J} \sim 1.2-1.7$ for the cases considered, indicating a nontrivial scale-dependent transition between the gravitationally dominated regime and the pressure-suppressed regime. A similar shift, $k_{_F}\approx 1.4 k_{_J}$, was reported by \cite{Shoji:2009} using a third-order perturbative calculation of the nonlinear matter power spectrum. However, our approach differs both conceptually and methodologically. Whereas \cite{Shoji:2009} inferred the effective filtering scale indirectly from the nonlinear matter power spectrum, our analysis derives it directly from the scale dependence of the baryonic filtering functions obtained through a second-order semi-analytical treatment. In particular, we focus on the total contribution encoded in $g^{(0)}_{_{1+2}}(k)$, which provides a direct characterization of how baryonic pressure modifies the evolution of fluctuations across different scales. The agreement between our estimates and those reported by \cite{Shoji:2009} is therefore noteworthy, as it emerges from two fundamentally different methodologies. This consistency suggests that the displacement of the effective filtering scale relative to the Jeans scale is a robust physical feature of the coupled baryon--CDM dynamics, rather than an artifact of a particular perturbative approach.

This range of shifts allows us to estimate, within our semi-analytical framework, that the temperature of the baryonic component—which scales as $T_{_B} \propto k_{_J}^{-2}$—may be underestimated \cite{Gnedin:1998}. In particular, our results suggest that the temperature inferred from the Ly$\alpha$ flux power spectrum \cite{Tegmark:2002, Gnedin:1998} may be higher by a factor of $1.46$–$2.85$ relative to the value obtained from linear theory. Similarly, \cite{Shoji:2009} found a factor of approximately $2$. This level of agreement between the two approaches is noteworthy, as both methods point to a shift in the inferred temperature. However, our approach avoids the need to compute the full matter power spectrum and provides an alternative framework for obtaining an approximate estimate of the temperature of the pressure-supported component. Furthermore, it is worth emphasizing that a more precise determination of this shift would enable improved estimates of quantities such as the filtering mass, $M_{_F} \propto k_{_F}^{-3}$, which in our case is reduced by a factor of approximately $1.77$–$3.24$ This result is consistent with the findings of \cite{Shoji:2009}, who reported that a shift of approximately $1.4$ in the filtering scale corresponds to a change of about a factor of $3$ in the filtering mass. A moderate increase in the filtering wavenumber therefore translates into a substantially larger reduction in the filtering mass owing to its cubic dependence. This makes nonlinear effects particularly relevant at low redshift and enables a more accurate description of structure formation in the Universe \cite{Gnedin:1998}.

As a key result of this work, while \cite{Shoji:2009} investigated higher-order baryonic effects through their impact on the nonlinear matter power spectrum, we explicitly compute the second-order velocity-divergence filtering functions $h^{(0)}_{_2}(\boldsymbol{k})$ and $h^{(0)}_{_{1+2}}(\boldsymbol{k})$ using the semi-analytical approximations given in Eqs.~\eqref{E62} and \eqref{E64}. This explicit computation allows us to directly characterize how baryonic pressure modifies the evolution of the velocity field beyond the linear regime. Our results indicate that nonlinear effects alter the scale dependence of the velocity field more strongly than that of the density field, yielding effective filtering-scale ratios $k_{_F}/k_{_J} \approx {1.67, 1.53, 1.70, 2.03}$ for the total contribution $h^{(0)}_{_{1+2}}(\boldsymbol{k})$ and for $k_{_J} = {1,2,3,6}\ h\mathrm{Mpc}^{-1}$, respectively. In particular, as illustrated in Figure~\ref{fig:4}, the nonlinear velocity-divergence filtering function exhibits an earlier and more pronounced suppression than its linear counterpart, indicating that the velocity field is more sensitive to baryonic pressure and nonlinear mode coupling. These results suggest that nonlinear baryonic effects leave a stronger imprint on velocity statistics than on density statistics, highlighting the importance of velocity-based observables as complementary probes of pressure-supported structure formation. Additionally, as a direction for future work, obtaining more general solutions for the evolution of the Jeans filtering function at second order will require incorporating the contribution of the decaying modes in \eqref{E42}. Under such conditions, an oscillatory component is expected to emerge at small scales, consistent with the behavior reported by \cite{Gnedin:1998, Shoji:2009, Fonseca:2024} in first-order analyses. Nevertheless, a complete solution of \eqref{E42} ultimately depends on determining the appropriate initial conditions for \eqref{E39}. Determining these initial conditions would allow for a more realistic treatment of baryonic pressure effects and, consequently, a more robust description of the evolution of baryonic fluctuations. Addressing these aspects would require a dedicated numerical analysis and therefore lies beyond the scope of the present work.

Furthermore, our results suggest that the non-linear suppression found in the velocity-divergence filtering function may have important consequences for velocity-based observables, such as the kinetic Sunyaev–Zel’dovich effect (kSZ) \cite{Sunyaev:1980}. If the magnitude of the non-linear corrections inferred from \eqref{E64} is confirmed at the level of velocity power spectra, this could significantly impact the interpretation of kSZ measurements. A full assessment of these effects and their impact on observables would require computing the corresponding velocity power spectrum, which we leave for future work.

Finally, we conclude that variations in the filtering scale cannot be neglected, and that incorporating baryonic effects through the JFF leads to improved estimates of key physical quantities, such as the filtering mass and the temperature of the baryonic gas, as well as a better assessment of their impact on observables inferred from both the matter and velocity power spectra. Moreover, the semi-analytical framework developed in this work provides an efficient and physically motivated approach for describing the scale-dependent effects of baryonic pressure on cosmological fluctuations. In particular, our analysis of the velocity-divergence field reveals damping regimes shifted toward larger scales relative to those found in the density field, highlighting the enhanced sensitivity of velocity fluctuations to pressure effects and nonlinear mode coupling. Looking forward, this framework can be extended to models in which the Jeans scale evolves with time, allowing for a more realistic description of the thermal history of baryonic matter and its impact on the growth of cosmic structures. Such extensions may improve the accuracy of predictions for both density- and velocity-based observables, thereby strengthening the connection between theoretical models and future cosmological surveys. More broadly, we emphasize that the Jeans filtering function, $g(k,z)$, should not be viewed merely as a tool for modeling cosmic gas. Rather, it provides a physical framework for understanding scale-dependent bias induced by baryonic processes. Future high-precision measurements of galaxy bias, particularly at high redshift and in gas-rich environments, may be sensitive to the imprint of this fundamental filtering mechanism, offering new opportunities to probe the interplay between linear cosmological evolution, baryonic physics, and the nonlinear processes that govern galaxy formation.
\appendix
\section{Standard Equations for an Ideal Fluid}
\label{app1}
We aim to describe the relations that govern the fluid equations. Thus, the standard equations for an ideal fluid, in physical coordinates $(t,{\boldsymbol{r}})$, are given by \cite{Hoyos:2015, Peebles:2020}:
\begin{align}
	\dfrac{\partial{\rho}}{\partial{t}}+\nabla_{_{\boldsymbol{r}}}\cdot(\rho {\boldsymbol{v}}) & = 0,\tag{A.1}\label{A1}\\
	\rho\left[\dfrac{\partial{{\boldsymbol{v}}}}{\partial{t}}+\bigg({\boldsymbol{v}}\cdot \nabla_{_{\boldsymbol{r}}} \bigg){\boldsymbol{v}}\right] & = -\nabla_{_{\boldsymbol{r}}}P-\rho\nabla_{_{\boldsymbol{r}}}\Phi,\tag{A.2}\label{A2}
\end{align}
by using Eulerian coordinates, with $\nabla_{_{\boldsymbol{r}}}=a^{-1}\nabla_{_{\boldsymbol{r}}}$, we can rewrite each term in the expressions \eqref{A1} and \eqref{A2}, as follows:
\begin{align*}
	\left(\dfrac{\partial}{\partial{t}}\right)_{{\boldsymbol{r}}} & = \left(\dfrac{\partial}{\partial{t}}\right)_{{\boldsymbol{x}}}+{\boldsymbol{r}}\dfrac{\partial}{\partial{t}}\left(\dfrac{1}{a}\right)\cdot \dfrac{\partial}{\partial{{\boldsymbol{x}}}}=\left(\dfrac{\partial}{\partial{t}}\right)_{{\boldsymbol{x}}}-\dfrac{\dot{a}}{a}{\boldsymbol{x}}\cdot \nabla_{_{\boldsymbol{x}}},\\
	\left(\dfrac{\partial{\rho}}{\partial{t}}\right)_{{\boldsymbol{r}}} & = \left(\dfrac{\partial{\rho}}{\partial{t}}\right)_{{\boldsymbol{x}}}-\dfrac{\dot{a}}{a}\left({\boldsymbol{x}}\cdot \nabla_{_{\boldsymbol{x}}}\right)\rho,\\
	\nabla_{_{\boldsymbol{r}}}\cdot (\rho {\boldsymbol{v}}) & = \dfrac{1}{a}\nabla_{_{\boldsymbol{x}}}\cdot\bigg[\rho \dot{a}{\boldsymbol{x}}+\rho{\boldsymbol{u}}\bigg]= 3\dfrac{\dot{a}}{a}\rho +\dfrac{\dot{a}}{a}{\boldsymbol{x}}\cdot \nabla_{_{\boldsymbol{x}}}\rho+\dfrac{1}{a}\nabla_{_{\boldsymbol{x}}}\cdot\left(\rho{\boldsymbol{u}}\right),
\end{align*}
where ${\boldsymbol{v}}=\dfrac{d{\boldsymbol{r}}}{dt}=\dot{a}{\boldsymbol{x}}+{\boldsymbol{u}}$. Thus, we get the continuity equation in Eulerian coordinates,
\begin{equation}\tag{A.3}\label{A3}
	\dfrac{\partial{\rho}}{\partial{t}}+3\dfrac{\dot{a}}{a}\rho+\dfrac{1}{a}\nabla_{_{\boldsymbol{x}}}\cdot(\rho{\boldsymbol{u}})=0,
\end{equation}
substituting the density contrast from equation \eqref{E6} into the previous expression, we obtain equation \eqref{E4}. Next, we will rewrite the Euler equation \eqref{A2} in the same way as we did for the mass conservation equation; therefore
\begin{multline*}
	\rho\left[\ddot{a}{\boldsymbol{x}}+\left(\dfrac{\partial{\boldsymbol{u}}}{\partial{t}}\right)-\dfrac{\dot{a}}{a}\bigg({\boldsymbol{x}}\cdot \nabla_{_{\boldsymbol{x}}}\bigg)\left(\dot{a}{\boldsymbol{x}}+{\boldsymbol{u}}\right)\right]\\+\rho\left[\left(\dot{a}{\boldsymbol{x}}+{\boldsymbol{u}}\right)\cdot\dfrac{\nabla_{_{\boldsymbol{x}}}}{a}\right]\left(\dot{a}{\boldsymbol{x}}+{\boldsymbol{u}}\right)=-\dfrac{1}{a}\nabla_{_{\boldsymbol{x}}}P-\dfrac{\rho}{a}\nabla_{_{\boldsymbol{x}}}\left[\phi-\dfrac{1}{2}a\ddot{a}x^{2}\right],
\end{multline*}
operating and simplifying,
\begin{multline*}
	\ddot{a}{\boldsymbol{x}}+\dfrac{\partial{\boldsymbol{u}}}{\partial{t}}-\dfrac{\dot{a}}{a}({\boldsymbol{x}}\cdot\nabla_{_{\boldsymbol{x}}})\dot{a}{\boldsymbol{x}}-\dfrac{\dot{a}}{a}({\boldsymbol{x}}\cdot\nabla_{_{\boldsymbol{x}}}){\boldsymbol{u}}+\dfrac{\dot{a}}{a}\bigg[(\dot{a}{\boldsymbol{x}}+{\boldsymbol{u}})\cdot\nabla_{_{\boldsymbol{x}}}\bigg]{\boldsymbol{x}}+\dfrac{1}{a}\bigg[(\dot{a}{\boldsymbol{x}}+{\boldsymbol{u}})\cdot\nabla_{_{\boldsymbol{x}}}\bigg]{\boldsymbol{u}}\\
	= -\dfrac{1}{\rho a}\nabla_{_{\boldsymbol{x}}}P-\dfrac{1}{a}\nabla_{_{\boldsymbol{x}}}\phi+\dfrac{1}{2a}(2a\ddot{a}){\boldsymbol{x}},
\end{multline*}
where $\Phi\equiv \phi-\dfrac{1}{2}a\ddot{a}x^{2}$ \cite{Bernardeau:2002, Peebles:2020}. We can obtain the Euler equation in conformal time \eqref{E5}. The Hubble parameter can be expressed as $\mathcal{H}(\tau)=\dfrac{d}{d\tau}\ln{a}=\dfrac{da}{dt}=\dot{a}$.

\section{Equations of Motion in Fourier Space}
\label{app2}
In this Appendix, we present the mathematical development used to derive the equations of motion \cite{Shoji:2009}, which we consider important and seldom detailed in the literature. We begin with the continuity equation, applying the Fourier transform to equation \eqref{E4}.
\begin{align}
	\mathcal{F}\bigg\{\dfrac{\partial\delta}{\partial\tau}\bigg\}+\mathcal{F}\{\theta\} & = -\mathcal{F}\bigg\{\nabla_{_{\boldsymbol{x}}}\cdot(\delta \boldsymbol{u})\bigg\},\tag{B.1}\label{B1}
\end{align}
where we have defined the velocity divergence as $\theta=\nabla_{_{\boldsymbol{x}}}\cdot \boldsymbol{u}$, and applied equation \eqref{E14} to each term,
\begin{align}
	\mathcal{F}\left\{ \dfrac{\partial}{\partial \tau} \int\limits_{-\infty}^{\infty} \dfrac{d^3 \boldsymbol{k}}{(2\pi)^3}
	e^{i \boldsymbol{k} \cdot \boldsymbol{x}} \tilde{\delta}(\boldsymbol{k}, \tau) \right\}
	& + \mathcal{F}\left\{ \int\limits_{-\infty}^{\infty} \dfrac{d^3 \boldsymbol{k}}{(2\pi)^3}
	e^{i \boldsymbol{k} \cdot \boldsymbol{x}} \tilde{\theta}(\boldsymbol{k}, \tau) \right\} \nonumber \\
	& = - \mathcal{F} \left\{ \nabla_{\boldsymbol{x}} \cdot
	\int\limits_{-\infty}^{\infty}\int\limits_{-\infty}^{\infty}\dfrac{d^3 \boldsymbol{k}_{_1}}{(2\pi)^3} \dfrac{d^3 \boldsymbol{k}_{_2}}{(2\pi)^3}
	e^{i(\boldsymbol{k}_{_1} + \boldsymbol{k}_{_2})\cdot \boldsymbol{x}}
	\tilde{\delta}(\boldsymbol{k}_{_1}, \tau) \tilde{\boldsymbol{u}}(\boldsymbol{k}_{_2}, \tau) \right\}, \nonumber \\
	\dfrac{\partial}{\partial \tau} \mathcal{F} \left\{
	\int\limits_{-\infty}^{\infty} \dfrac{d^3 \boldsymbol{k}}{(2\pi)^3}
	e^{i \boldsymbol{k} \cdot \boldsymbol{x}} \tilde{\delta}(\boldsymbol{k}, \tau) \right\}
	& + \mathcal{F} \left\{ \int\limits_{-\infty}^{\infty} \dfrac{d^3 \boldsymbol{k}}{(2\pi)^3}
	e^{i \boldsymbol{k} \cdot \boldsymbol{x}} \tilde{\theta}(\boldsymbol{k}, \tau) \right\} \nonumber \\
	& \hspace{-2.7cm} = - \mathcal{F} \left\{ \int\limits_{-\infty}^{\infty}\int\limits_{-\infty}^{\infty} \dfrac{d^3 \boldsymbol{k}_{_1}}{(2\pi)^3}
	\dfrac{d^3 \boldsymbol{k}_{_2}}{(2\pi)^3} \, i(\boldsymbol{k}_{_1} + \boldsymbol{k}_{_2})
	e^{i(\boldsymbol{k}_{_1} + \boldsymbol{k}_{_2})\cdot \boldsymbol{x}}
	\tilde{\delta}(\boldsymbol{k}_{_1}, \tau)
	\cdot \left( -i \dfrac{\boldsymbol{k}_{_2} \tilde{\theta}(\boldsymbol{k}_{_2}, \tau)}{{k}_{_2}^2} \right) \right\}\nonumber.
\end{align}
Recall that the velocity field can be written as $\tilde{\boldsymbol{u}}(\boldsymbol{k},\tau) = -i \boldsymbol{k}\dfrac{\tilde{\theta}(\boldsymbol{k}, \tau)}{k^2}$ \cite{Somogyi:2010}.
\begin{multline}
	\dfrac{\partial}{\partial \tau} \tilde{\delta}(\boldsymbol{k}, \tau)
	+ \tilde{\theta}(\boldsymbol{k}, \tau) = -\int\limits_{-\infty}^{\infty} d^3\boldsymbol{x} e^{-i \boldsymbol{k} \cdot \boldsymbol{x}}\int\limits_{-\infty}^{\infty} \dfrac{d^3 \boldsymbol{k}_{_1}}{(2\pi)^3} \dfrac{d^3 \boldsymbol{k}_{_2}}{(2\pi)^3}
	\dfrac{\boldsymbol{k}_{_2} \cdot (\boldsymbol{k}_{_1} + \boldsymbol{k}_{_2})}{{k}_{_2}^2}
	e^{i(\boldsymbol{k}_{_1} + \boldsymbol{k}_{_2})\cdot \boldsymbol{x}}
	\tilde{\delta}(\boldsymbol{k}_{_1}, \tau) \tilde{\theta}(\boldsymbol{k}_{_2}, \tau).\nonumber
\end{multline}
Organizing the terms to identify the definition of the three-dimensional Dirac delta function $\displaystyle{\int\limits_{-\infty}^{\infty}} \dfrac{d^3 \boldsymbol{x}}{(2\pi)^3} \, e^{i(\boldsymbol{k}_{_1} + \boldsymbol{k}_{_2} - \boldsymbol{k}) \cdot \boldsymbol{x}}=\delta^{\mbox{\tiny D}}(\boldsymbol{k}_{_1} + \boldsymbol{k}_{_2} - \boldsymbol{k})$,
\begin{equation*}
	\dfrac{\partial}{\partial \tau} \tilde{\delta}(\boldsymbol{k}, \tau) + \tilde{\theta}(\boldsymbol{k}, \tau) = -\dfrac{1}{(2\pi)^3} \int\limits_{-\infty}^{\infty} d^3 \boldsymbol{k}_{_1} \, d^3 \boldsymbol{k}_{_2} \,
	\dfrac{\boldsymbol{k}_{_2} \cdot (\boldsymbol{k}_{_1} + \boldsymbol{k}_{_2})}{{k}_{_2}^2} \,
	\delta^{\mbox{\tiny D}}(\boldsymbol{k}_{_1} + \boldsymbol{k}_{_2} - \boldsymbol{k}) \,
	\tilde{\delta}(\boldsymbol{k}_{_1}, \tau) \, \tilde{\theta}(\boldsymbol{k}_{_2}, \tau).
\end{equation*}
Finally, the continuity equation takes the form:
\begin{multline}
	\dfrac{\partial}{\partial{\tau}}\tilde{\delta}(\boldsymbol{k},\tau)+\tilde{\theta}(\boldsymbol{k},\tau)
	=-\dfrac{1}{(2\pi)^{3}}\int\limits_{-\infty}^{\infty}\int\limits_{-\infty}^{\infty}d^{3}{\boldsymbol{k}_{_1}}d^{3}{\boldsymbol{k}_{_2}}\dfrac{\boldsymbol{k}_{_2}\cdot \boldsymbol{k}}{k_{_2}^{2}}\delta^{D}(\boldsymbol{k}_{_1}+\boldsymbol{k}_{_2}-\boldsymbol{k})\tilde{\delta}(\boldsymbol{k}_{_1},\tau)\tilde{\theta}(\boldsymbol{k}_{_2},\tau),\tag{B.2}\label{B2}
\end{multline}
where $\boldsymbol{k}=\boldsymbol{k}_{_1}+\boldsymbol{k}_{_2}$. We now write the Euler equation for CDM. To do so, we compute the divergence of equation \eqref{E9}:
\begin{align*}
	\dfrac{\partial}{\partial\tau}\bigg[\nabla_{_{\boldsymbol{x}}}\cdot\boldsymbol{u}_{\mbox{\tiny C}}\bigg]+\mathcal{H}(\tau)\nabla_{_{\boldsymbol{x}}}\cdot\boldsymbol{u}_{\mbox{\tiny C}}+\dfrac{6}{\tau^{2}}\delta(\boldsymbol{x},\tau) & = -\nabla_{_{\boldsymbol{x}}}\cdot \bigg[(\boldsymbol{u}_{\mbox{\tiny C}}\cdot \nabla_{_{\boldsymbol{x}}})\boldsymbol{u}_{\mbox{\tiny C}}\bigg],\\
	\dfrac{\partial}{\partial\tau}\theta_{\mbox{\tiny{C}}}(\boldsymbol{x},\tau)+\mathcal{H}(\tau)\theta_{\mbox{\tiny{C}}}(\boldsymbol{x},\tau)+\dfrac{6}{\tau^{2}}\delta(\boldsymbol{x},\tau) & = -\nabla_{_{\boldsymbol{x}}}\cdot \bigg[(\boldsymbol{u}_{\mbox{\tiny C}}\cdot \nabla_{_{\boldsymbol{x}}})\boldsymbol{u}_{\mbox{\tiny C}}\bigg].
\end{align*}
Here, we have assumed that $\nabla^2_{\boldsymbol{x}} \phi = \dfrac{6}{\tau^2}$. Applying the Fourier transform to each term, we obtain:
\begin{equation}
	\dfrac{\partial}{\partial{\tau}}\tilde{\theta}_{\mbox{\tiny{C}}}(\boldsymbol{k},\tau)+\mathcal{H}(\tau)\tilde{\theta}_{\mbox{\tiny{C}}}(\boldsymbol{k},\tau)+\dfrac{6}{\tau^{2}}\tilde{\delta}(\boldsymbol{k},\tau) = -\mathcal{F}\bigg\{\nabla_{_{\boldsymbol{x}}}\cdot \bigg[(\boldsymbol{u}_{\mbox{\tiny C}}\cdot \nabla_{_{\boldsymbol{x}}})\boldsymbol{u}_{\mbox{\tiny C}}\bigg]\bigg\}.\tag{B.3}\label{B3}
\end{equation}
Working with the last term on the right-hand side of equation \eqref{B3}, and using the relevant property,
\begin{align*}
	\nabla_{_{\boldsymbol{x}}} \big( \boldsymbol{u}_{\mbox{\tiny C}} \cdot \boldsymbol{u}_{\mbox{\tiny C}} \big) 
	&= \boldsymbol{u}_{\mbox{\tiny C}} \times \big( \nabla_{_{\boldsymbol{x}}} \times \boldsymbol{u}_{\mbox{\tiny C}} \big)
	+ \big( \boldsymbol{u}_{\mbox{\tiny C}} \cdot \nabla_{_{\boldsymbol{x}}} \big) \boldsymbol{u}_{\mbox{\tiny C}}
	+ \boldsymbol{u}_{\mbox{\tiny C}} \times \big( \nabla_{_{\boldsymbol{x}}} \times \boldsymbol{u}_{\mbox{\tiny C}} \big)
	+ \big( \boldsymbol{u}_{\mbox{\tiny C}} \cdot \nabla_{_{\boldsymbol{x}}} \big) \boldsymbol{u}_{\mbox{\tiny C}}, \\
	\nabla_{_{\boldsymbol{x}}} \big( \boldsymbol{u}_{\mbox{\tiny C}} \cdot \boldsymbol{u}_{\mbox{\tiny C}} \big)
	&= 2 \boldsymbol{u}_{\mbox{\tiny C}} \times \boldsymbol{w} + 2 \big( \boldsymbol{u}_{\mbox{\tiny C}} \cdot \nabla_{_{\boldsymbol{x}}} \big) \boldsymbol{u}_{\mbox{\tiny C}},\\
	\big( \boldsymbol{u}_{\mbox{\tiny C}} \cdot \nabla_{_{\boldsymbol{x}}} \big) \boldsymbol{u}_{\mbox{\tiny C}}
	& = \dfrac{1}{2} \nabla_{_{\boldsymbol{x}}} \big( u_{\mbox{\tiny C}}^{2} \big) - \boldsymbol{u}_{\mbox{\tiny C}} \times \boldsymbol{w}.
\end{align*}
It is important to recall that the vorticity, $\boldsymbol{w}=\nabla_{_{\boldsymbol{x}}}\times \boldsymbol{u}$, decays with the expansion, as discussed in \cite{Bernardeau:2002}, therefore
\begin{equation}
	\big(\boldsymbol{u}_{\mbox{\tiny C}}\cdot \nabla_{_{\boldsymbol{x}}}\big)\boldsymbol{u}_{\mbox{\tiny C}} = \dfrac{1}{2}\nabla_{_{\boldsymbol{x}}}\big(u_{\mbox{\tiny C}}^{2}\big).\tag{B.4}\label{B4}
\end{equation}
Applying relation \eqref{B4} to equation \eqref{B3}, we obtain:
\begin{align*}
	& \dfrac{\partial}{\partial{\tau}}\tilde{\theta}_{\mbox{\tiny{C}}}(\boldsymbol{k},\tau)+\mathcal{H}(\tau)\tilde{\theta}_{\mbox{\tiny{C}}}(\boldsymbol{k},\tau)+\dfrac{6}{\tau^{2}}\tilde{\delta}(\boldsymbol{k},\tau)\\
	& = -\dfrac{1}{2}\mathcal{F}\left\{\nabla_{_{\boldsymbol{x}}}\cdot\nabla_{_{\boldsymbol{x}}}\left[\int\limits_{-\infty}^{\infty}\int\limits_{-\infty}^{\infty}\dfrac{d^{3}\boldsymbol{k}_{_1}}{(2\pi)^{3}}\dfrac{d^{3}\boldsymbol{k}_{_2}}{(2\pi)^{3}}e^{i\boldsymbol{k}_{_1}\cdot \boldsymbol{x}}e^{i\boldsymbol{k}_{_2}\cdot \boldsymbol{x}}\tilde{\boldsymbol{u}}_{\mbox{\tiny{C}}}(\boldsymbol{k}_{_1},\tau)\cdot \tilde{\boldsymbol{u}}_{\mbox{\tiny{C}}}(\boldsymbol{k}_{_2},\tau)\right]\right\},\\
	& =  -\dfrac{1}{2}\mathcal{F}\left\{\nabla_{_{\boldsymbol{x}}}\cdot\left[\int\limits_{-\infty}^{\infty}\int\limits_{-\infty}^{\infty}\dfrac{d^{3}\boldsymbol{k}_{_1}}{(2\pi)^{3}}\dfrac{d^{3}\boldsymbol{k}_{_2}}{(2\pi)^{3}}i(\boldsymbol{k}_{_1}+\boldsymbol{k}_{_2})e^{i(\boldsymbol{k}_{_1}+\boldsymbol{k}_{_2})\cdot \boldsymbol{x}}\tilde{\boldsymbol{u}}_{\mbox{\tiny{C}}}(\boldsymbol{k}_{_1},\tau)\cdot \tilde{\boldsymbol{u}}_{\mbox{\tiny{C}}}(\boldsymbol{k}_{_2},\tau)\right]\right\},\\
	& = -\dfrac{1}{2}\mathcal{F}\left\{\nabla_{_{\boldsymbol{x}}}\cdot\left[\int\limits_{-\infty}^{\infty}\int\limits_{-\infty}^{\infty}\dfrac{d^{3}\boldsymbol{k}_{_1}}{(2\pi)^{3}}\dfrac{d^{3}\boldsymbol{k}_{_2}}{(2\pi)^{3}}i(\boldsymbol{k}_{_1}+\boldsymbol{k}_{_2})e^{i(\boldsymbol{k}_{_1}+\boldsymbol{k}_{_2})\cdot \boldsymbol{x}} \right.\right.\\
	& \hspace{6.3cm}\left.\left. \times\left(-i\dfrac{\boldsymbol{k}_{_1}\tilde{\theta}_{\mbox{\tiny{C}}}(\boldsymbol{k}_{_1},\tau)}{k_{_1}^{2}}\right)\cdot \left(-i\dfrac{\boldsymbol{k}_{_2}\tilde{\theta}_{\mbox{\tiny{C}}}(\boldsymbol{k}_{_2},\tau)}{k_{_2}^{2}}\right)\right]\right\},\\
	& = \dfrac{1}{2}\mathcal{F}\left\{\int\limits_{-\infty}^{\infty}\int\limits_{-\infty}^{\infty}\dfrac{d^{3}\boldsymbol{k}_{_1}}{(2\pi)^{3}}\dfrac{d^{3}\boldsymbol{k}_{_2}}{(2\pi)^{3}}i^{2}\dfrac{(\boldsymbol{k}_{_1}+\boldsymbol{k}_{_2})(\boldsymbol{k}_{_1}+\boldsymbol{k}_{_2})\boldsymbol{k}_{_1}\cdot \boldsymbol{k}_{_2}}{k_{_1}^{2}k_{_2}^{2}}e^{i(\boldsymbol{k}_{_1}+\boldsymbol{k}_{_2})\cdot \boldsymbol{x}}\tilde{\theta}_{\mbox{\tiny{C}}}(\boldsymbol{k}_{_1},\tau)\tilde{\theta}_{\mbox{\tiny{C}}}(\boldsymbol{k}_{_2},\tau)\right\},\\
	& = -\dfrac{1}{2}\int\limits_{-\infty}^{\infty}d^{3}\boldsymbol{x}e^{-i\boldsymbol{k}\cdot \boldsymbol{x}}\left[\int\limits_{-\infty}^{\infty}\int\limits_{-\infty}^{\infty}\dfrac{d^{3}\boldsymbol{k}_{_1}}{(2\pi)^{3}}\dfrac{d^{3}\boldsymbol{k}_{_2}}{(2\pi)^{3}}\dfrac{\vert \boldsymbol{k}_{_1}+\boldsymbol{k}_{_2}\vert^{2}\boldsymbol{k}_{_1}\cdot \boldsymbol{k}_{_2}}{k_{_1}^{2}k_{_2}^{2}}e^{i(\boldsymbol{k}_{_1}+\boldsymbol{k}_{_2})\cdot \boldsymbol{x}}\tilde{\theta}_{\mbox{\tiny{C}}}(\boldsymbol{k}_{_1},\tau)\tilde{\theta}_{\mbox{\tiny{C}}}(\boldsymbol{k}_{_2},\tau)\right].
\end{align*}
Finally, for CDM, we have: 
\begin{multline}
	\dfrac{\partial}{\partial{\tau}}\tilde{\theta}_{\mbox{\tiny{C}}}(\boldsymbol{k},\tau)+\mathcal{H}(\tau)\tilde{\theta}_{\mbox{\tiny{C}}}(\boldsymbol{k},\tau)+\dfrac{6}{\tau^{2}}\tilde{\delta}(\boldsymbol{k},\tau)\\
	= -\dfrac{1}{(2\pi)^{3}}\int\limits_{-\infty}^{\infty}\int\limits_{-\infty}^{\infty}d^{3}\boldsymbol{k}_{_1}d^{3}\boldsymbol{k}_{_2}\delta^{\mbox{\tiny D}}(\boldsymbol{k}_{_1}+\boldsymbol{k}_{_2}-\boldsymbol{k})k^{2}\dfrac{\boldsymbol{k}_{_1}\cdot \boldsymbol{k}_{_2}}{2k_{_1}^{2}k_{_2}^{2}}\tilde{\theta}_{\mbox{\tiny{C}}}(\boldsymbol{k}_{_1},\tau)\tilde{\theta}_{\mbox{\tiny{C}}}(\boldsymbol{k}_{_2},\tau).\tag{B.5}\label{B5}
\end{multline}
For baryonic matter, we have a similar equation to \eqref{B5}; however, we must deal with the pressure term appearing in the last term on the left-hand side of equation \eqref{E10}. Therefore \cite{Fonseca:2024,Shoji:2009},
\begin{equation*}
	\dfrac{\nabla_{_{\boldsymbol{x}}}[P_{\mbox{\tiny B}}(\rho_{\mbox{\tiny{B}}})]}{\rho_{\mbox{\tiny{B}}}(\tau,\boldsymbol{x})} = -c_{_s}^{2}(\tau,\boldsymbol{x})\dfrac{\nabla_{_{\boldsymbol{x}}}\delta_{\mbox{\tiny B}}(\tau,\boldsymbol{x})}{1+\delta_{\mbox{\tiny B}}(\tau,\boldsymbol{x})},
\end{equation*}
where $c_{_s}$ is the speed of sound, defined as $c^{2}_{_s}=\dfrac{\partial{P}}{\partial{\rho}}$ \cite{Gnedin:1998, Knobel:2013}. Thus, we obtain the equation:
\begin{multline}
	\dfrac{\partial}{\partial{\tau}}\tilde{\theta}_{\mbox{\tiny{B}}}(\boldsymbol{k},\tau)+\mathcal{H}(\tau)\tilde{\theta}_{\mbox{\tiny{B}}}(\boldsymbol{k},\tau)+\dfrac{6}{\tau^{2}}\tilde{\delta}(\boldsymbol{k},\tau)\\
	= -\dfrac{1}{(2\pi)^{3}}\int\limits_{-\infty}^{\infty}\int\limits_{-\infty}^{\infty}d^{3}\boldsymbol{k}_{_1}d^{3}\boldsymbol{k}_{_2}\delta^{\mbox{\tiny D}}(\boldsymbol{k}_{_1}+\boldsymbol{k}_{_2}-\boldsymbol{k})k^{2}\dfrac{\boldsymbol{k}_{_1}\cdot \boldsymbol{k}_{_2}}{2k_{_1}^{2}k_{_2}^{2}}\tilde{\theta}_{\mbox{\tiny{B}}}(\boldsymbol{k}_{_1},\tau)\tilde{\theta}_{\mbox{\tiny{B}}}(\boldsymbol{k}_{_2},\tau)\\
	+\mathcal{F}\bigg\{\nabla_{_{\boldsymbol{x}}}\cdot \left(c_{_s}^{2}(\tau,\boldsymbol{x})\dfrac{\nabla_{_{\boldsymbol{x}}}\delta_{\mbox{\tiny B}}(\tau,\boldsymbol{x})}{1+\delta_{\mbox{\tiny B}}(\tau,\boldsymbol{x})}\right)\bigg\}.\tag{B.6}\label{B6}
\end{multline}
Developing the last term under the assumption that the sound speed is scale-independent \cite{Shoji:2009}, we obtain the following expression:
\begin{align*}
	\mathcal{F}\bigg\{\nabla_{_{\boldsymbol{x}}}\cdot \left(c_{_s}^{2}(\tau,\boldsymbol{x})\dfrac{\nabla_{_{\boldsymbol{x}}}\delta_{\mbox{\tiny B}}(\tau,\boldsymbol{x})}{1+\delta_{\mbox{\tiny B}}(\tau,\boldsymbol{x})}\right)\bigg\} & =c^{2}_{_s}(\tau)i\boldsymbol{k}\cdot\mathcal{F}\bigg\{\dfrac{\nabla_{_{\boldsymbol{x}}}\delta_{\mbox{\tiny B}}(\tau,\boldsymbol{x})}{1+\delta_{\mbox{\tiny B}}(\tau,\boldsymbol{x})}\bigg\}
\end{align*}
Expanding and using the linearity property of the transform on the right-hand side,
\begin{equation*}
	= c_{_s}^{2}(\tau)i\boldsymbol{k}\cdot \bigg[\mathcal{F}\bigg\{\nabla_{_{\boldsymbol{x}}}\delta_{\mbox{\tiny B}}(\tau,\boldsymbol{x})\bigg\}-\mathcal{F}\bigg\{\bigg(\nabla_{_{\boldsymbol{x}}}\delta_{\mbox{\tiny B}}(\tau,\boldsymbol{x})\bigg)\delta_{\mbox{\tiny B}}(\tau,\boldsymbol{x})\bigg\}
	+\mathcal{F}\bigg\{\bigg(\nabla_{_{\boldsymbol{x}}}\delta_{\mbox{\tiny B}}(\tau,\boldsymbol{x})\bigg)\delta^{2}_{\mbox{\tiny B}}(\tau,\boldsymbol{x})\bigg\}\bigg],
\end{equation*}
For each term in the last expression, for the first and second terms:
\begin{align}
	\mathcal{F}\bigg\{\nabla_{_{\boldsymbol{x}}}\delta_{\mbox{\tiny B}}(\tau,\boldsymbol{x})\bigg\} & = i\boldsymbol{k}\tilde{\delta}_{\mbox{\tiny B}}(\boldsymbol{k},\tau),\tag{B.7}\label{B7}\\
	\mathcal{F}\bigg\{\bigg(\nabla_{_{\boldsymbol{x}}}\delta_{\mbox{\tiny B}}(\tau,\boldsymbol{x})\bigg)\delta_{\mbox{\tiny B}}(\tau,\boldsymbol{x})\bigg\} & = \dfrac{1}{2}	\mathcal{F}\bigg\{\nabla_{_{\boldsymbol{x}}}\bigg(\delta_{\mbox{\tiny B}}(\tau,\boldsymbol{x})\delta_{\mbox{\tiny B}}(\tau,\boldsymbol{x})\bigg)\bigg\}, \nonumber
\end{align}
where we have used $\nabla_{_{\boldsymbol{x}}}\big(\delta\delta\big) = 2\big(\nabla_{_{\boldsymbol{x}}}\delta\big)\delta\rightarrow \big(\nabla_{_{\boldsymbol{x}}}\delta\big)\delta=\dfrac{1}{2}\nabla_{_{\boldsymbol{x}}}\big(\delta\delta\big)$, thus
\begin{align*}
	& = \dfrac{1}{2}\mathcal{F}\left\{\nabla_{_{\boldsymbol{x}}}\left(\int\limits_{-\infty}^{\infty}\int\limits_{-\infty}^{\infty}\dfrac{d^{3}\boldsymbol{k}_{_1}}{(2\pi)^{3}}\dfrac{d^{3}\boldsymbol{k}_{_2}}{(2\pi)^{3}}e^{i(\boldsymbol{k}_{_1}+\boldsymbol{k}_{_2})\cdot \boldsymbol{x}}\tilde{\delta}_{\mbox{\tiny B}}(\boldsymbol{k}_{_1},\tau)\tilde{\delta}_{\mbox{\tiny B}}(\boldsymbol{k}_{_2},\tau)\right)\right\},\\
	& = \dfrac{1}{2}\mathcal{F}\left\{\int\limits_{-\infty}^{\infty}\int\limits_{-\infty}^{\infty}\dfrac{d^{3}\boldsymbol{k}_{_1}}{(2\pi)^{3}}\dfrac{d^{3}\boldsymbol{k}_{_2}}{(2\pi)^{3}}i(\boldsymbol{k}_{_1}+\boldsymbol{k}_{_2})e^{i(\boldsymbol{k}_{_1}+\boldsymbol{k}_{_2})\cdot \boldsymbol{x}}\tilde{\delta}_{\mbox{\tiny B}}(\boldsymbol{k}_{_1},\tau)\tilde{\delta}_{\mbox{\tiny B}}(\boldsymbol{k}_{_2},\tau)\right\},\\
	& = \dfrac{1}{2}\int\limits_{-\infty}^{\infty}d^{3}\boldsymbol{x}e^{-i\boldsymbol{k}\cdot \boldsymbol{x}}\left[\int\limits_{-\infty}^{\infty}\int\limits_{-\infty}^{\infty}\dfrac{d^{3}\boldsymbol{k}_{_1}}{(2\pi)^{3}}\dfrac{d^{3}\boldsymbol{k}_{_2}}{(2\pi)^{3}}i(\boldsymbol{k}_{_1}+\boldsymbol{k}_{_2})e^{i(\boldsymbol{k}_{_1}+\boldsymbol{k}_{_2})\cdot \boldsymbol{x}}\tilde{\delta}_{\mbox{\tiny B}}(\boldsymbol{k}_{_1},\tau)\tilde{\delta}_{\mbox{\tiny B}}(\boldsymbol{k}_{_2},\tau)\right],
\end{align*}
therefore
\begin{multline}
	\mathcal{F}\bigg\{\bigg(\nabla_{_{\boldsymbol{x}}}\delta_{\mbox{\tiny B}}(\tau,\boldsymbol{x})\bigg)\delta_{\mbox{\tiny B}}(\tau,\boldsymbol{x})\bigg\} = \dfrac{i\boldsymbol{k}}{2(2\pi)^{3}}\int\limits_{-\infty}^{\infty}\int\limits_{-\infty}^{\infty}d^{3}\boldsymbol{k}_{_1}d^{3}\boldsymbol{k}_{_2}\delta^{\mbox{\tiny D}}(\boldsymbol{k}_{_1}+\boldsymbol{k}_{_2}-\boldsymbol{k})\\
	\times\tilde{\delta}_{\mbox{\tiny B}}(\boldsymbol{k}_{_1},\tau)\tilde{\delta}_{\mbox{\tiny B}}(\boldsymbol{k}_{_2},\tau).\tag{B.8}\label{B8}
\end{multline}
For the last term (the third term), we follow a similar procedure,
\begin{align*}
	\nabla_{_{\boldsymbol{x}}}\big(\delta\delta^{2}\big) & = \delta \nabla_{_{\boldsymbol{x}}}\big(\delta^{2}\big)+\delta^{2}\nabla_{_{\boldsymbol{x}}}\big(\delta\big) = 3\delta^{2}\big(\nabla_{_{\boldsymbol{x}}}\delta\big) \rightarrow \delta^{2}\big(\nabla_{_{\boldsymbol{x}}}\delta\big) = \dfrac{1}{3}\nabla_{_{\boldsymbol{x}}}\big(\delta\delta^{2}\big).
\end{align*}
Hence, we have:
\begin{multline*}
	\mathcal{F}\bigg\{\dfrac{1}{3}\nabla_{_{\boldsymbol{x}}}\bigg(\delta_{\mbox{\tiny B}}(\boldsymbol{x},\tau)\delta_{\mbox{\tiny B}}(\boldsymbol{x},\tau)\delta_{\mbox{\tiny B}}(\boldsymbol{x},\tau)\bigg)\bigg\}\\= \dfrac{1}{3}\mathcal{F}\left\{\nabla_{_{\boldsymbol{x}}}\int\limits_{-\infty}^{\infty}\int\limits_{-\infty}^{\infty}\int\limits_{-\infty}^{\infty}\dfrac{d^{3}\boldsymbol{k}_{_1}}{(2\pi)^{3}}\dfrac{d^{3}\boldsymbol{k}_{_2}}{(2\pi)^{3}}\dfrac{d^{3}\boldsymbol{k}_{_3}}{(2\pi)^{3}}e^{i(\boldsymbol{k}_{_1}+\boldsymbol{k}_{_2}+\boldsymbol{k}_{_3})\cdot \boldsymbol{x}}\tilde{\delta}_{\mbox{\tiny B}}(\boldsymbol{k}_{_1},\tau)\tilde{\delta}_{\mbox{\tiny B}}(\boldsymbol{k}_{_2},\tau)\tilde{\delta}_{\mbox{\tiny B}}(\boldsymbol{k}_{_3},\tau)\right\},\\
	=\dfrac{1}{3}\int\limits_{-\infty}^{\infty}d^{3}\boldsymbol{x}e^{-i\boldsymbol{k}\cdot \boldsymbol{x}}\left[\int\limits_{-\infty}^{\infty}\int\limits_{-\infty}^{\infty}\int\limits_{-\infty}^{\infty}\dfrac{d^{3}\boldsymbol{k}_{_1}}{(2\pi)^{3}}\dfrac{d^{3}\boldsymbol{k}_{_2}}{(2\pi)^{3}}\dfrac{d^{3}\boldsymbol{k}_{_3}}{(2\pi)^{3}}i(\boldsymbol{k}_{_1}+\boldsymbol{k}_{_2}+\boldsymbol{k}_{_3})\right.\\
	\left.\times e^{i(\boldsymbol{k}_{_1}+\boldsymbol{k}_{_2}+\boldsymbol{k}_{_3})\cdot \boldsymbol{x}}\tilde{\delta}_{\mbox{\tiny B}}(\boldsymbol{k}_{_1},\tau)\tilde{\delta}_{\mbox{\tiny B}}(\boldsymbol{k}_{_2},\tau)\tilde{\delta}_{\mbox{\tiny B}}(\boldsymbol{k}_{_3},\tau)\right.\bigg],
\end{multline*}
with $\boldsymbol{k}=\boldsymbol{k}_{_1}+\boldsymbol{k}_{_2}+\boldsymbol{k}_{_3}$, we obtain,
\begin{multline}
	\mathcal{F}\bigg\{\bigg(\nabla_{_{\boldsymbol{x}}}\delta_{\mbox{\tiny B}}(\boldsymbol{x},\tau)\bigg)\delta^{2}_{\mbox{\tiny B}}(\boldsymbol{x},\tau)\bigg\}\\=\dfrac{i\boldsymbol{k}}{3(2\pi)^{6}}\int\limits_{-\infty}^{\infty}\int\limits_{-\infty}^{\infty}\int\limits_{-\infty}^{\infty}d^{3}\boldsymbol{k}_{_1}d^{3}\boldsymbol{k}_{_2}d^{3}\boldsymbol{k}_{_3}\delta^{\mbox{\tiny D}}(\boldsymbol{k}_{_1}+\boldsymbol{k}_{_2}+\boldsymbol{k}_{_3})
	\tilde{\delta}_{\mbox{\tiny B}}(\boldsymbol{k}_{_1},\tau)\tilde{\delta}_{\mbox{\tiny B}}(\boldsymbol{k}_{_2},\tau)\tilde{\delta}_{\mbox{\tiny B}}(\boldsymbol{k}_{_3},\tau).\tag{B.9}\label{B9}
\end{multline}
Finally, we incorporate \eqref{B7}–\eqref{B9} into \eqref{B6} and obtain the Euler equation for baryonic matter \eqref{E19}.

\section{Equations for Baryonic Matter Fluctuations}
\label{app3}
Here, we present the mathematical details of how equations \eqref{E35} and \eqref{E36} are derived. Then, by applying the JFF, along with equations \eqref{E32} and \eqref{E34}, to equation \eqref{E17}, we obtain the following expression on the left-hand side:
\begin{equation}
	\dfrac{\partial}{\partial{\tau}}\tilde{\delta}_{\mbox{\tiny{B}}}(\boldsymbol{k},\tau) + \tilde{\theta}_{\mbox{\tiny{B}}}(\boldsymbol{k},\tau) = \dfrac{\partial}{\partial{\tau}}\left(\displaystyle{\sum_{n=1}^{\infty}}a^{n}(\tau)g_{_n}(\boldsymbol{k},\tau)\tilde{\delta}_{_{n,\mbox{\tiny C}}}(\boldsymbol{k})\right)
	+\displaystyle{\sum_{n=1}^{\infty}}\dot{a}(\tau)a^{n-1}(\tau)h_{_n}(\boldsymbol{k},\tau)\tilde{\theta}_{_{n,\mbox{\tiny C}}}(\boldsymbol{k}),\nonumber
\end{equation}
thus, we can write:
\begin{multline}
	\displaystyle{\sum_{n=1}^{\infty}}\left[na^{n-1}(\tau)\dot{a}(\tau)\tilde{\delta}_{_{n,\mbox{\tiny C}}}(\boldsymbol{k})g_{_n}(\boldsymbol{k},\tau)+a^{n}(\tau)\tilde{\delta}_{_{n,\mbox{\tiny C}}}(\boldsymbol{k})\dot{g}_{_n}(\boldsymbol{k},\tau)\right]
	+\displaystyle{\sum_{n=1}^{\infty}}\dot{a}(\tau)a^{n-1}(\tau)h_{_n}(\boldsymbol{k},\tau)\tilde{\theta}_{_{n,\mbox{\tiny C}}}(\boldsymbol{k})\\
    =-\dfrac{1}{(2\pi)^{3}}
    \int\limits_{-\infty}^{\infty}\int\limits_{-\infty}^{\infty} d^{3}{\boldsymbol{k}}_{_1}d^{3}{\boldsymbol{k}}_{_2}\dfrac{{\boldsymbol{k}}\cdot {\boldsymbol{k}}_{_2}}{k_{_2}^{2}}\delta^{\mbox{\tiny{D}}}({\boldsymbol{k}}_{_1}+{\boldsymbol{k}}_{_2}-{\boldsymbol{k}})\left(\displaystyle{\sum_{\ell=1}^{\infty}}a^{\ell}(\tau)g_{_\ell}(\boldsymbol{k}_{_1},\tau)\tilde{\delta}_{_{\ell,\mbox{\tiny C}}}(\boldsymbol{k}_{_1})\right) \\
	\times\left(\displaystyle{\sum_{m=1}^{\infty}}\dot{a}(\tau)a^{m-1}(\tau)h_{_m}(\boldsymbol{k}_{_2},\tau)\tilde{\theta}_{_{m,\mbox{\tiny C}}}(\boldsymbol{k}_{_2})\right).\tag{C.1}\label{C1}
\end{multline}
By reorganizing the sums and some terms in the previous expression, we obtain equation \eqref{E35}. Similarly, for equation \eqref{E36}, we follow the same procedure on the left-hand side of equation \eqref{E19}:
\begin{align*}
	& \dfrac{\partial}{\partial{\tau}}\tilde{\theta}_{\mbox{\tiny{B}}}({\boldsymbol{k}},\tau)+\mathcal{H}(\tau)\tilde{\theta}_{\mbox{\tiny{B}}}({\boldsymbol{k}},\tau)+\dfrac{6}{\tau^{2}}\tilde{\delta}({\boldsymbol{k}},\tau)
	= \dfrac{\partial}{\partial{\tau}}\left[\displaystyle{\sum_{n=1}^{\infty}}\dot{a}(\tau)a^{n-1}(\tau)h_{_n}(\boldsymbol{k},\tau)\tilde{\theta}_{_{n,\mbox{\tiny C}}}(\boldsymbol{k})\right]\\
	& +\dfrac{2}{\tau^{2}}\displaystyle{\sum_{n=1}^{\infty}}\dot{a}(\tau)a^{n-1}(\tau)h_{_n}(\boldsymbol{k},\tau)\tilde{\theta}_{_{n,\mbox{\tiny C}}}(\boldsymbol{k})
	+\dfrac{6}{\tau^{2}}\bigg[f_{\mbox{\tiny C}}\tilde{\delta}_{\mbox{\tiny{C}}}(\boldsymbol{k},\tau)+f_{\mbox{\tiny B}}\tilde{\delta}_{\mbox{\tiny{B}}}(\boldsymbol{k},\tau)\bigg]. \nonumber
\end{align*}
It is important to highlight that $\tilde{\delta}(\boldsymbol{k},\tau)=f_{\mbox{\tiny C}}\tilde{\delta}_{\mbox{\tiny{C}}}(\boldsymbol{k},\tau)+f_{\mbox{\tiny B}}\tilde{\delta}_{\mbox{\tiny{B}}}(\boldsymbol{k},\tau)$. Here, $f_{\mbox{\tiny C}}$ and $f_{\mbox{\tiny B}}$ denote dimensionless quantities related to CDM and baryonic densities \cite{Naoz:2005,Shoji:2009}. Thus, 
\begin{multline*}
	= \displaystyle{\sum_{n=1}^{\infty}}\ddot{a}(\tau)a^{n-1}(\tau)h_{_n}(\boldsymbol{k},\tau)\tilde{\theta}_{_{n,\mbox{\tiny C}}}(\boldsymbol{k})+\dot{a}(\tau)\dfrac{\partial}{\partial{\tau}}\left[a^{n-1}(\tau)h_{n}(\boldsymbol{k},\tau)\tilde{\theta}_{_{n,\mbox{\tiny C}}}(\boldsymbol{k})\right]\\
	+\dfrac{2}{\tau^{2}}\displaystyle{\sum_{n=1}^{\infty}}\dot{a}(\tau)a^{n-1}(\tau)h_{_n}(\boldsymbol{k},\tau)\tilde{\theta}_{_{n,\mbox{\tiny C}}}(\boldsymbol{k})+\dfrac{6}{\tau^{2}}\left[f_{\mbox{\tiny C}}\displaystyle{\sum_{n=1}^{\infty}}a^{n}(\tau)\delta_{_{n,\mbox{\tiny C}}}(\boldsymbol{k})
	f_{\mbox{\tiny B}}\displaystyle{\sum_{n=1}^{\infty}}a^{n}(\tau)\delta_{_{n,\mbox{\tiny C}}}(\boldsymbol{k})g_{_n}(\boldsymbol{k},\tau)\right].
\end{multline*}
By grouping terms, we obtain:
\begin{multline}
	= \displaystyle{\sum_{n=1}^{\infty}}\bigg\{\bigg[\ddot{a}(\tau)a^{n-1}(\tau)+\dot{a}^{2}(\tau)a^{n-1}(\tau)(n-1)\bigg]h_{_n}(\boldsymbol{k},\tau)\tilde{\theta}_{_{n,\mbox{\tiny C}}}(\boldsymbol{k})+\dot{a}(\tau)a^{n-1}(\tau)\tilde{\theta}_{_{n,\mbox{\tiny C}}}(\boldsymbol{k})\dot{h}_{_n}(\boldsymbol{k},\tau)\bigg.\\
	\bigg.+\dfrac{2}{\tau^{2}}\dot{a}(\tau)a^{n-1}(\tau)h_{_n}(\boldsymbol{k},\tau)\tilde{\theta}_{_{n,\mbox{\tiny C}}}(\boldsymbol{k})+\dfrac{6}{\tau^{2}}a^{n}(\tau)\bigg[f_{\mbox{\tiny C}}+f_{\mbox{\tiny B}}g_{_n}(\boldsymbol{k},\tau)\bigg]\delta_{_{n,\mbox{\tiny C}}}(\boldsymbol{k})\bigg\}.
	\tag{C.2}\label{C2}
\end{multline} 
Now, on the right-hand side of equation \eqref{E19}, we consider the first term:
\begin{multline}
	= -\dfrac{1}{(2\pi)^{3}}\int\limits_{-\infty}^{\infty}\int\limits_{-\infty}^{\infty} d^{3}{\boldsymbol{k}}_{_1}d^{3}{\boldsymbol{k}}_{_2}k^{2}\dfrac{{\boldsymbol{k}_{_1}}\cdot {\boldsymbol{k}}_{_2}}{2k_{_1}^{2}k_{_2}^{2}}\delta^{\mbox{\tiny{D}}}({\boldsymbol{k}}_{_1}+{\boldsymbol{k}}_{_2}-{\boldsymbol{k}})\\
	\times\displaystyle{\sum_{\ell=1}^{\infty}}\dot{a}(\tau)a^{\ell-1}(\tau)h_{_\ell}(\boldsymbol{k}_{_1},\tau)\tilde{\theta}_{_{\ell,\mbox{\tiny C}}}(\boldsymbol{k}_{_1})\displaystyle{\sum_{m=1}^{\infty}}\dot{a}(\tau)a^{m-1}(\tau)h_{_m}(\boldsymbol{k}_{_2},\tau)\tilde{\theta}_{_{m,\mbox{\tiny C}}}(\boldsymbol{k}_{_2}).\tag{C.3}\label{C3}
\end{multline}
Focusing on the second term on the right-hand side of the same equation,
\begin{multline}
	+c_{_s}^{2}(\tau)k^{2}\left[\displaystyle{\sum_{n=1}^{\infty}}a^{n}(\tau)\tilde{\delta}_{n,\mbox{\tiny C}}(\boldsymbol{k})g_{_n}(\boldsymbol{k},\tau)-\dfrac{1}{2(2\pi)^{3}}\int\limits_{-\infty}^{\infty}\int\limits_{-\infty}^{\infty} d^{3}{\boldsymbol{k}}_{_1}d^{3}{\boldsymbol{k}}_{_2}\delta^{\mbox{\tiny{D}}}({\boldsymbol{k}}_{_1}+{\boldsymbol{k}}_{_2}-{\boldsymbol{k}})\right.\\
	\times\displaystyle{\sum_{\ell=1}^{\infty}}a^{\ell}(\tau)\tilde{\delta}_{_{\ell,\mbox{\tiny C}}}(\boldsymbol{k}_{_1})g_{_\ell}(\boldsymbol{k}_{_1},\tau)\displaystyle{\sum_{m=1}^{\infty}}a^{m}(\tau)\tilde{\delta}_{_{m,\mbox{\tiny C}}}(\boldsymbol{k}_{_2})g_{_m}(\boldsymbol{k}_{2},\tau)\\
	+\dfrac{1}{3(2\pi)^{6}}\int\limits_{-\infty}^{\infty}\int\limits_{-\infty}^{\infty} \int\limits_{-\infty}^{\infty} d^{3}{\boldsymbol{k}}_{_1}d^{3}{\boldsymbol{k}}_{_2}d^{3}{\boldsymbol{k}}_{_3}\delta^{\mbox{\tiny{D}}}({\boldsymbol{k}}_{_1}+{\boldsymbol{k}}_{_2}+{\boldsymbol{k}}_{_3}-{\boldsymbol{k}})\\
	\left.\times\displaystyle{\sum_{\ell=1}^{\infty}}a^{\ell}(\tau)\tilde{\delta}_{_{\ell,\mbox{\tiny C}}}(\boldsymbol{k}_{_1})g_{_\ell}(\boldsymbol{k}_{_1},\tau)\displaystyle{\sum_{m=1}^{\infty}}a^{m}(\tau)\tilde{\delta}_{_{m,\mbox{\tiny C}}}(\boldsymbol{k}_{_2})g_{_m}(\boldsymbol{k}_{_2},\tau)\displaystyle{\sum_{p=1}^{\infty}}a^{p}(\tau)\tilde{\delta}_{_{p,\mbox{\tiny C}}}(\boldsymbol{k}_{_3})g_{_p}(\boldsymbol{k}_{_3},\tau)\right].\tag{C.4}\label{C4}
\end{multline}
Mixing the results from equations \eqref{C2} to \eqref{C4}, we arrive at equation \eqref{E36}.

\section{Equations Describing Baryonic Matter Fluctuations to Second Order}
\label{app4}
To describe the evolution to second order, we set $n=2$ $(\ell=m=1)$ in eqs. \eqref{E35} and \eqref{E36}. For the continuity equation, this gives:
\begin{align*}
	& \bigg[2a(\tau)\dot{a}(\tau)g_{_2}(\boldsymbol{k},\tau)+a^{2}(\tau)\dot{g}_{_2}(\boldsymbol{k},\tau)\bigg]\tilde{\delta}_{_{2,\mbox{\tiny C}}}(\boldsymbol{k})+\dot{a}(\tau)a(\tau)\tilde{\theta}_{_{2,\mbox{\tiny C}}}(\boldsymbol{k})h_{_2}(\boldsymbol{k},\tau)\\
	& = -\dfrac{1}{(2\pi)^{3}}\int\limits_{-\infty}^{\infty}\int\limits_{-\infty}^{\infty} d^{3}\boldsymbol{k}_{_1}d^{3}\boldsymbol{k}_{_2}\dfrac{\boldsymbol{k}\cdot \boldsymbol{k}_{_2}}{k_{_2}^{2}}\delta^{\mbox{\tiny D}}(\boldsymbol{k}_{_1}+\boldsymbol{k}_{_2}-\boldsymbol{k})a(\tau)\dot{a}(\tau) g_{_1}(\boldsymbol{k}_{_1})h_{_1}(\boldsymbol{k}_{_2})\tilde{\delta}_{_{1,\mbox{\tiny C}}}(\boldsymbol{k}_{_1})\tilde{\theta}_{_{1,\mbox{\tiny C}}}(\boldsymbol{k}_{_2}).
\end{align*}
If we divide by $a^{2}(\tau)$, consider $\tilde{\theta}_{1,\mbox{\tiny C}}(\boldsymbol{k})=-\tilde{\delta}_{1, \mbox{\tiny C}}(\boldsymbol{k})$ \cite{Bernardeau:2002} and $h_{_1}(\boldsymbol{k})=g_{_1}(\boldsymbol{k})$ in linear regime, the equation simplifies to:
\begin{align}
	& \dfrac{4}{\tau}g_{_2}(\boldsymbol{k},\tau)\tilde{\delta}_{_{2,\mbox{\tiny C}}}(\boldsymbol{k})+\dot{g}_{_2}(\boldsymbol{k},\tau)\tilde{\delta}_{_{2,\mbox{\tiny C}}}(\boldsymbol{k})+\dfrac{2}{\tau}\tilde{\theta}_{_{2,\mbox{\tiny C}}}(\boldsymbol{k})h_{_2}(\boldsymbol{k},\tau)\nonumber\\
	& = \dfrac{2}{\tau}\dfrac{1}{(2\pi)^{3}}\int\limits_{-\infty}^{\infty}\int\limits_{-\infty}^{\infty} d^{3}\boldsymbol{k}_{_1}d^{3}\boldsymbol{k}_{_2}\dfrac{\boldsymbol{k}\cdot \boldsymbol{k}_{_2}}{k_{_2}^{2}}\delta^{\mbox{\tiny D}}(\boldsymbol{k}_{_1}+\boldsymbol{k}_{_2}-\boldsymbol{k})g_{_1}(\boldsymbol{k}_{_1})g_{_1}(\boldsymbol{k}_{_2})\tilde{\delta}_{_{1,\mbox{\tiny C}}}(\boldsymbol{k}_{_1})\tilde{\delta}_{_{1,\mbox{\tiny C}}}(\boldsymbol{k}_{_2}).\tag{D.1}\label{D1}
\end{align}
This latter equation has the same form as \eqref{E35}. Now, for the Euler equation, we find on the left-hand side:
\begin{align*}
	\bigg[\ddot{a}(\tau)a(\tau)+\dot{a}(\tau)\bigg]\tilde{\theta}_{_{2,\mbox{\tiny C}}}(\boldsymbol{k})h_{_2}(\boldsymbol{k},\tau)+\dot{a}(\tau)a(\tau)\tilde{\theta}_{_{2,\mbox{\tiny C}}}(\boldsymbol{k})\dot{h}_{_2}(\boldsymbol{k},\tau)\\
	+\dot{a}(\tau)a(\tau)\dfrac{2}{\tau}\tilde{\theta}_{_{2,\mbox{\tiny C}}}(\boldsymbol{k})h_{_2}(\boldsymbol{k},\tau)+\dfrac{6}{\tau^{2}}a^{2}(\tau)\bigg[f_{\mbox{\tiny C}}+f_{\mbox{\tiny B}}g_{_2}(\boldsymbol{k},\tau)\bigg]\tilde{\delta}_{_{2,\mbox{\tiny C}}}(\boldsymbol{k}). \tag{D.2}\label{D2}
\end{align*}
For the first and third terms, recall that $\dot{a}(\tau) = \dfrac{2}{\tau}a(\tau)$. Dividing by $a^{2}(\tau)$, we obtain:
\begin{align}
	& \dfrac{1}{a^{2}(\tau)}\bigg[\ddot{a}(\tau)a(\tau)+\dot{a}(\tau)+\dfrac{2}{\tau}\dot{a}(\tau)a(\tau)\bigg]\tilde{\theta}_{_{2,\mbox{\tiny C}}}(\boldsymbol{k})h_{_2}(\boldsymbol{k},\tau) \nonumber\\
	& = \dfrac{1}{a^{2}(\tau)}\left[\dfrac{d}{d{\tau}}\big(\dot{a}(\tau)a(\tau)\big)+\dfrac{2}{\tau}\dot{a}(\tau)a(\tau)\right]\tilde{\theta}_{_{2,\mbox{\tiny C}}}(\boldsymbol{k})h_{_2}(\boldsymbol{k},\tau),\nonumber\\
	& = \dfrac{1}{a^{2}(\tau)}\left[\dfrac{d}{d{\tau}}\left(\dfrac{2}{\tau^{2}}a^{2}(\tau)\right)+\dfrac{2}{\tau}\dot{a}(\tau)a(\tau)\right]\tilde{\theta}_{_{2,\mbox{\tiny C}}}(\boldsymbol{k})h_{_2}(\boldsymbol{k},\tau) = \dfrac{10}{\tau^{2}}\tilde{\theta}_{_{2,\mbox{\tiny C}}}(\boldsymbol{k})h_{_2}(\boldsymbol{k},\tau).\tag{D.3}\label{D3}
\end{align}
For the second term of \eqref{D2}, we proceed as follows:
\begin{equation}
	\dfrac{\dot{a}(\tau)a(\tau)}{a^{2}(\tau)}\tilde{\theta}_{_{2,\mbox{\tiny C}}}(\boldsymbol{k})\dot{h}_{_2}(\boldsymbol{k},\tau) = \dfrac{\dot{a}(\tau)}{a(\tau)}\tilde{\theta}_{_{2,\mbox{\tiny C}}}(\boldsymbol{k})\dot{h}_{_2}(\boldsymbol{k},\tau)= \dfrac{2}{\tau}\tilde{\theta}_{_{2,\mbox{\tiny C}}}(\boldsymbol{k})\dot{h}_{_2}(\boldsymbol{k},\tau). \tag{D.4}\label{D4}
\end{equation}
As for the fourth term of \eqref{D2}, recalling that we are dividing by $a^{2}(\tau)$, we obtain:
\begin{align}
	\dfrac{6}{\tau^{2}}\bigg[f_{\mbox{\tiny C}}+f_{\mbox{\tiny B}}g_{_2}(\boldsymbol{k},\tau)\bigg]\tilde{\delta}_{_{2,\mbox{\tiny C}}}(\boldsymbol{k}) & = \dfrac{6}{\tau^{2}}\bigg[1-f_{\mbox{\tiny B}}+f_{\mbox{\tiny B}}g_{_2}(\boldsymbol{k},\tau)\bigg]\tilde{\delta}_{_{2,\mbox{\tiny C}}}(\boldsymbol{k}),\nonumber\\
	= \dfrac{6}{\tau^{2}}\tilde{\delta}_{_{2,\mbox{\tiny C}}}(\boldsymbol{k})&-\dfrac{6}{\tau^{2}}f_{\mbox{\tiny B}}\tilde{\delta}_{_{2,\mbox{\tiny C}}}(\boldsymbol{k}) +\dfrac{6}{\tau^{2}}f_{\mbox{\tiny B}}g_{_2}(\boldsymbol{k},\tau)\tilde{\delta}_{_{2,\mbox{\tiny C}}}(\boldsymbol{k})-\underbrace{k^{2}c^{2}_{_s}(\tau)\tilde{\delta}_{_{2,\mbox{\tiny C}}}(\boldsymbol{k})g_{_2}(\boldsymbol{k},\tau)}_{\shortstack{\tiny The second term on the\\ \tiny right-hand side of Eq.\eqref{E36}}}\nonumber,\\
	= \dfrac{6}{\tau^{2}}\tilde{\delta}_{_{2,\mbox{\tiny C}}}(\boldsymbol{k})& -\dfrac{6}{\tau^{2}}f_{\mbox{\tiny B}}\tilde{\delta}_{_{2,\mbox{\tiny C}}}(\boldsymbol{k})+\dfrac{6}{\tau^{2}}f_{\mbox{\tiny B}}g_{_2}(\boldsymbol{k},\tau)\tilde{\delta}_{_{2,\mbox{\tiny C}}}(\boldsymbol{k})-\dfrac{6}{\tau^{2}}\dfrac{k^{2}}{k^{2}_{\mbox{\tiny J}}}\tilde{\delta}_{_{2,\mbox{\tiny C}}}(\boldsymbol{k})g_{_2}(\boldsymbol{k},\tau),\nonumber\\
	\dfrac{6}{\tau^{2}}\bigg[f_{\mbox{\tiny C}}+f_{\mbox{\tiny B}}g_{_2}(\boldsymbol{k},\tau)\bigg]\tilde{\delta}_{_{2,\mbox{\tiny C}}}(\boldsymbol{k}) & = \dfrac{6}{\tau^{2}}(1-f_{\mbox{\tiny B}})\tilde{\delta}_{_{2,\mbox{\tiny C}}}(\boldsymbol{k})+\dfrac{6}{\tau^{2}}\left(f_{\mbox{\tiny B}}-\dfrac{k^{2}}{k^{2}_{_J}}\right)\tilde{\delta}_{_{2,\mbox{\tiny C}}}(\boldsymbol{k})g_{_2}(\boldsymbol{k},\tau). \nonumber
	\intertext{Assuming zeroth-order iteration, where $f_{_c} = 1$, we obtain \cite{Shoji:2009}:}
	\dfrac{6}{\tau^{2}}\bigg[f_{\mbox{\tiny C}}+f_{\mbox{\tiny B}}g_{_2}(\boldsymbol{k},\tau)\bigg]\tilde{\delta}_{_{2,\mbox{\tiny C}}}(\boldsymbol{k}) & = \dfrac{6}{\tau^{2}}\tilde{\delta}_{_{2,\mbox{\tiny C}}}(\boldsymbol{k})-\dfrac{6}{\tau^{2}}\dfrac{k^{2}}{k^{2}_{_J}}\tilde{\delta}_{_{2,\mbox{\tiny C}}}(\boldsymbol{k})g_{_2}(\boldsymbol{k},\tau). \tag{D.5}\label{D5}
\end{align}
Finally, we consider only the first and third terms on the right-hand side of \eqref{E36}
\begin{align*}
	&
	-\dfrac{1}{(2\pi)^{3}} \int\limits_{-\infty}^{\infty}\int\limits_{-\infty}^{\infty} d^{3}{\boldsymbol{k}}_{_1} d^{3}{\boldsymbol{k}}_{_2}
	k^{2} \dfrac{{\boldsymbol{k}_{_1}}\cdot {\boldsymbol{k}}_{_2}}{2k_{_1}^{2}k_{_2}^{2}}
	\delta^{\mbox{\tiny D}}({\boldsymbol{k}}_{_1}+{\boldsymbol{k}}_{_2}-{\boldsymbol{k}})
	\dfrac{\dot{a}^{2}(\tau)}{a^{2}(\tau)} h_{_1}(\boldsymbol{k}_{_1},\tau) h_{_1}(\boldsymbol{k}_{_2},\tau)
	\tilde{\theta}_{_{1,\mbox{\tiny C}}}(\boldsymbol{k}_{_1})
	\tilde{\theta}_{_{1,\mbox{\tiny C}}}(\boldsymbol{k}_{_2})\\
	&  \quad -\dfrac{k^{2} c_{_s}^{2}(\tau)}{2(2\pi)^{3}} \int\limits_{-\infty}^{\infty}\int\limits_{-\infty}^{\infty} d^{3}{\boldsymbol{k}}_{_1} d^{3}{\boldsymbol{k}}_{_2} \,
	\delta^{\mbox{\tiny D}}({\boldsymbol{k}}_{_1}+{\boldsymbol{k}}_{_2}-{\boldsymbol{k}}) \,
	\dfrac{{a}^{2}(\tau)}{a^{2}(\tau)} \,
	\tilde{\delta}_{_{1,\mbox{\tiny C}}}(\boldsymbol{k}_{_1})
	\tilde{\delta}_{_{1,\mbox{\tiny C}}}(\boldsymbol{k}_{_2})g_{_1}({\boldsymbol{k}_{_1}},\tau) g_{_1}({\boldsymbol{k}_{_2}},\tau) \\
	& =
	-\dfrac{4}{(2\pi)^{3}\tau^{2}} \int\limits_{-\infty}^{\infty}\int\limits_{-\infty}^{\infty} d^{3}{\boldsymbol{k}}_{_1} d^{3}{\boldsymbol{k}}_{_2} \,
	k^{2} \dfrac{{\boldsymbol{k}_{_1}}\cdot {\boldsymbol{k}}_{_2}}{2k_{_1}^{2}k_{_2}^{2}} \,
	\delta^{\mbox{\tiny D}}({\boldsymbol{k}}_{_1}+{\boldsymbol{k}}_{_2}-{\boldsymbol{k}}) \,
	g_{_1}(\boldsymbol{k}_{_1}) g_{_1}(\boldsymbol{k}_{_2})\tilde{\delta}_{_{1,\mbox{\tiny C}}}(\boldsymbol{k}_{_1})
	\tilde{\delta}_{_{1,\mbox{\tiny C}}}(\boldsymbol{k}_{_2})\\
	& \quad -\dfrac{6k^{2}}{2\tau^{2}k^{2}_{\mbox{\tiny J}} (2\pi)^{3}} \int\limits_{-\infty}^{\infty}\int\limits_{-\infty}^{\infty} d^{3}{\boldsymbol{k}}_{_1} d^{3}{\boldsymbol{k}}_{_2} \,
	\delta^{\mbox{\tiny D}}({\boldsymbol{k}}_{_1}+{\boldsymbol{k}}_{_2}-{\boldsymbol{k}}) \,
	\tilde{\delta}_{_{1,\mbox{\tiny C}}}(\boldsymbol{k}_{_1})
	\tilde{\delta}_{_{1,\mbox{\tiny C}}}(\boldsymbol{k}_{_2})g_{_1}({\boldsymbol{k}_{_1}}) g_{_1}({\boldsymbol{k}_{_2}}), \\
	& =
	\dfrac{4}{(2\pi)^{3} \tau^{2}} \int\limits_{-\infty}^{\infty}\int\limits_{-\infty}^{\infty} d^{3}{\boldsymbol{k}}_{_1} d^{3}{\boldsymbol{k}}_{_2} \,
	\delta^{\mbox{\tiny D}}({\boldsymbol{k}}_{_1}+{\boldsymbol{k}}_{_2}-{\boldsymbol{k}})\left[
	-\dfrac{3}{4} \dfrac{k^{2}}{k^{2}_{_J}}
	- k^{2} \dfrac{{\boldsymbol{k}_{_1}}\cdot {\boldsymbol{k}}_{_2}}{2k_{_1}^{2}k_{_2}^{2}} \right]  \\
	& \qquad \quad \times \tilde{\delta}_{_{1,\mbox{\tiny C}}}(\boldsymbol{k}_{_1})
	\tilde{\delta}_{_{1,\mbox{\tiny C}}}(\boldsymbol{k}_{_2})
	g_{_1}({\boldsymbol{k}_{_1}}) g_{_1}({\boldsymbol{k}_{_2}}).
	\tag{D.6}\label{D6}
\end{align*} 
By putting together \eqref{D3} to \eqref{D6} into \eqref{E36}, we obtain \eqref{E38}. Using \eqref{E37} to \eqref{E38}, we can derive \eqref{E39}, as follows: from \eqref{E37}, we isolate the term
\begin{align*}
	\dfrac{2}{\tau}\tilde{\theta}_{_{2,\mbox{\tiny C}}}(\boldsymbol{k})h_{_2}(\boldsymbol{k},\tau) & 
	= \dfrac{2}{\tau}A_{_2}(\boldsymbol{k}) -\dot{g}_{_2}(\boldsymbol{k},\tau)\tilde{\delta}_{_{2,\mbox{\tiny C}}}(\boldsymbol{k})
	- \dfrac{4}{\tau}g_{_2}(\boldsymbol{k},\tau)\tilde{\delta}_{_{2,\mbox{\tiny C}}}(\boldsymbol{k}),\\
	\tilde{\theta}_{_{2,\mbox{\tiny C}}}(\boldsymbol{k})\dot{h}_{_2}(\boldsymbol{k},\tau) & = -\dfrac{1}{2}\dot{g}_{_2}(\boldsymbol{k},\tau)\tilde{\delta}_{_{2,\mbox{\tiny C}}}(\boldsymbol{k})-\dfrac{\tau}{2}\ddot{g}_{_2}(\boldsymbol{k},\tau)\tilde{\delta}_{_{2,\mbox{\tiny C}}}(\boldsymbol{k})-2\dot{g}_{_2}(\boldsymbol{k},\tau)\tilde{\delta}_{_{2,\mbox{\tiny C}}}(\boldsymbol{k}).
\end{align*}
and substituting it into \eqref{E38},
\begin{multline*}
	\dfrac{10}{\tau^{2}}\left[\dfrac{2}{\tau}A_{_2}(\boldsymbol{k}) -\dot{g}_{_2}(\boldsymbol{k},\tau)\tilde{\delta}_{_{2,\mbox{\tiny C}}}(\boldsymbol{k})
	- \dfrac{4}{\tau}g_{_2}(\boldsymbol{k},\tau)\tilde{\delta}_{_{2,\mbox{\tiny C}}}(\boldsymbol{k})\right]\\
	+\dfrac{2}{\tau}\left[-\dfrac{1}{2}\dot{g}_{_2}(\boldsymbol{k},\tau)\tilde{\delta}_{_{2,\mbox{\tiny C}}}(\boldsymbol{k})-\dfrac{\tau}{2}\ddot{g}_{_2}(\boldsymbol{k},\tau)\tilde{\delta}_{_{2,\mbox{\tiny C}}}(\boldsymbol{k})-2\dot{g}_{_2}(\boldsymbol{k},\tau)\tilde{\delta}_{_{2,\mbox{\tiny C}}}(\boldsymbol{k})\right] + \dfrac{6}{\tau^{2}}\tilde{\delta}_{_{2,\mbox{\tiny C}}}(\boldsymbol{k})
	\\
	- \dfrac{k^{2}}{k_{_J}^{2}}\tilde{\delta}_{_{2,\mbox{\tiny C}}}(\boldsymbol{k})g_{_2}(\boldsymbol{k},\tau) = \dfrac{4}{\tau^{2}}B_{_2}(\boldsymbol{k}). 
\end{multline*}
After a few algebraic simplification steps, we obtain expression \eqref{E39}.

\section{Solution to the Differential Equation for the JFF of the Density Field}
\label{app5}
The homogeneous equation for \eqref{E39} is a Cauchy-Euler equation. Therefore, we can propose the solution $g_{_2}(\boldsymbol{k},\tau)=\tau^{\eta}$. Substituting this into the homogeneous expression, we obtain:
\begin{align*}
	\tau^{2}\ddot{g}^{(0)}_{_2}(\boldsymbol{k},\tau)+10\tau\dot{g}^{(0)}_{_2}(\boldsymbol{k},\tau)+\left[20+6\dfrac{k^{2}}{k^{2}_{_J}}\right]g_{_2}^{(0)}(\boldsymbol{k},\tau) & = 0,\\
	\eta(\eta-1)\tau^{\eta-2}\tau^{2}+10\eta\tau^{\eta-1}\tau+\left[20+6\dfrac{k^{2}}{k^{2}_{_J}}\right]\tau^{\eta} & = 0,\\
	\left(\eta^{2}+9\eta+\left[20+6\dfrac{k^{2}}{k^{2}_{_J}}\right]\right)\tau^{\eta} & = 0.
\end{align*}
By solving the auxiliary equation, we find the roots of the form
\begin{equation*}
	\eta=-\dfrac{9}{2}\left(1\pm \sqrt{1-\dfrac{4}{81}\left(20+\dfrac{k^{2}}{k^{2}_{_J}}\right)}\right).
\end{equation*}
The general solution with the non-homogeneous term can be obtained through the method of undetermined coefficients. Thus, we propose a particular solution of the form $g_{_2}^{(0)}(\boldsymbol{k},\tau)=A$, where $E$ is a constant, by substituting this into \eqref{E39}, we obtain:
\begin{align}
	\left[20+6\dfrac{k^{2}}{k^{2}_{_J}}\right]A & = 6+10\dfrac{A_{_2}(\boldsymbol{k})}{\tilde{\delta}_{_{2,\mbox{\tiny C}}}(\boldsymbol{k})}-\dfrac{4B_{_2}(\boldsymbol{k})}{\tilde{\delta}_{_{2,\mbox{\tiny C}}}(\boldsymbol{k})} \longrightarrow 
	A  = \dfrac{6+10\dfrac{A_{_2}(\boldsymbol{k})}{\tilde{\delta}_{_{2,\mbox{\tiny C}}}(\boldsymbol{k})}-4\dfrac{B_{_2}(\boldsymbol{k})}{\tilde{\delta}_{_{2,\mbox{\tiny C}}}(\boldsymbol{k})}}{20+6\dfrac{k^{2}}{k^{2}_{_J}}}.\tag{E.1}\label{e1}
\end{align}
Therefore, we can write the general solution given in \eqref{E42}. Finally, for this last term, we can rewrite it using the expressions for $A(\boldsymbol{k})$ and $B(\boldsymbol{k})$. To do so, we can follow the next steps in the algebra.
\begin{align*}
    A & = \dfrac{6+10\dfrac{A_{_2}(\boldsymbol{k})}{\tilde{\delta}_{_{2,\mbox{\tiny C}}}(\boldsymbol{k})}-4\dfrac{B_{_2}(\boldsymbol{k})}{\tilde{\delta}_{_{2,\mbox{\tiny C}}}(\boldsymbol{k})}}{20+6\dfrac{k^{2}}{k^{2}_{_J}}}  = \dfrac{1+\dfrac{5A_{_2}(\boldsymbol{k})-2B_{_2}(\boldsymbol{k})}{\tilde{\delta}_{_{2,\mbox{\tiny C}}}(\boldsymbol{k})}}{\dfrac{10}{3}+\dfrac{k^{2}}{k^{2}_{_J}}}.
\end{align*}
We expand only the term $5A_{_2}(\boldsymbol{k})-2B_{_2}(\boldsymbol{k})$ as follows:
\begin{multline}
    5A_{_2}(\boldsymbol{k})-2B_{_2}(\boldsymbol{k})
    = \dfrac{5}{(2\pi)^{3}}\int\limits_{-\infty}^{\infty}\int\limits_{-\infty}^{\infty} d^{3}\boldsymbol{k}_{_1} d^{3}\boldsymbol{k}_{_2}
	\dfrac{\boldsymbol{k}\cdot\boldsymbol{k}_{_2}}{k_{_2}^{2}} 
	\delta^{\mbox{\tiny D}}(\boldsymbol{k}_{_1} + \boldsymbol{k}_{_2} - \boldsymbol{k})
	\, g_{_1}(\boldsymbol{k}_{_1}) g_{_1}(\boldsymbol{k}_{_2}) \tilde{\delta}_{_{1,\mbox{\tiny C}}}(\boldsymbol{k}_{_1})
	\tilde{\delta}_{_{1,\mbox{\tiny C}}}(\boldsymbol{k}_{_2})\\
    -\dfrac{2}{(2\pi)^{3}}
	\int\limits_{-\infty}^{\infty}\int\limits_{-\infty}^{\infty} d^{3}\boldsymbol{k}_{_1} d^{3}\boldsymbol{k}_{_2}
	\delta^{\mbox{\tiny D}}(\boldsymbol{k}_{_1} + \boldsymbol{k}_{_2} - \boldsymbol{k})\left[-\dfrac{3}{4} \dfrac{k^{2}}{k^{2}_{_J}}
	- k^{2} \dfrac{\boldsymbol{k}_{_1} \cdot \boldsymbol{k}_{_2}}{2k_{_1}^{2} k_{_2}^{2}} \right] g_{_1}(\boldsymbol{k}_{_1}) g_{_1}(\boldsymbol{k}_{_2})
	\tilde{\delta}_{_{1,\mbox{\tiny C}}}(\boldsymbol{k}_{_1}) 
	\tilde{\delta}_{_{1,\mbox{\tiny C}}}(\boldsymbol{k}_{_2}),\\
    =\dfrac{1}{(2\pi)^{3}}\int\limits_{-\infty}^{\infty}\int\limits_{-\infty}^{\infty} d^{3}\boldsymbol{k}_{_1} d^{3}\boldsymbol{k}_{_2} \,
	\delta^{\mbox{\tiny D}}(\boldsymbol{k}_{_1} + \boldsymbol{k}_{_2} - \boldsymbol{k})\left[5\dfrac{\boldsymbol{k}\cdot\boldsymbol{k}_{_2}}{k_{_2}^{2}} + \dfrac{3}{2} \dfrac{k^{2}}{k^{2}_{_J}}
	+ k^{2} \dfrac{\boldsymbol{k}_{_1} \cdot \boldsymbol{k}_{_2}}{2k_{_1}^{2} k_{_2}^{2}}\right]\\
    \times g_{_1}(\boldsymbol{k}_{_1}) g_{_1}(\boldsymbol{k}_{_2}) \tilde{\delta}_{_{1,\mbox{\tiny C}}}(\boldsymbol{k}_{_1})\tilde{\delta}_{_{1,\mbox{\tiny C}}}(\boldsymbol{k}_{_2}).\tag{E.2}\label{e2}
\end{multline}
It is possible to demostrate that: $7\tilde{\delta}'_{_{2,\mbox{\tiny C}}}(\boldsymbol{k})=5A_{_2}(\boldsymbol{k})-2B_{_2}(\boldsymbol{k})$ \cite{Shoji:2009},
\begin{multline*}
    \tilde{\delta}'_{_{2,\mbox{\tiny C}}}(\boldsymbol{k}) = \dfrac{5}{7}A_{_2}(\boldsymbol{k})-\dfrac{2}{7}B_{_2}(\boldsymbol{k}),\\
    = \dfrac{1}{(2\pi)^{3}}\int\limits_{-\infty}^{\infty}\int\limits_{-\infty}^{\infty} d^{3}\boldsymbol{k}_{_1} d^{3}\boldsymbol{k}_{_2} \,
	\delta^{\mbox{\tiny D}}(\boldsymbol{k}_{_1} + \boldsymbol{k}_{_2} - \boldsymbol{k})\left[\dfrac{5}{7}\dfrac{\boldsymbol{k}\cdot\boldsymbol{k}_{_2}}{k_{_2}^{2}} + \dfrac{3}{14} \dfrac{k^{2}}{k^{2}_{_J}}
	+ \dfrac{1}{7}k^{2} \dfrac{\boldsymbol{k}_{_1} \cdot \boldsymbol{k}_{_2}}{2k_{_1}^{2} k_{_2}^{2}}\right]\\
    \times g_{_1}(\boldsymbol{k}_{_1}) g_{_1}(\boldsymbol{k}_{_2}) \tilde{\delta}_{_{1,\mbox{\tiny C}}}(\boldsymbol{k}_{_1})\tilde{\delta}_{_{1,\mbox{\tiny C}}}(\boldsymbol{k}_{_2}).\tag{E.3}\label{e3}
\end{multline*}
Therefore, working with the term inside the square brackets in \eqref{e3}, we have
\begin{align*}
    \dfrac{5}{7}\dfrac{\boldsymbol{k}\cdot\boldsymbol{k}_{_2}}{k_{_2}^{2}} + \dfrac{3}{14} \dfrac{k^{2}}{k^{2}_{_J}}
	+ \dfrac{1}{7}k^{2} \dfrac{\boldsymbol{k}_{_1} \cdot \boldsymbol{k}_{_2}}{2k_{_1}^{2} k_{_2}^{2}} & = \dfrac{5}{7}\dfrac{(\boldsymbol{k}_{_1}+\boldsymbol{k}_{_2})\cdot \boldsymbol{k}_{_2}}{k_{_2}^{2}} + \dfrac{3}{14}\dfrac{k^{2}}{k^{2}_{_J}}+\dfrac{1}{7}(k^{2}_{_1}+2\boldsymbol{k}_{_1}\cdot \boldsymbol{k}_{_2}+k^{2}_{_2})\dfrac{\boldsymbol{k}_{_1} \cdot \boldsymbol{k}_{_2}}{2k_{_1}^{2} k_{_2}^{2}},\\
    & = \dfrac{5}{7}\dfrac{\boldsymbol{k}_{_1}\cdot \boldsymbol{k}_{_2}}{k^{2}_{_2}}+\dfrac{5}{7}+ \dfrac{3}{14}\dfrac{k^{2}}{k^{2}_{_J}}+\dfrac{1}{7}\dfrac{\boldsymbol{k}_{_1}\cdot \boldsymbol{k}_{_2}}{k^{2}_{_2}}+\dfrac{2}{7}\dfrac{(\boldsymbol{k}_{_1}\cdot \boldsymbol{k}_{_2})^{2}}{k^{2}_{_1}k^{2}_{_2}}+\dfrac{1}{7}\dfrac{\boldsymbol{k}_{_1}\cdot \boldsymbol{k}_{_2}}{k^{2}_{_1}},\\
    & = \dfrac{6}{7}\dfrac{\boldsymbol{k}_{_1}\cdot \boldsymbol{k}_{_2}}{k^{2}_{_2}}+\dfrac{1}{7}\dfrac{\boldsymbol{k}_{_1}\cdot \boldsymbol{k}_{_2}}{k^{2}_{_2}}+\dfrac{5}{7}+\dfrac{3}{14}\dfrac{k^{2}}{k^{2}_{_J}}+\dfrac{2}{7}\dfrac{(\boldsymbol{k}_{_1}\cdot \boldsymbol{k}_{_2})^{2}}{k^{2}_{_1}k^{2}_{_2}}.
    \intertext{Adding and subtracting the term $\dfrac{5}{14}\dfrac{\boldsymbol{k}_{_1}\cdot \boldsymbol{k}_{_2}}{k^{2}_{_2}}$}
    & = \dfrac{1}{2}\dfrac{\boldsymbol{k}_{_1}\cdot \boldsymbol{k}_{_2}}{k^{2}_{_2}}+\dfrac{1}{2}\dfrac{\boldsymbol{k}_{_1}\cdot \boldsymbol{k}_{_2}}{k^{1}_{_2}}+\dfrac{5}{7}+\dfrac{3}{14}\dfrac{k^{2}}{k^{2}_{_J}}+\dfrac{2}{7}\dfrac{(\boldsymbol{k}_{_1}\cdot \boldsymbol{k}_{_2})^{2}}{k^{2}_{_1}k^{2}_{_2}},\\
    \dfrac{5}{7}\dfrac{\boldsymbol{k}\cdot\boldsymbol{k}_{_2}}{k_{_2}^{2}} + \dfrac{3}{14} \dfrac{k^{2}}{k^{2}_{_J}}
	+ \dfrac{1}{7}k^{2} \dfrac{\boldsymbol{k}_{_1} \cdot \boldsymbol{k}_{_2}}{2k_{_1}^{2} k_{_2}^{2}} & = \underbrace{\dfrac{5}{7}+\dfrac{2}{7}\dfrac{(\boldsymbol{k}_{_1}\cdot \boldsymbol{k}_{_2})^{2}}{k^{2}_{_1}k^{2}_{_2}}+\dfrac{1}{2}\dfrac{(\boldsymbol{k}_{_1}\cdot \boldsymbol{k}_{_2})(k^{2}_{_1}+k^{2}_{_2})}{k^{2}_{_1}k^{2}_{_2}}}_{F^{(s)}_{2}(\boldsymbol{k}_{_1},\boldsymbol{k}_{_2})}+\dfrac{3}{14}\dfrac{k^{2}}{k^{2}_{_J}}.
\end{align*}
Finally, we find the expression
\begin{multline}
    \tilde{\delta}'_{_{2,\mbox{\tiny C}}}(\boldsymbol{k}) = \dfrac{1}{(2\pi)^{3}}\int\limits_{-\infty}^{\infty}\int\limits_{-\infty}^{\infty} d^{3}\boldsymbol{k}_{_1} d^{3}\boldsymbol{k}_{_2}
	\delta^{\mbox{\tiny D}}(\boldsymbol{k}_{_1} + \boldsymbol{k}_{_2} - \boldsymbol{k})\left[F^{(s)}_{_2}(\boldsymbol{k}_{_1},\boldsymbol{k}_{_2})+\dfrac{3}{14}\dfrac{k^{2}}{k^{2}_{_J}}\right]\\
    \times g_{_1}(\boldsymbol{k}_{_1}) g_{_1}(\boldsymbol{k}_{_2}) \tilde{\delta}_{_{1,\mbox{\tiny C}}}(\boldsymbol{k}_{_1})\tilde{\delta}_{_{1,\mbox{\tiny C}}}(\boldsymbol{k}_{_2}).\tag{E.4}\label{e4}
\end{multline}

\acknowledgments
This research was supported by the Universidad Nacional de Colombia through a Faculty of Sciences Grant, and by the Universidad Antonio Nariño through Grant VCTI2024211. D.F.F. also expresses his sincere gratitude to the ICTP South American Institute for Fundamental Research (ICTP-SAIFR) for the financial support provided to attend the V Joint ICTP-Trieste/ICTP-SAIFR School on Cosmology, where several of the ideas developed in this work were first presented and discussed. We acknowledge the use of the open-source Python packages NumPy (https://numpy.org/), SymPy (https://www.sympy.org/), and Matplotlib (https://matplotlib.org/). The authors would like to thank the referee for their constructive comments and valuable suggestions, which helped improve the manuscript and motivated additional results.

\end{document}